\documentclass[pra,showpacs,twocolumn]{revtex4}
\usepackage{amsmath}
\usepackage{amssymb}
\usepackage{amscd}
\usepackage{amsthm}
\usepackage{wasysym}
\usepackage[ansinew]{inputenc}
\usepackage[T1]{fontenc}
\usepackage{ae,aecompl}
\usepackage{dsfont}
\usepackage[dvips]{graphicx}
\usepackage{pslatex}
\DeclareGraphicsExtensions{.eps}

\newcommand{\be}{\begin{equation}}
\newcommand{\ee}{\end{equation}}

\newcommand{\nn}{\nonumber}

\newcommand{\rmi}{{\rm i }}
\newcommand{\rme}{{ \rm e }}
\newcommand{\rmd}{{ \rm d }}

\newcommand{\J}{v}
\newcommand{\U}{c}
\newcommand{\g}{g}
\newcommand{\s}{s}

\newcommand{\p}{\partial}

\newcommand{\cP}{\mathcal{P}}
\newcommand{\cH}{\mathcal{H}}
\newcommand{\cT}{\mathcal{T}}

\def\half{\tfrac{1}{2}}
\def\RR{\mathds{R}}

\newcommand{\ra}{\rangle}

\newcommand{\ba}{\begin{eqnarray}}
\newcommand{\ea}{\end{eqnarray}}
\newcommand{\rf}[1]{(\ref{#1})}
\def\ra{\rangle}

\begin{document}
\bibliographystyle{apsrev}
\title[]{Quantum Classical Correspondence for a non-Hermitian Bose-Hubbard Dimer}
\author{Eva-Maria Graefe$^{1,2}$}
\author{Hans J\"urgen Korsch$^2$}
\author{Astrid Elisa Niederle$^{2,3}$}
\address{${}^1$ School of Mathematics, University of Bristol, Bristol, BS8 1TW, UK\\
${}^2$ FB Physik, TU Kaiserslautern, D--67653 Kaiserslautern, Germany\\
${}^3$ Theoretical Physics, Saarland University, D--66041 Saarbr\"ucken, Germany}
\date{\today}

\begin{abstract} 
We investigate the many-particle and mean-field correspondence for a non-Hermitian $N$-particle Bose-Hubbard dimer where a complex onsite energy describes an effective decay from one of the modes. Recently a generalized mean-field approximation for this non-Hermitian many-particle system yielding an alternative complex nonlinear Schr\"odinger equation was introduced. Here we give details of this mean-field approximation and show that the resulting dynamics can be expressed in a generalized canonical form that includes a metric gradient flow. 
The interplay of nonlinearity and non-Hermiticity introduces a qualitatively new behavior to the mean-field dynamics: The presence of the non-Hermiticity promotes the self-trapping transition, while damping the self-trapping oscillations, and the nonlinearity introduces a strong sensitivity to the initial conditions in the decay of the normalization. Here we present a complete characterization of the mean-field dynamics and the fixed point structure. We also investigate the full many-particle dynamics, which shows a rich variety of breakdown and revival as well as tunneling phenomena on top of the mean-field structure. 
\end{abstract}

\pacs{03.65.-w, 03.75.Kk, 05.30.Jp}

\maketitle

\section{Introduction}
In the past decade the theoretical investigation of Bose-Einstein condensates 
led to a widespread interest in nonlinear quantum theories such as the 
nonlinear Schr\"odinger equation of Gross-Pitaevskii type \cite{Pita03}. 
In contrast to nonlinear generalizations of quantum mechanics 
at a fundamental level \cite{Wein89}, in the context of ultracold atoms the 
nonlinearity arises as the consequence of an effective single particle 
description in a mean-field approximation of an initially linear many-particle 
quantum system. This limit is formally similar to the classical limit of standard 
single particle quantum mechanics. In this spirit the mean-field approximation is often formulated as a replacement of the particle creation and annihilation 
operators with c-numbers that describe the amplitudes of the effective single 
particle wave function. The time evolution is then governed by canonical equations of motion based on the fact that nonlinear as well as linear quantum dynamics can be formulated as special cases of classical canonical dynamics on the phase space 
of pure states, the projective Hilbert space. Thus, for Hermitian systems, 
the correspondence between the many-particle description and the 
mean-field approximation can be investigated in analogy with 
the usual quantum classical correspondence for a single particle system \cite{Fran00,Moss06,Kolo07a,Trim08,Trim09,Hill09}. In particular the Bose-Hubbard dimer that models $N$ bosons in only two modes, became a standard example many of whose features can be analytically understood \cite{Milb97,Steel98,Fran01,Wu03,Mahm05,Wu06,07semiMP,Bouk09}.  

For both many-particle and single-particle quantum mechanics, 
the Hamiltonian is usually demanded to be Hermitian 
for the description of closed systems. 
However, there is a rapidly growing interest 
in the use of non-Hermitian Hamiltonians arising from different 
areas. The first is the field of open quantum systems 
where complex energies with negative imaginary parts are 
used to describe an overall probability decrease that models 
decay, transport or scattering phenomena 
(see, e.g., \cite{Gamo28,Datt90b,Okol03,Mois98,Berr04,09nhclass} 
and references therein). Although in most cases these non-Hermitian 
Hamiltonians are introduced heuristically, they can be derived in a mathematically satisfactory way starting from a system coupled to a continuum of states (see, e.g., 
\cite{Maha69,Okol03} and references cited therein). 
It is interesting to note that within the past decade a somewhat
orthogonal motivation also generated
considerable interest in the physics of non-Hermitian operators.
This is based on the observation that a class of non-Hermitian
Hamiltonians respecting  a certain antilinear symmetry, often
referred to as $\cP\cT$-symmetry, yields purely real
eigenvalues in some parameter regions \cite{Bend98}. Further, with the
introduction of an appropriate inner product they can be used to
define a fully consistent quantum theory for closed systems \cite{Bend02b}. 
The so-called $\cP\cT$-symmetric Hamiltonians
have been the subject of extensive studies in the past decade 
see, e.g. \cite{PT06a}. Recently there is increasing interest in 
$\cP\cT$-symmetric systems in the context of optics \cite{Elga07,Muss08,Klai08,Long09,Bend10,West10,Scho10}, 
where first experimental results could be obtained \cite{Guo09,Ruet10}. 
Non-Hermitian quantum dynamics differ drastically from 
their unitary counterparts, and their generic features are far from being fully understood. In particular, the investigation of the quantum classical 
correspondence for non-Hermitian systems is only at its beginning \cite{09nhclass,Scho05,Bend10b,Jone10,Keat06,Kopp10}.

Recently, considerable attention has been paid to non-Hermitian extensions 
of the Gross-Pitaevskii equation including an imaginary potential, in the context 
of scattering and transport behavior of BECs
\cite{Mois05,Mois03,Schl06a,Paul07b,08nlLorentz,09ddshell}, as well as the
implications of decay or leaking boundary conditions in partially open traps
\cite{06nlnh,Livi06,Fran07,Ng08}. The corresponding 
non-Hermitian nonlinear Schr\"odinger equations have been
formulated in an \textit{ad hoc} manner as a complex generalization of the
mean-field description in the Hermitian case. However, for a many-particle system 
the generalization of the mean-field approximation in the presence of a complex potential is nontrivial and intimately related to the semiclassical limit of non-Hermitian single particle quantum theories. Recently, a derivation starting from 
a non-Hermitian many-particle system has been presented in \cite{08nhbh_s} 
for an open Bose-Hubbard dimer \cite{Hill06,08PT} described by the
Hamiltonian
\begin{eqnarray}
\hat{\cal H}&=&\epsilon(\hat a_1^\dagger\hat a_1-\hat
a_2^\dagger\hat a_2 )-2\rmi\gamma\hat a_1^\dagger\hat a_1 + \J
(\hat a_1^\dagger \hat a_2 + \hat a_1 \hat a_2^\dagger) \nn\\
&&+\frac{\U
}{2}(\hat a_1^\dagger\hat a_1  - \hat a_2^\dagger\hat a_2)^2.
\label{eqn-BH-Hamiltonian}
\end{eqnarray}
Here $\hat a_j$ and $\hat a_j^\dagger$ are bosonic annihilation 
and creation operators for mode $j$,  $\J $ is the coupling constant, and $\U $ is the 
strength of the onsite interaction. For convenience we assume 
both $v$ and $\U$ to be positive in the following. 
The system is opened by making the onsite energy of 
mode 1 complex. Note that the expectation value of the particle number 
$\hat N=\hat a_1^\dagger\hat a_1+\hat a_2^\dagger\hat a _2$ 
is conserved and the opening describes a decay of the 
overall probability encoded in the normalization of the 
many-particle wave function. A direct experimental 
realization of the Hamiltonian (\ref{eqn-BH-Hamiltonian}) 
can be achieved by using ultracold bosonic atoms in 
a finite double-well trap, confined by a small 
tunneling barrier on one side and an approximately infinite barrier 
on the other. The parameter $\gamma$ can then be tuned in 
the experiment by lowering or raising the tunnel barrier.
An imaginary energy shift $\hat \cH=\hat\cH_{\cP\cT}-\rmi\gamma \hat N$ transforms this non-Hermitian Bose-Hubbard dimer 
into a model that is $\cP\cT$-symmetric in the unbiased case ($\epsilon=0$):
\begin{eqnarray}
\hat \cH_{\cP\cT} &=& (\epsilon-\rmi\gamma)(\hat
a_1^{\dagger}\hat a_1 -\hat a_2^{\dagger}\hat a_2) +
v(\hat a_1^{\dagger}\hat a_2 + \hat a_2^{\dagger}\hat
a_1)\nn\\
&& + \frac{c}{2} ( \hat a_1^{\dagger}\hat a_1 - \hat
a_2^{\dagger}\hat a_2)^2.\label{Ham1}
\end{eqnarray}

In the present article we provide a detailed description of the 
mean-field approximation for this non-Hermitian 
many-particle system introduced in \cite{08nhbh_s}. 
Furthermore, we show that the mean-field dynamics 
can be formulated in terms of 
generalized canonical evolution equations on 
the classical phase space given by the Bloch sphere. 
These equations consist of a combination of a familiar 
Hamiltonian flow and an additional gradient flow that 
accounts for damping. This structure was recently introduced 
as the classical limit of non-Hermitian quantum theories on a flat phase 
space \cite{09nhclass} and it is likely that it holds for 
arbitrary phase space geometries. It is closely related also 
to canonical formulations of classical dissipative dynamics that have been investigated in the past two decades \cite{Kauf84,Morr86,Bloc92,Bloc96,Holm07}. The full many-particle dynamics can be understood 
as quantum behavior on top of the generalized classical structure, 
incorporating breakdown and revival phenomena as well as tunneling effects. 

The article is organized as follows: In section \ref{sec-nherm2x2} 
we provide the background of the non-Hermitian single particle two-level 
system, and introduce a renormalized Bloch representation for the dynamics. 
Further, some concepts of non-Hermitian quantum mechanics 
that are of relevance in the following are provided. In section 
\ref{sec-nhermBH} the non-Hermitian Bose-Hubbard dimer is introduced 
as a many-boson generalization of the non-Hermitian two-level system. 
In section \ref{sec-genMF} we review the generalized mean-field approximation introduced in \cite{08nhbh_s} and show that it can be expressed in a canonical form of dissipative classical mechanics suggested in \cite{09nhclass}. We analyze the resulting mean-field dynamics in detail in section \ref{sec-MF-dyn} and 
compare it to the full many-particle system in section \ref{sec-MP-MF}. 
We end with a brief summary and an outlook. 

\section{The non-Hermitian two-level system}
\label{sec-nherm2x2}
The non-Hermitian Bose-Hubbard dimer \rf{eqn-BH-Hamiltonian} 
can be regarded as an $N$ boson 
generalization of a single particle two-level system with an imaginary energy term 
modeling a decay from one of the states, which can be described 
by the $2\times2$ Hamiltonian 
\begin{equation}\label{eqn-2times2-nherm}
\hat H=\left(\begin{array}{cc}
\epsilon-2\rmi\gamma & v\\
v & -\epsilon
\end{array}\right),\quad \epsilon,v,\gamma\in\RR,\ \gamma>0.
\end{equation}
Here the state with the lower onsite energy is assumed to
be stable and the other one to decay with a width $\gamma$.
The general case of two decaying states differs from this model 
only by an imaginary energy offset. Despite its simplicity 
the system \rf{eqn-2times2-nherm} incorporates 
many of the generic features of non-Hermitian quantum 
mechanics and was the subject of many studies in the past (see, e.g., \cite{Mond93,Heis00,Berr04,03crossing,Guen07}). 
In this section we briefly review some features of 
this system and a related $\cP\cT$-symmetric model. Furthermore, we
present a less familiar representation of the Bloch dynamics. 

The non-Hermitian two-level system
(\ref{eqn-2times2-nherm}) is intimately related to a prominent 
$\cP\cT$-symmetric toy-model. Applying a constant energy shift 
$\hat H\to \hat H+\rmi\gamma{\mathds 1}$, that is,
$\psi\to\psi\rme^{\gamma t}$, the system
(\ref{eqn-2times2-nherm}) can be mapped onto the Hamiltonian 
\begin{equation}\label{eqn-2times2-PT}
\hat H_{\cP\cT}=\left(\begin{array}{cc}
\epsilon-\rmi\gamma & v\\
v & -\epsilon+\rmi\gamma
\end{array}\right),
\end{equation}
which is $\cP\cT$-symmetric for $\epsilon=0$.
Introducing the discrete parity operator
\begin{equation}
\cP=\left(\begin{array}{cc}
0 & 1\\
1 & 0
\end{array}\right)
\end{equation}
that interchanges the two levels and the time reversal operator
$\cT:\rmi\to-\rmi$ that performs a complex conjugation, we see that
$\hat H$ commutes with $\cP\cT$, whereas it 
commutes neither with $\cP$ nor with $\cT$ alone. 
Although in the general case for $\epsilon\neq0$ 
the Hamiltonian \rf{eqn-2times2-PT} is not $\cP\cT$-symmetric, 
to distinguish it from the purely decaying 
system \rf{eqn-2times2-nherm} we shall refer to it as $\cP\cT$-symmetric in 
the following.

\begin{figure}[tb]
\centering
\includegraphics[width=4cm]{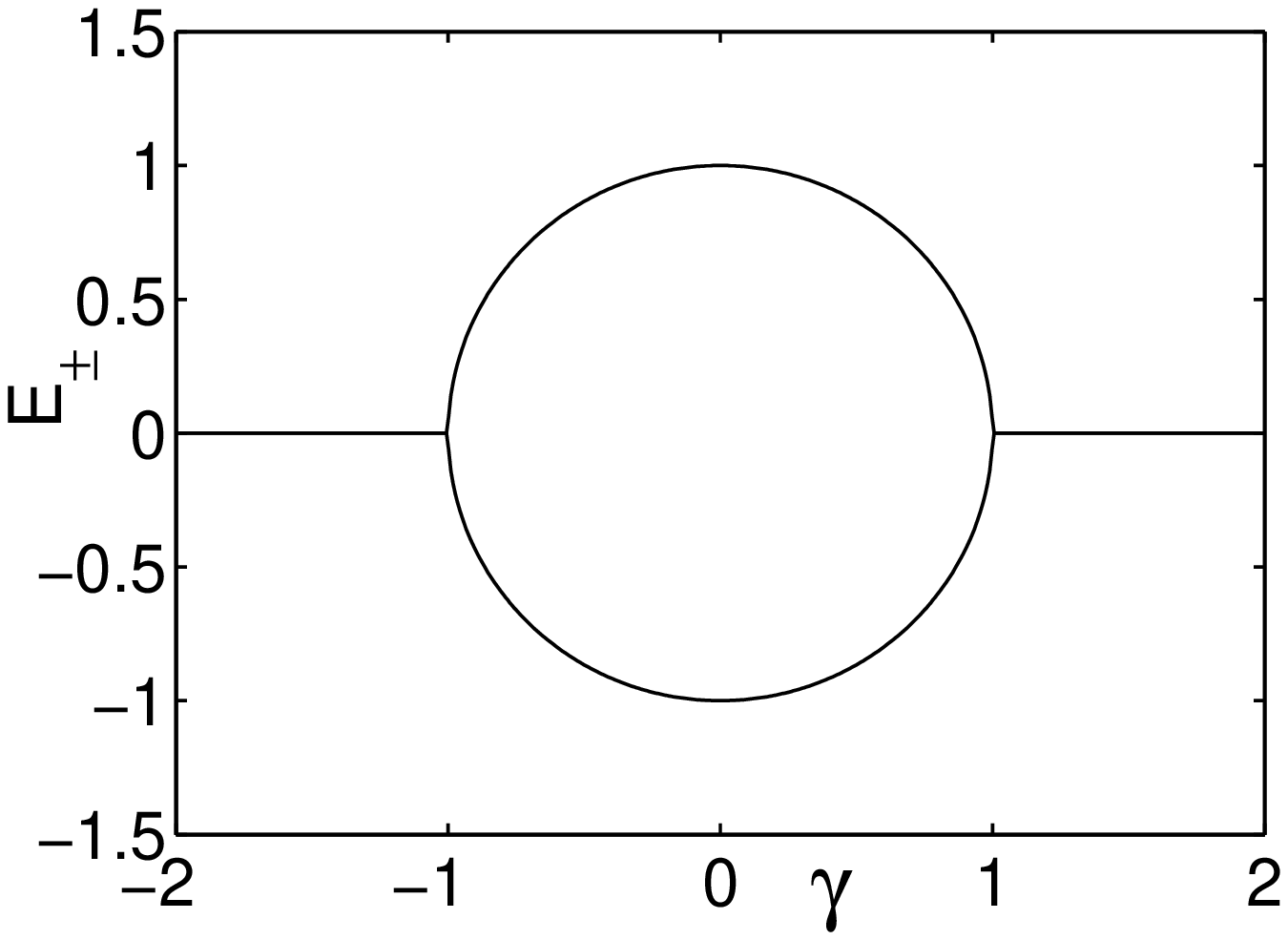}
\includegraphics[width=4cm]{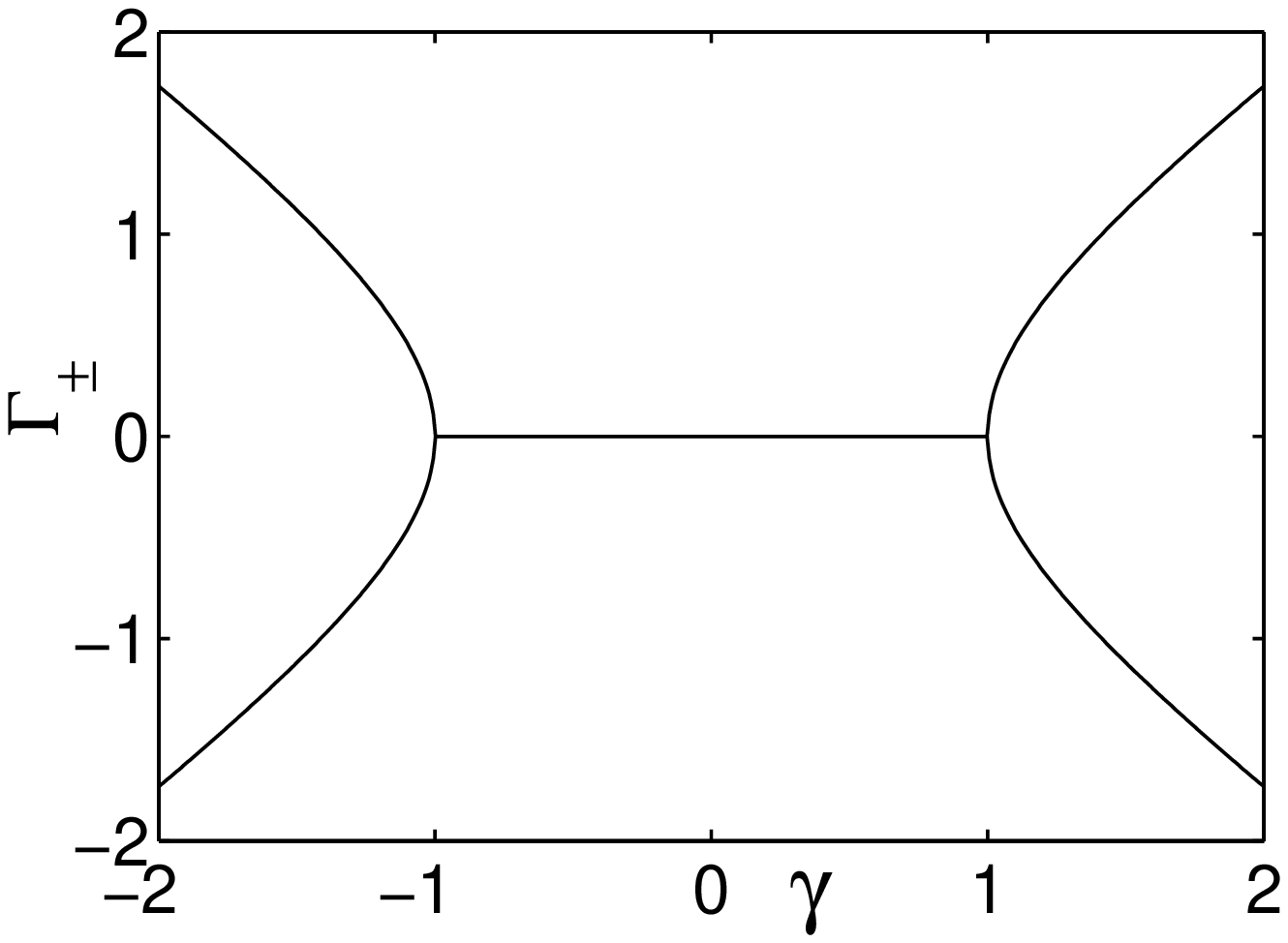}
\caption{\label{fig-Ew_PT_2times2} Real (left) and imaginary (right) parts of the eigenvalues~(\ref{eqn-eval-2by2-PT}) of the $\cP\cT$-symmetric two level system in dependence on the parameter $\gamma$ for $v=1$.}
\end{figure}

The eigenvalues of the $\cP\cT$-symmetric two-level system are given by  
\begin{equation}\label{eqn-eval-2by2-PT}
\lambda_{\pm}=\pm\sqrt{(\epsilon-\rmi\gamma)^2+v^2}=E_\pm-\rmi\Gamma_\pm.
\end{equation}
Thus, although the Hamiltonian is not Hermitian, 
for certain parameters it has a purely real spectrum. 
In fact in the unbiased case $\epsilon=0$ 
there is a whole region in parameter space $|\gamma|\leq |v|$ in which the
spectrum is real. This is illustrated in Fig.~\ref{fig-Ew_PT_2times2}, 
which shows the eigenvalues of $\hat H_{\cP\cT}$
as a function of $\gamma$ for $\epsilon=0$ and $v=1$. 
In the regions of purely real eigenvalues 
all eigenvectors are simultaneous eigenvectors of the $\cP\cT$-operator; 
this is often denoted as unbroken $\cP\cT$-symmetry. 
The eigenvalues of the decaying system \rf{eqn-2times2-nherm} are always 
complex with a negative imaginary part, which is degenerate for both eigenvalues 
in the regions were the $\cP\cT$-symmetric system has a purely real spectrum. 

The eigenvalues of both the $\cP\cT$-symmetric and the decaying 
systems degenerate along lines in the parameter space which are specified by
\begin{equation}\label{eqn-exceptional-lines}
\epsilon=0\quad {\rm and}\quad v=\pm\gamma.
\end{equation}
For $\gamma=0$ this reduces to the so-called diabolical point of the Hermitian
two level system \cite{Berr84a}. At the complex degeneracies 
for $\gamma\neq0$, the exceptional points (EP) 
\cite{Kato66book,Berr04,Heis00,08PT}, 
the essence of the peculiar behavior of non-Hermitian systems 
becomes apparent. At an EP not only the eigenvalues, 
but also the eigenvectors coincide. Thus, while the eigenvectors
build a basis of the Hilbert space outside the EP when they
coincide at the EP they are not sufficient to span the Hilbert space. 
In other words, along the exceptional lines \rf{eqn-exceptional-lines} 
the Hamiltonian is not diagonalizable but equivalent to a Jordan block. 
The occurrence of EPs can have crucial impact on the physical behavior 
of a system (see, e.g., \cite{Demb01,Cart07,Berr04,Muss08,Wier08,Klai08}).
For the $\cP\cT$-symmetric system \rf{eqn-2times2-PT} the EPs mark 
the border to the region of broken $\cP\cT$-symmetry where the eigenvalues 
are complex \cite{08PT}. 

In the region of unbroken $\cP\cT$-symmetry the 
system \rf{eqn-2times2-PT} shows a pseudo-closed behavior. This means 
that with the introduction of an appropriate inner product 
the time evolution can be expressed in a unitary way. However,
this should not be confused with the conservation of the usual
probability as it is given by the normalization of the wave function
in the original inner product space $||\psi||^2$. While this is conserved 
for the time evolution in an eigenstate with real energy, 
this is in general not true for an arbitrary initial
state, due to the nonorthogonality of the eigenfunctions. 

The dynamics of a two-level quantum system can easily be 
expressed in closed form. For a time independent Hamiltonian 
$\hat H$ the Schr\"odinger equation $\rmi\dot\psi=\hat H\psi$ 
with the initial condition $\psi(t=0)=\psi_0$ is solved by
$\psi(t)=\hat U(t)\psi_0$, where $\hat
U(t)=\exp(-\rmi \hat H t)$ is the time evolution operator. 
For the $\cP\cT$-symmetric two-level system
(\ref{eqn-2times2-PT}) outside the EP one finds:
\begin{equation}
\hat U(t)=\left( \begin{array}{cc} {\cos(\omega t)-{\rm i} \zeta \frac{\sin(\omega t)}{\omega}} & {-{\rm i} v \frac{\sin(\omega t)}{\omega}}\\
{-{\rm i} v \frac{\sin(\omega t)}{\omega}} & {\cos(\omega t)+{\rm i} \zeta \frac{\sin(\omega t)}{\omega}} \end{array} \right),
\end{equation}
with the complex energy $\zeta=\epsilon-\rmi\gamma$, 
and accordingly the complex frequency
$\omega=\sqrt{\zeta^2+v^2}$, which is determined by the
eigenvalue difference $\omega=\half(\lambda_+ -\lambda_-)$.
However, at the EP ($\zeta=-\rmi v$) the
frequency goes to zero. In this limit the time evolution 
operator is given by
\begin{equation}
\hat U_{\rm EP}(t)=\left( \begin{array}{cc} 1-vt & -\rmi\,vt\\
-\rmi\, vt & 1+vt \end{array} \right).
\end{equation}
The time evolution of the normalization $n=|\psi_1|^2+|\psi_2|^2$ 
is determined by the population imbalance according to the relation
\begin{equation}
\dot n=-2\gamma(|\psi_1|^2-|\psi_2|^2).
\end{equation}
From the behavior of the $\cP\cT$-symmetric system the 
dynamics of the non-Hermitian two-level system
(\ref{eqn-2times2-nherm}) can be found by applying the time dependent
transformation $\psi(t)\to\rme^{-\gamma t}\psi(t)$.
\begin{figure}[tb]
\centering
\includegraphics[width=4cm]{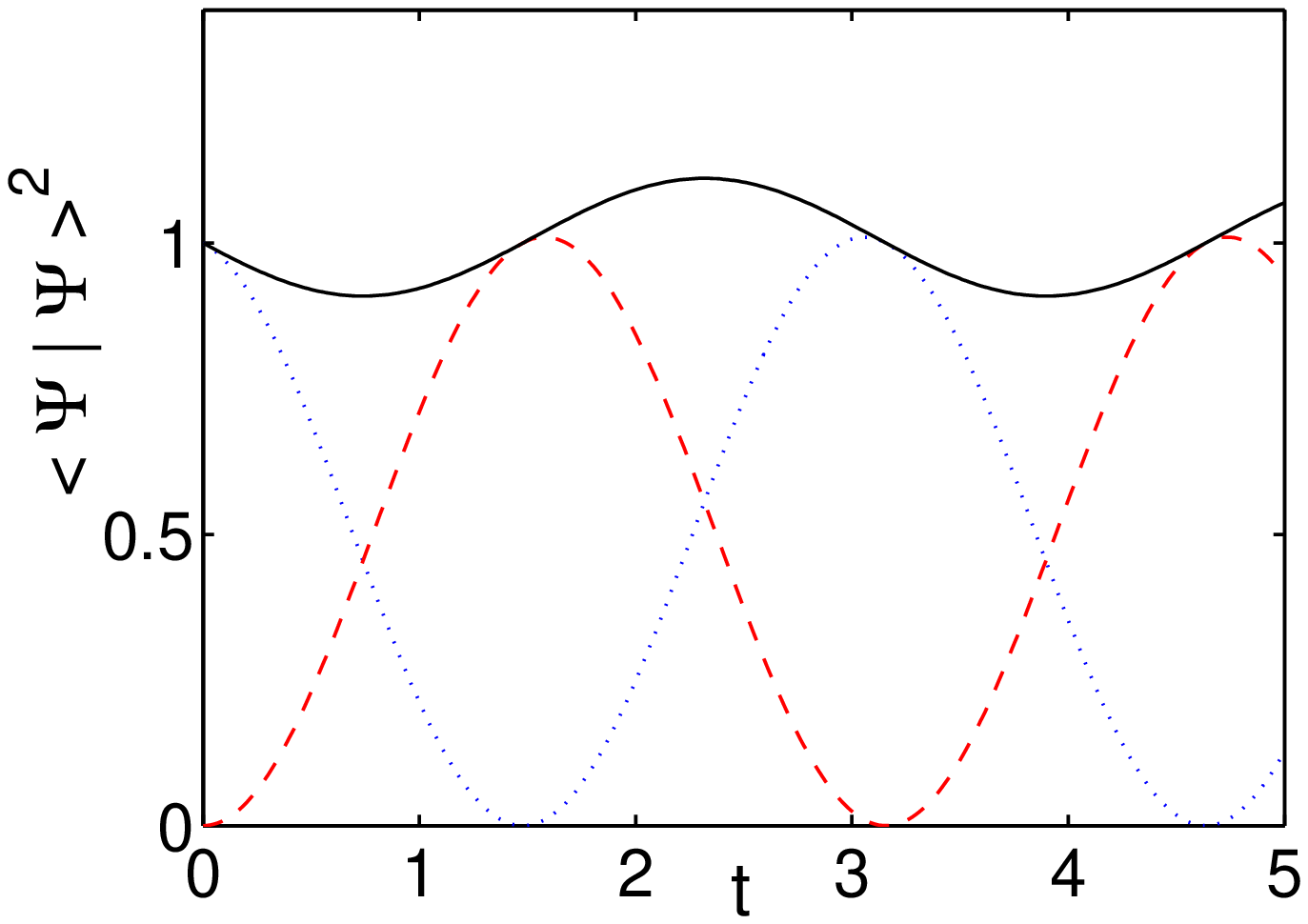}
\includegraphics[width=4cm]{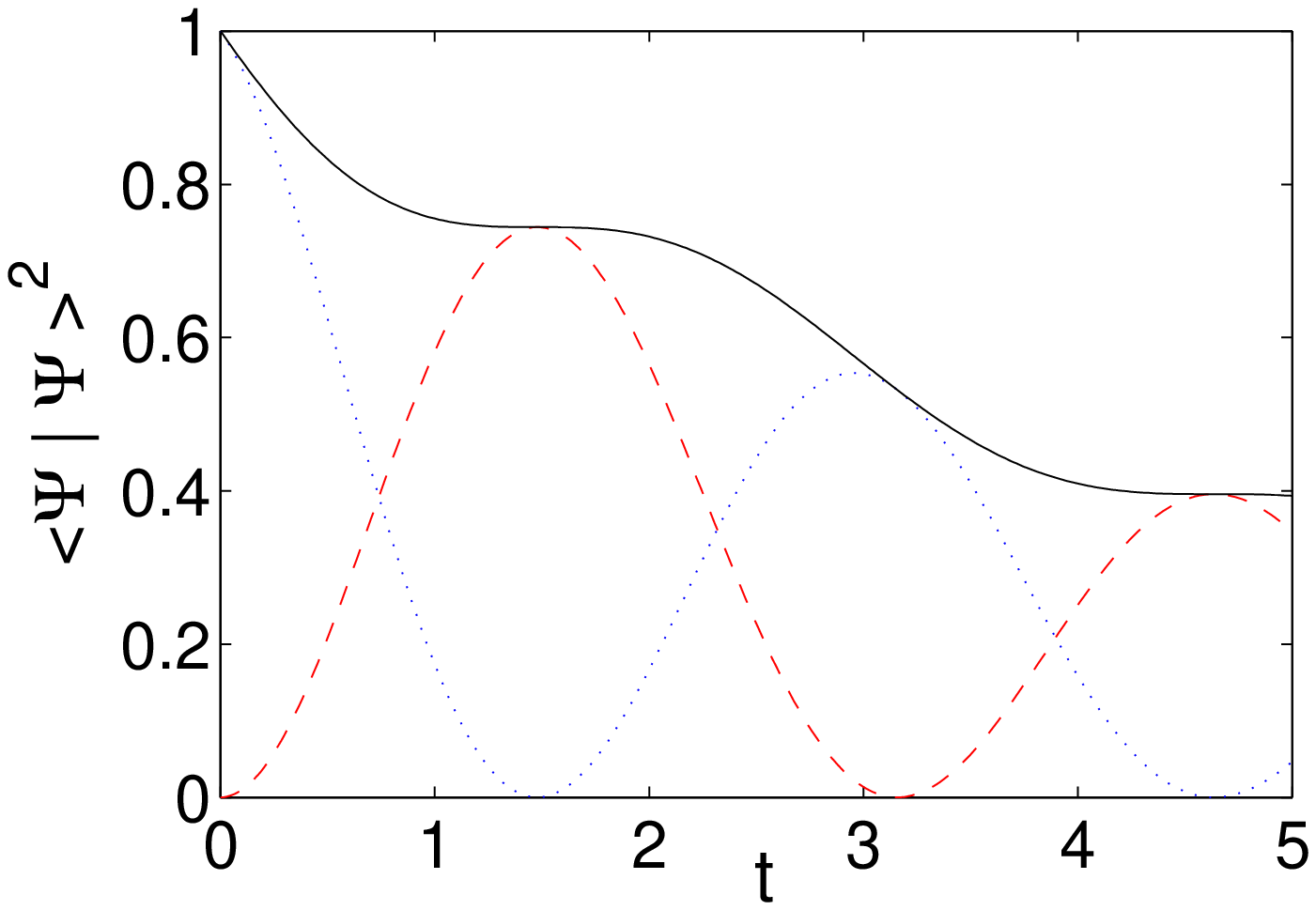}
\includegraphics[width=4cm]{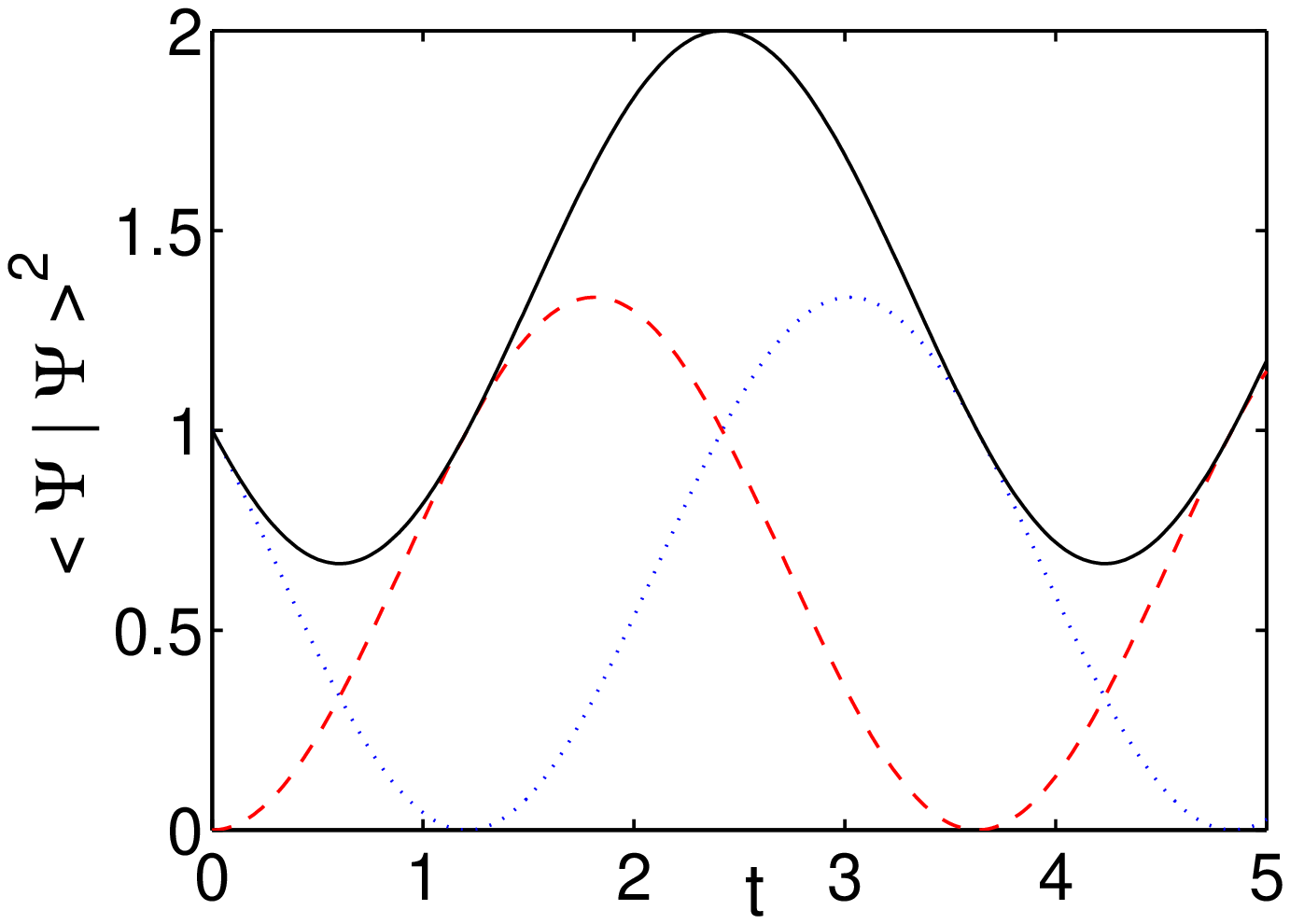}
\includegraphics[width=4cm]{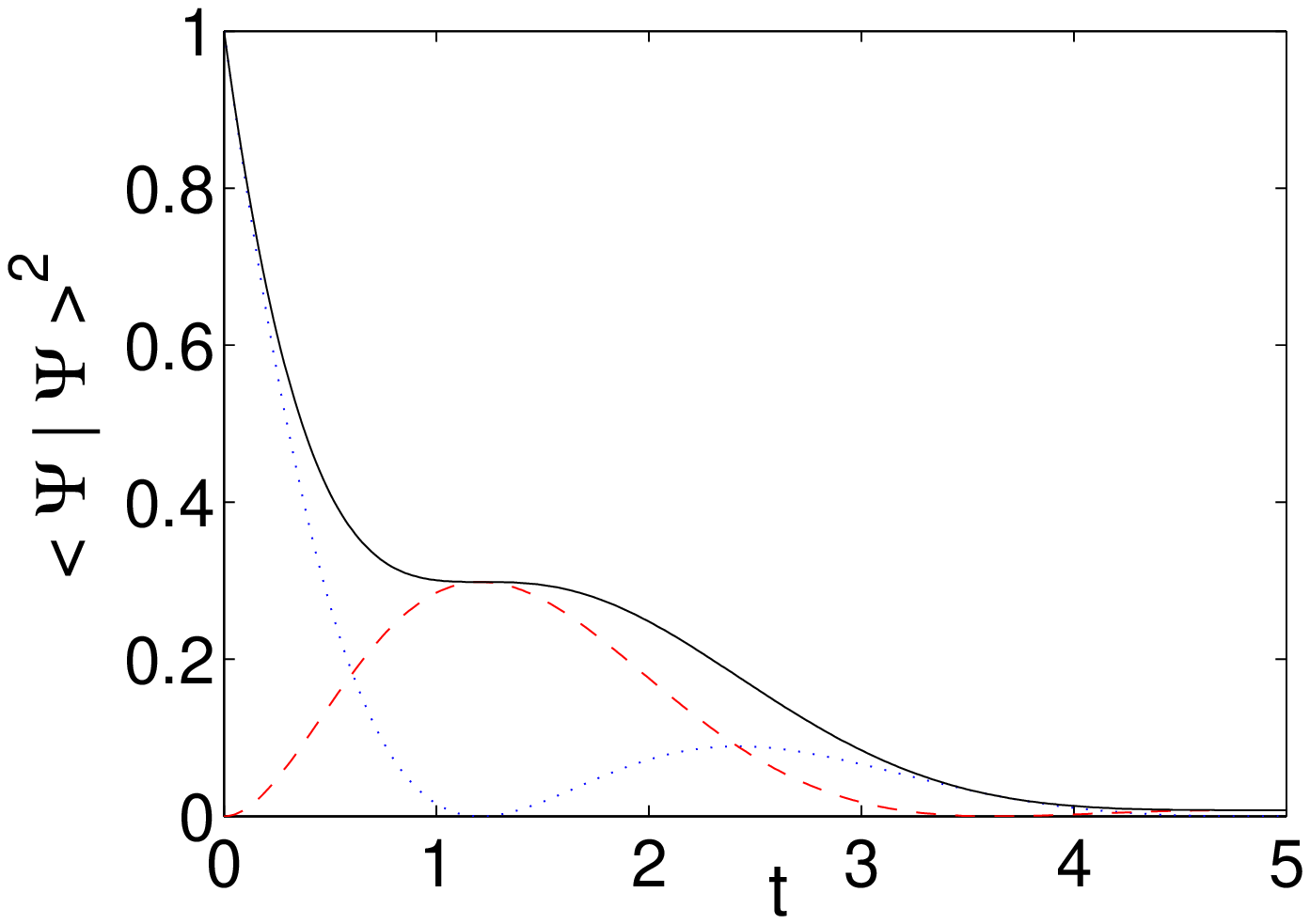}
\includegraphics[width=4cm]{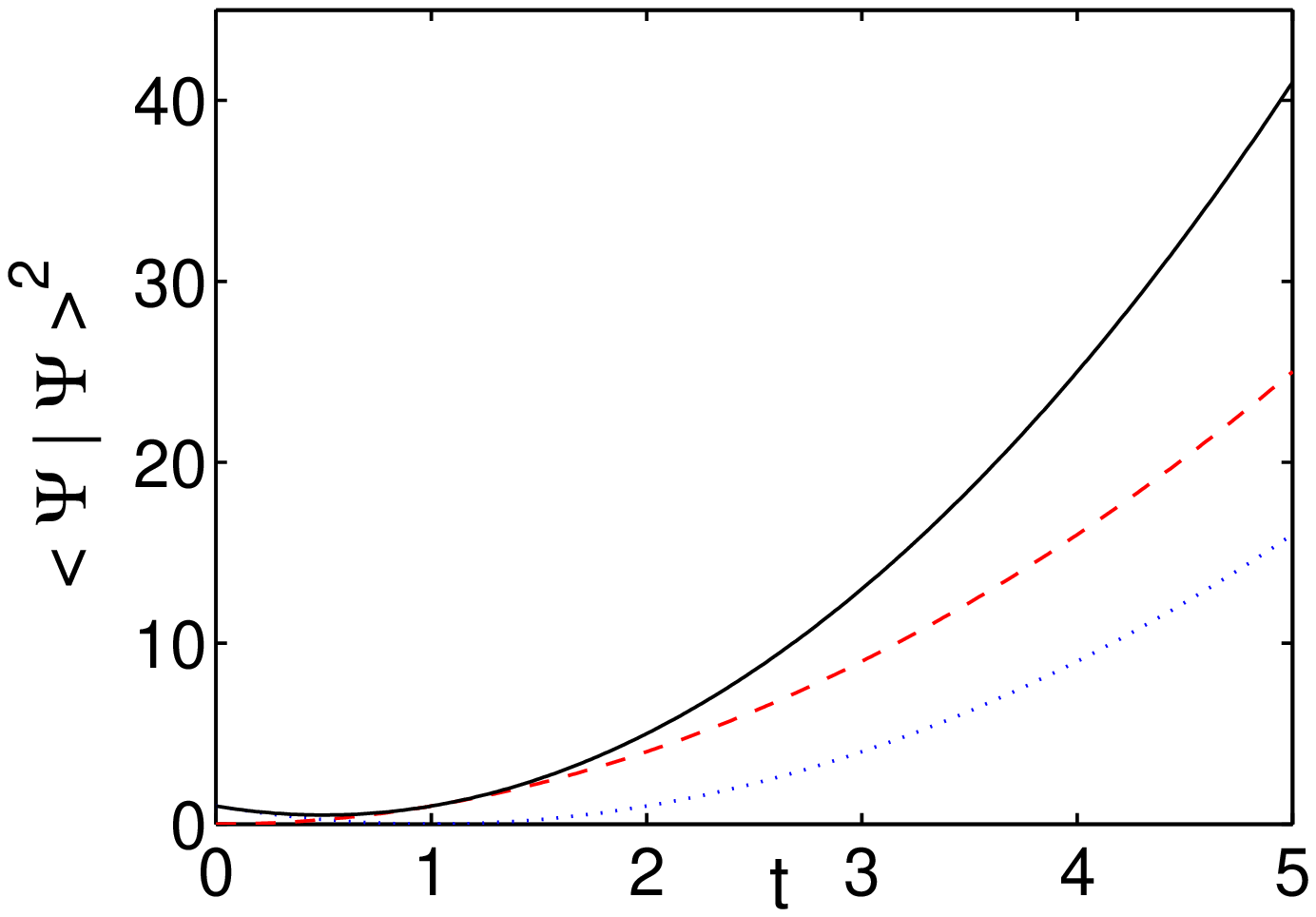}
\includegraphics[width=4cm]{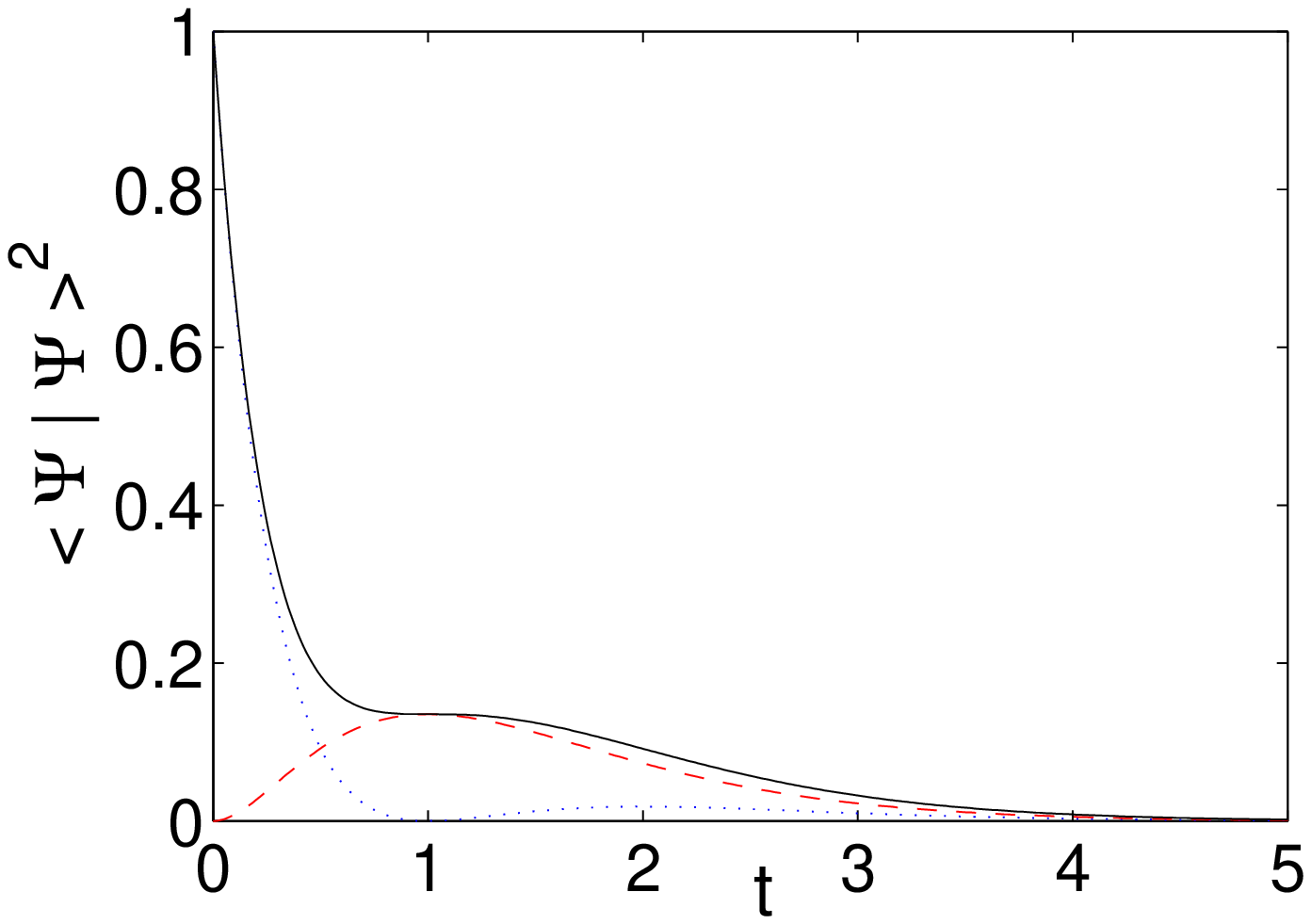}
\includegraphics[width=4cm]{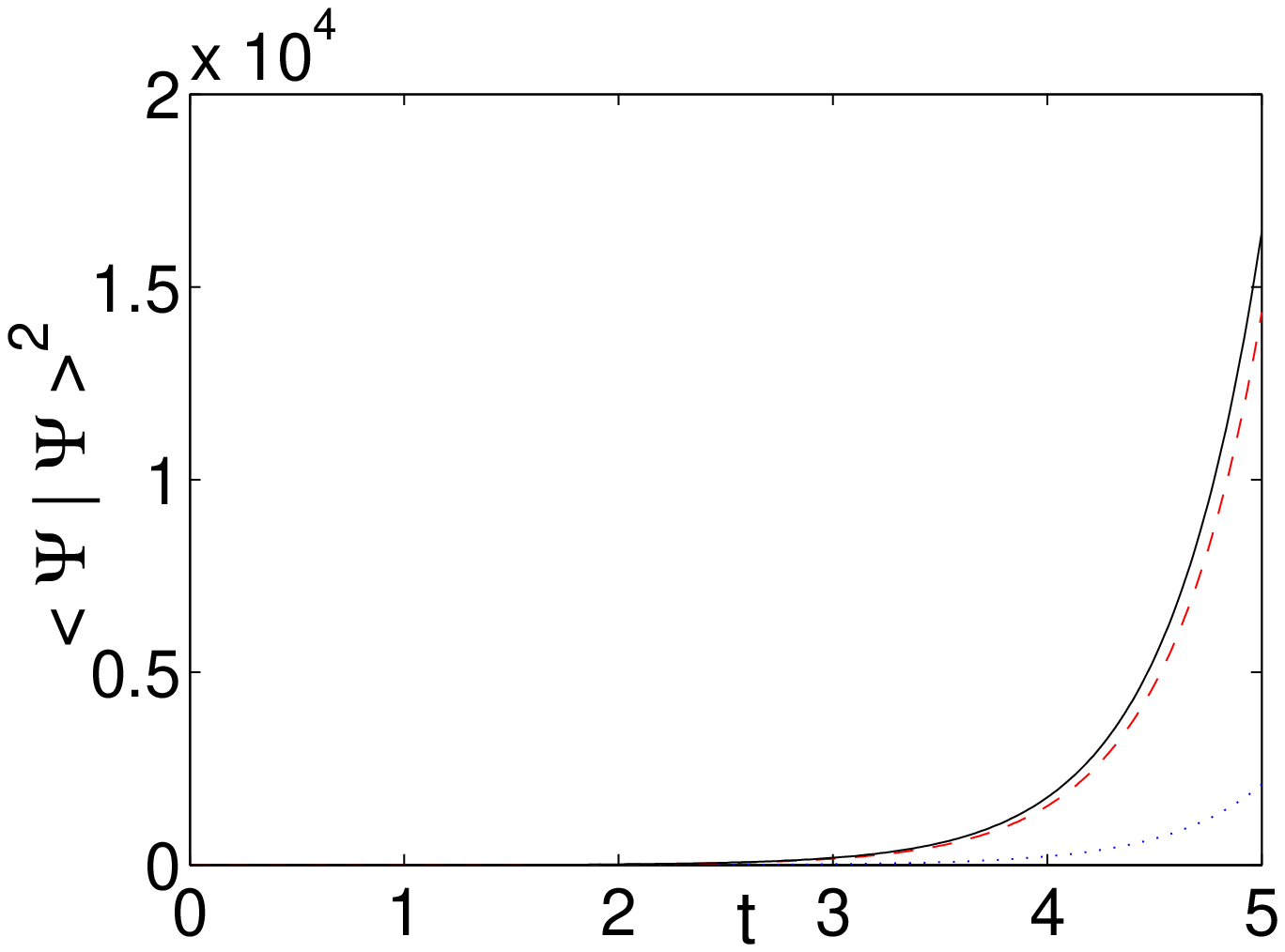}
\includegraphics[width=4cm]{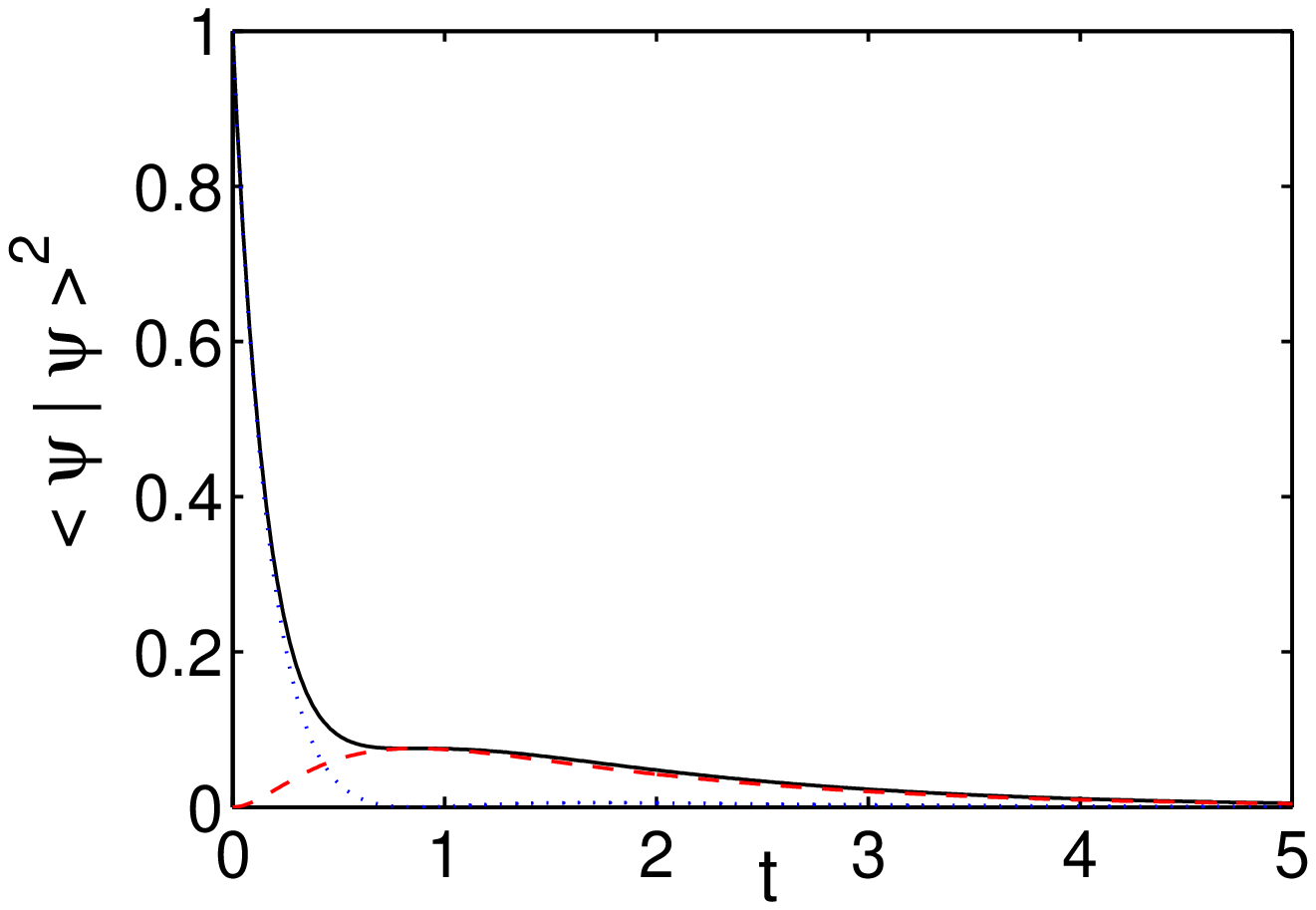}
\caption{\label{fig_norm_PT} (Color online) Dynamics of the $\cP\cT$-symmetric 
(\ref{eqn-2times2-PT}) (left) and the decaying (\ref{eqn-2times2-nherm}) 
(right) non-Hermitian two-level system with $\epsilon=0$, $v=1$, and different
values of $\gamma$ (from top to bottom: $\gamma=0.1,\, 0.5, \, 1,\, 1.5$) for an initial state in level 1. Shown here are the absolute 
values of the components of the wave function
$|\psi_1|^2$ (blue dotted line) and $|\psi_2|^2$ 
(red dashed line) as well as the total probability
$n=|\psi_1|^2+|\psi_2|^2$ (black solid line).}
\end{figure}

Figure \ref{fig_norm_PT} shows some examples of the 
dynamics for different non-Hermiticities $\gamma$ 
with $\epsilon=0$, $v=1$, and for an initial state in level $1$. 
The left column shows the dynamics for the $\cP\cT$-symmetric 
system \rf{eqn-2times2-PT} and the right column for the 
decaying system \rf{eqn-2times2-nherm} for the same parameter 
values. It can be seen that for the $\cP\cT$-symmetric 
system the normalization oscillates for $\gamma<v$ 
with a period that increases with increasing $\gamma$ and 
diverges to infinity as $\gamma$ approaches the EP, 
$\gamma=v$. The rate of decrease of the normalization takes its
maximum value when the population is in the first level; growth and
decrease rates are balanced when both levels are equally populated;
and the growth rate is maximal when the population in the second
level is maximal. We observe that while for small values of $\gamma$
the system performs Rabi-type oscillations between the two levels,
the population oscillations within each level become parallel when
the EP is approached. This nicely illustrates the fact that a
complex term in the energy cannot be regarded as an overall
modulation of the normalization of the system, but rather changes the
full dynamics in a dramatic way. The oscillatory behavior breaks
down completely at the EP where the period diverges. Instead we
observe an algebraic growth of the probability. This can be
obtained analytically from $\psi(t)=\hat U_{\rm EP}(t)\psi(0)$, with the initial 
state in level 1, as
\begin{equation}
n(t)=1-2vt+2v^2t^2.
\end{equation}
For larger values of $\gamma$, the $\cP\cT$-symmetry is broken and
so is the balance between growth and decrease -- the normalization
grows exponentially. 

For the purely decaying system \rf{eqn-2times2-nherm}, on the other hand, 
we observe a monotonic decrease of the normalization. The decay behavior is
not exponential, which is intuitively understood by recalling that
the population only decays from one of the levels. Therefore, the
decrease is determined by the population of this level, which varies
in time if the system is not in an eigenstate. It is interesting to note that in
contrast to the $\cP\cT$-symmetric system, we cannot detect an
obvious trace of the presence of the EP in the decay dynamics for
the non-Hermitian system (\ref{eqn-2times2-nherm}). 

The similarity of the optical wave equations in waveguide structures to the 
Schr\"odinger equation makes it possible to observe the described dynamics 
and the $\cP\cT$-related phase transition in optical waveguide structures with gain and loss. This has not only
been investigated theoretically \cite{Elga07,Klai08}, but has recently been realized experimentally \cite{Guo09,Ruet10}. 

Although the non-Hermitian Schr\"odinger equation does
not preserve the normalization, it is possible to
describe the dynamics of the system consistently in terms of a Bloch
vector that stays confined to the surface of the Bloch sphere
throughout the time evolution. For this
purpose we first define the renormalized state vector with the components 
\begin{equation}
\varphi_j=\frac{\psi_j}{\sqrt{|\psi_1|^2+|\psi_2|^2}}.
\end{equation}
For both the decaying \rf{eqn-2times2-nherm} and the 
$\cP\cT$-symmetric system \rf{eqn-2times2-PT} the 
dynamics are then governed by the non-Hermitian (and nonlinear)
effective Schr\"odinger equation:
\begin{equation}\label{eqn-2times2-nherm-normalized}
\rmi\frac{\rmd}{\rmd\, t}\begin{pmatrix} {\varphi}_1 \\{\varphi}_2 \end{pmatrix}
=\begin{pmatrix} \epsilon - \rmi \gamma(1-\kappa) & v \\
   v & - \epsilon +\rmi\gamma(1+\kappa)  \end{pmatrix}
   \begin{pmatrix}\varphi_1\\\varphi_2\end{pmatrix}, 
\end{equation}
with $\kappa=|\varphi_1|^2-|\varphi_2|^2$. 
This dynamics by definition conserves the normalization 
$|\varphi_1|^2+|\varphi_2|^2=1$. We can then define the components of the 
normalized Bloch vector in the familiar way with respect to 
the renormalized wave function $\varphi$:
\begin{eqnarray}
\s_x&=\half(\varphi_1^*\varphi_2+\varphi_1\varphi_2^*)&=\frac{1}{2}\frac{\psi_1^*\psi_2+\psi_1\psi_2^*}{\psi_1^*\psi_1+\psi_2^*\psi_2}\nn\\
\s_y&=\tfrac{1}{2\rmi}(\varphi_1^*\varphi_2-\varphi_1\varphi_2^*)&=\frac{1}{2\rmi}\frac{\psi_1^*\psi_2-\psi_1\psi_2^*}{\psi_1^*\psi_1+\psi_2^*\psi_2}
\label{eqn-bloch-vector}\\
\s_z&=\half(\varphi_1^*\varphi_1-\varphi_2^*\varphi_2)&=\frac{1}{2}\frac{\psi_1^*\psi_1-\psi_2^*\psi_2}{\psi_1^*\psi_1+\psi_2^*\psi_2}.\nn
\end{eqnarray}
Using this definition we can obtain the generalized Bloch equations
of motion from \rf{eqn-2times2-nherm-normalized} as
\begin{eqnarray}
\dot \s_x &=& -2\epsilon \s_y+4\gamma\s_x\s_z\nn\\
\dot \s_y &=& 2\epsilon \s_x-2v \s_z+4\gamma\s_y\s_z\label{eqn_bloch_nherm}\\
\dot \s_z &=& 2v \s_y-\gamma(1-4\s_z^2)\nn.
\end{eqnarray}
Here again the normalization $\s_x^2+s_y^2+s_z^2=\tfrac{1}{4}$ is
conserved by construction. 

The dynamics of the renormalized quantities decouple from the 
time dependency of the normalization $n=|\psi_1|^2+|\psi_2|^2$ of 
the state vector which can be obtained from the Bloch dynamics via
\begin{equation}
\dot{n}=\left\{\begin{array}{ll}
-4\gamma(\s_z+\half)n, &\quad {\rm for}\ \eqref{eqn-2times2-nherm}\\
-4\gamma\s_z n,&\quad {\rm for}\ \eqref{eqn-2times2-PT}.
\end{array}\right. 
\end{equation}
This allows a separate investigation of both dynamics.
\begin{figure}[tb]
\centering
\includegraphics[width=4cm]{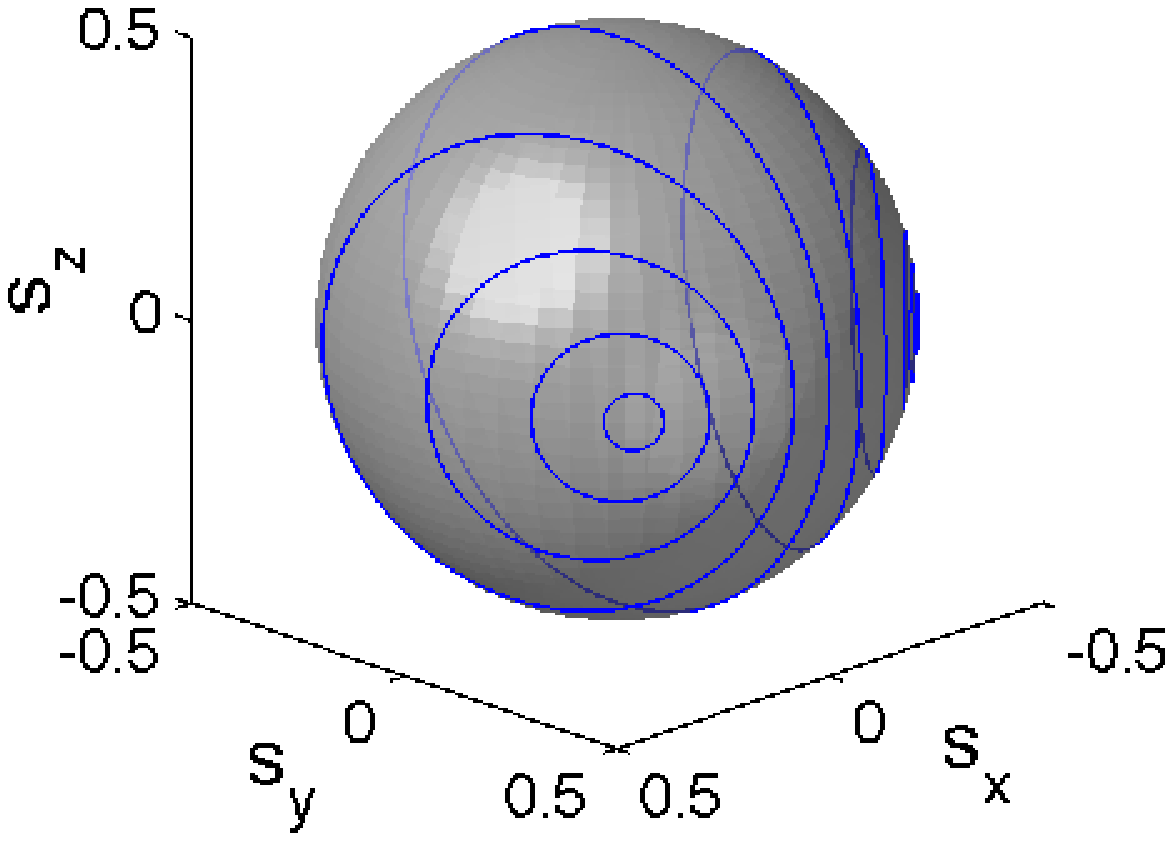}
\includegraphics[width=4cm]{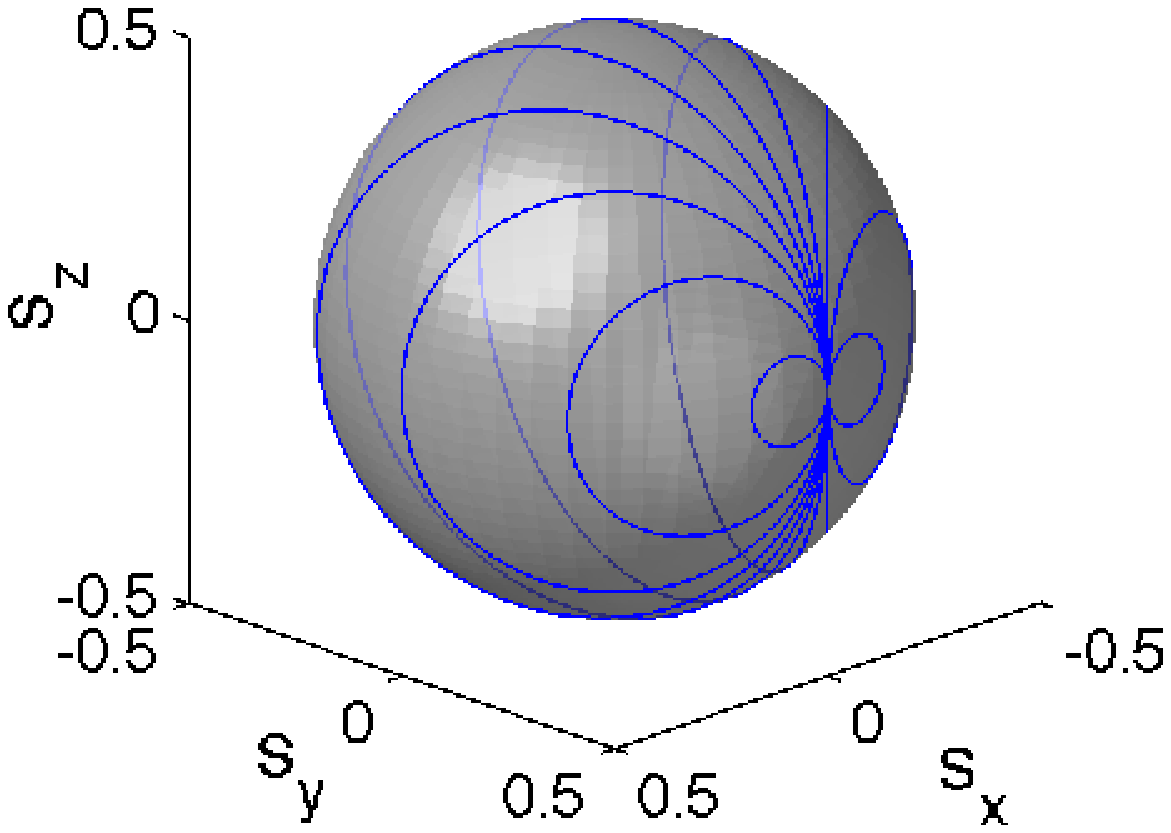}
\includegraphics[width=4cm]{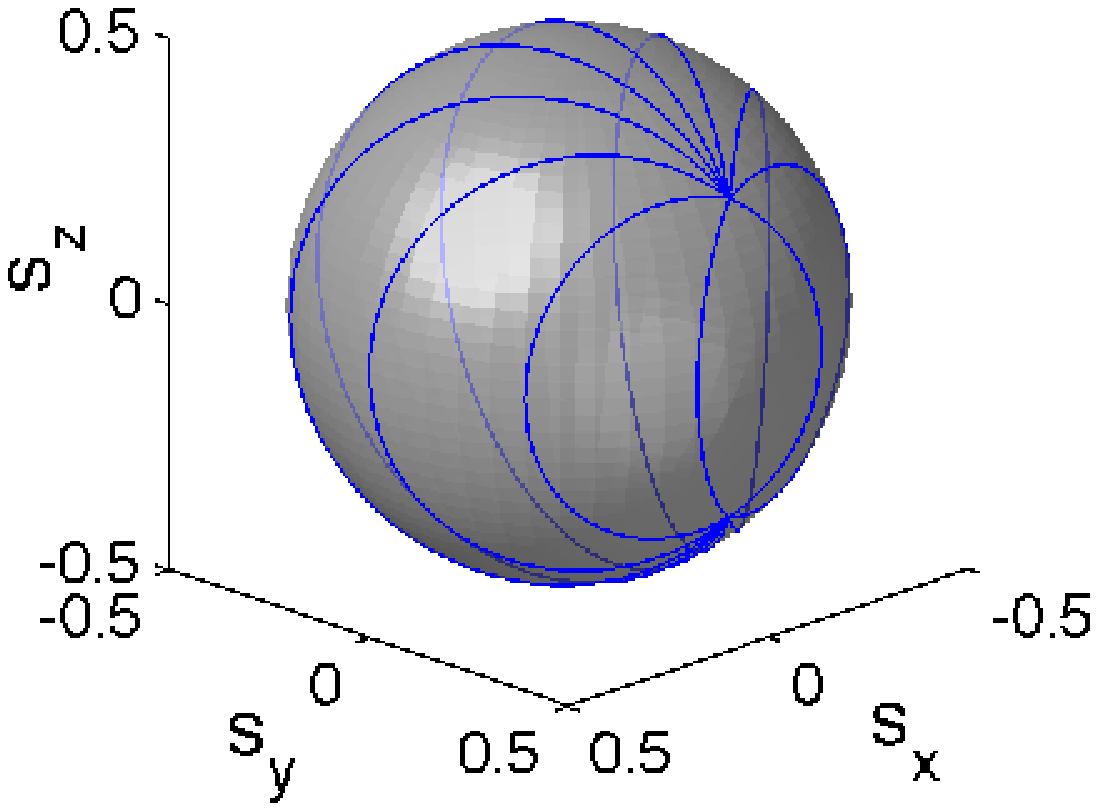}
\includegraphics[width=4cm]{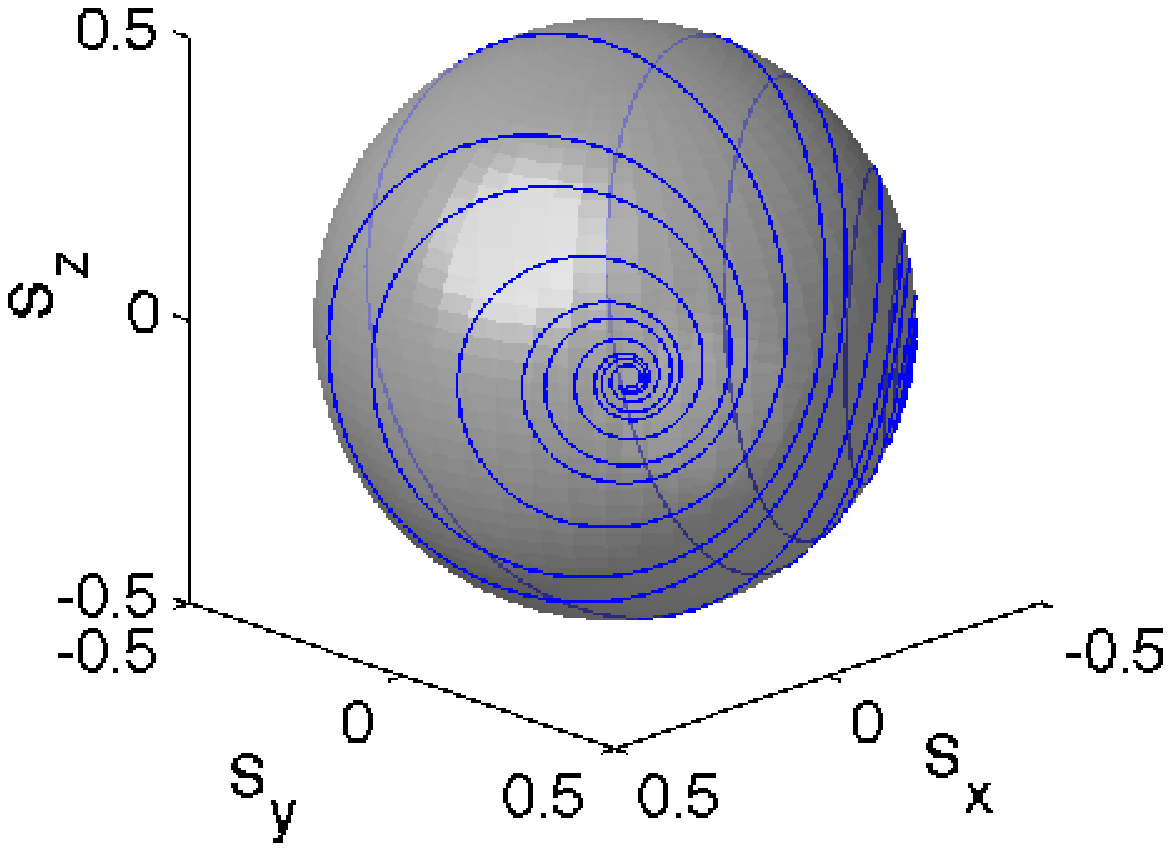}
\caption{\label{fig_bloch_nhermI} Effective Bloch dynamics of the 
non-Hermitian two level system~(\ref{eqn_bloch_nherm}) with $v=1$
and different values of $\epsilon$ and $\gamma$. 
The two plots on the top and the left plot on the bottom are for 
the unbiased system with $\epsilon=0$ and increasing values 
of $\gamma=0.75,\, 1,\, 1.25$, respectively. The right plot on the bottom shows the dynamics for a biased systems with 
$\epsilon=0.1$ and $\gamma=0.75$.}
\end{figure}

The Bloch dynamics is organized according to the 
fixed points (the stationary states), which can be obtained analytically 
from  the real roots of the fourth order polynomial
\begin{equation}\label{eqn-fixed-points-nherm-lin-s}
16\gamma^2\s_z^4+4(\epsilon^2+v^2-\gamma^2)\s_z^2-\epsilon^2=0,
\end{equation}
where the corresponding $\s_y$ and $\s_x$ values are given by 
$\s_y=\tfrac{\gamma}{2v}(1-4\s_z^2)$ and the normalization 
condition. For every parameter set there are only two fixed 
points which can be of different types, including sinks and sources. 
In general the type of the fixed points can be identified from the 
behavior of the surrounding vector field in a systematic manner, 
which we postpone to the discussion of the general nonlinear case in section \ref{sec-MF-dyn}. 

In Fig.~\ref{fig_bloch_nhermI} we show four examples of the 
Bloch dynamics, three for an unbiased system with $\epsilon=0$ and 
different values of $\gamma$, and one where all 
parameters are nonzero. In the first plot (top on the left) in 
Fig.~\ref{fig_bloch_nhermI}, where $\gamma<v$, we observe Rabi-type
oscillations surrounding one of the two fixed points located at
$\s_z=0$. However, compared to the Hermitian case the picture is
deformed. The two fixed points are not centered at $\s_y=0$
corresponding to a phase difference of zero and $\pi$ between the
amplitudes in the two levels, respectively, but with increasing
$\gamma$ they approach each other along the equator toward
$\s_y=\half$ and $\s_x=0$. This is connected to the fact that the strict 
$\cP$-symmetry which enforces both $\s_z$ and $\s_y$ to be zero 
in the Hermitian case, with $\epsilon,\gamma=0$, is broken 
for $\gamma\neq0$ and first replaced by the 
$\cP\cT$-symmetry, which only demands that $\s_z=0$. 
At the EP $\gamma=v$ (shown in the right plot on top in Fig.~\ref{fig_bloch_nhermI}) the two fixed points meet 
and the symmetry is broken. For even larger
values of $\gamma$ one of the fixed points becomes a sink of the
dynamics, and the other a source, both located at $\s_z\neq0$, 
that is, they belong to configurations where one of the levels is 
favored despite the symmetry of the system. 
This could be denoted as a \textit{decay-trapping}.
With increasing values of $\gamma$ the sink 
approaches the south pole of the Bloch sphere 
(corresponding to the stable level) and
the source approaches the north pole (corresponding to the level from
which the decay happens). This is due to the fact that 
the Bloch dynamics describes the mean values of the
remaining part of the population which moves away from the
center of the decay. For nonvanishing $\epsilon$ the system 
is not $\cP\cT$-symmetric, and the situation is changed. In this case
the fixed points change into a sink and a source for
arbitrary small values of $\gamma$. An example of the non-Hermitian
Bloch dynamics for $\epsilon\neq0$ is depicted in the lower right plot in
Fig.~\ref{fig_bloch_nhermI}. 

\section{The non-Hermitian Bose-Hubbard model}
\label{sec-nhermBH}
The non-Hermitian Bose-Hubbard dimer \rf{eqn-BH-Hamiltonian} 
can now be defined as the single particle non-Hermitian two-level 
system \rf{eqn-2times2-nherm} populated with $N$ bosons, with the bosonic
particle creation and annihilation operators $\hat a_j^\dagger,\
\hat a_j$ for the two levels that fulfill the usual bosonic commutation relations 
$[\hat a_j,\hat a_k^\dagger]=\delta_{jk}$\,, $[\hat a_1,\hat a_2]=0$. 

A non-Hermitian many-particle Hamiltonian of the present type 
does not describe the loss of individual
particles. Rather, it describes the decrease in time of the
probability to find the entire many-particle ensemble in the two
modes. This information is completely encoded in the normalization
of the many-particle wave function $|\Psi\rangle$. The expectation
value of the particle number operator $\langle \Psi|\hat
N|\Psi\rangle/\langle \Psi|\Psi\rangle$ stays constant in time. In
other words, the ``decay'' is regarded as a feature of the state,
rather than of the particles. The fact that the Hamiltonian 
(\ref{eqn-BH-Hamiltonian}) commutes with the
number operator $\hat{N}$ implies that the matrix representation
in the Fock (particle number) basis has a block diagonal
structure, that is, it does not induce coupling between subspaces
associated with different particle numbers. Therefore, in what
follows we shall restrict our discussion to these subspaces of fixed $N$.

In analogy with the Bloch representation of the 
single particle system, the Hamiltonian (\ref{eqn-BH-Hamiltonian}) 
can also be expressed in the form of an angular momentum system. 
Introducing the angular momentum operators 
$\hat L_x$, $\hat L_y$ and $\hat L_z$ according to the Schwinger 
representation
\begin{gather}
\nn \hat L_x ={\textstyle \frac12}(\hat a_1^\dagger\hat a_2+\hat a_1\hat a_2^\dagger)\,,\qquad
\hat L_y={\textstyle \frac{1}{2\rmi}}(\hat a_1^\dagger\hat a_2-\hat a_1\hat a_2^\dagger)\,,\\
\hat L_z={\textstyle \frac12}(\hat a_1^\dagger\hat a_1-\hat a_2^\dagger\hat a_2),
\label{lxlylz}
\end{gather}
which obey the usual $SU(2)$ commutation relation
 \be
 \label{comm}
 [\hat L_x, \hat L_y]={\rm i}\hat L_z,
 \ee
and its cyclic permutations, the Hamiltonian 
(\ref{eqn-BH-Hamiltonian}) can be reformulated in the form
\begin{eqnarray}
\hat{\cal H}=2 (\epsilon-\rmi \gamma )\hat L_z+2\J \hat L_x+2\U \hat
L^2_z-\rmi \gamma \hat N \,. \label{BH-hamiltonian-L}
\end{eqnarray}
The conservation of $\hat N$ appears as the conservation of $\hat
L^2=\frac{\hat N}{2}\big(\frac{\hat N}{2}+1\big)$, i.e. the
rotational quantum number $l =N/2$.

In the standard basis of the angular momentum algebra 
$|l,m\ra$, which can be defined by the relations
\ba
\hat L_\pm|l,m\ra &=&\sqrt{(l\mp m)(l\pm m+1)}|l,m\pm 1\ra,\nn\\
\hat L_z|l,m\ra&=&m|l,m\ra\,\label{ang-mom-4}
\ea
with $l=N/2$, the Hamiltonian $\hat\cH$ takes the form of a 
tridiagonal $(N+1)\times(N+1)$-matrix. Special features of the spectrum 
of the present model and a corresponding $\cP\cT$-symmetric 
model are discussed in \cite{Hill06,08PT}. 
\begin{figure}[tb]
\centering
\includegraphics[width=4cm]{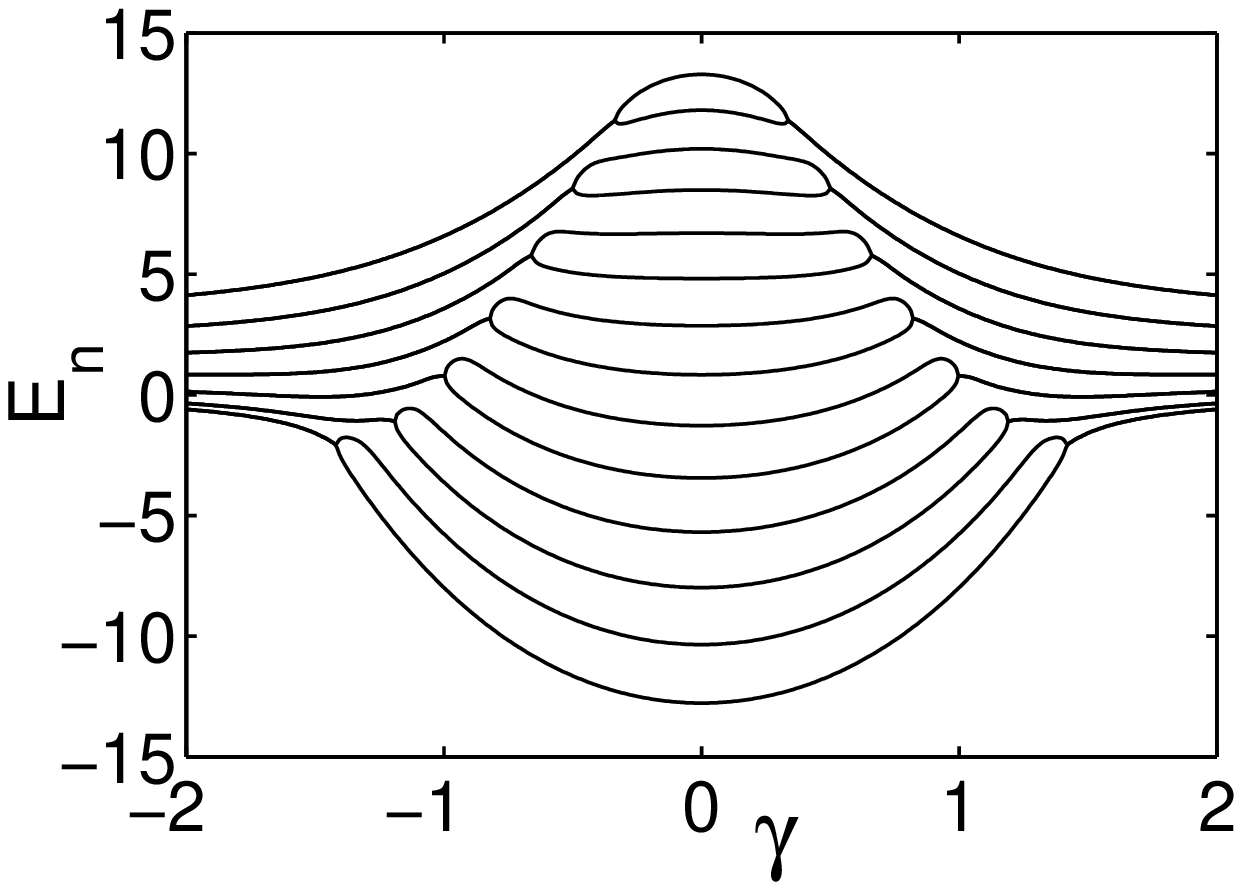}
\includegraphics[width=4cm]{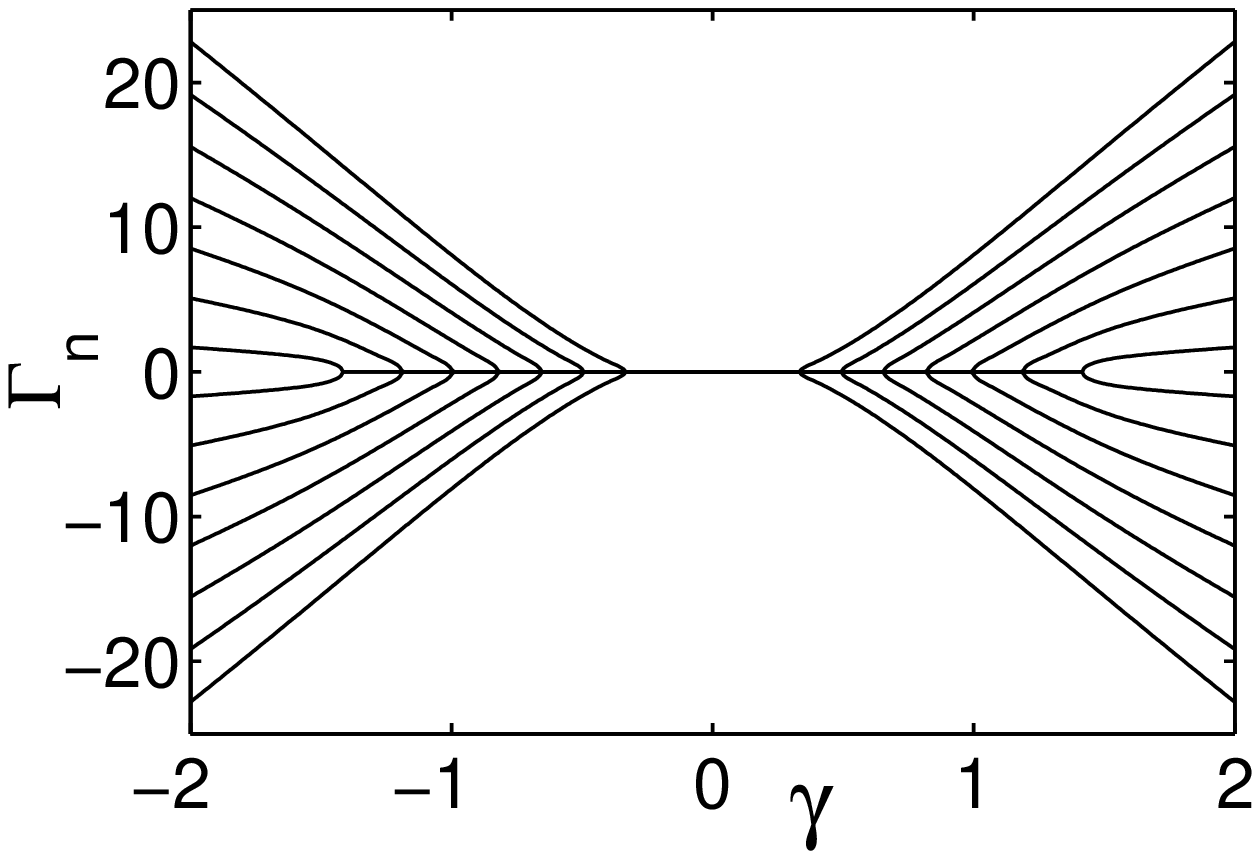}
\includegraphics[width=4cm]{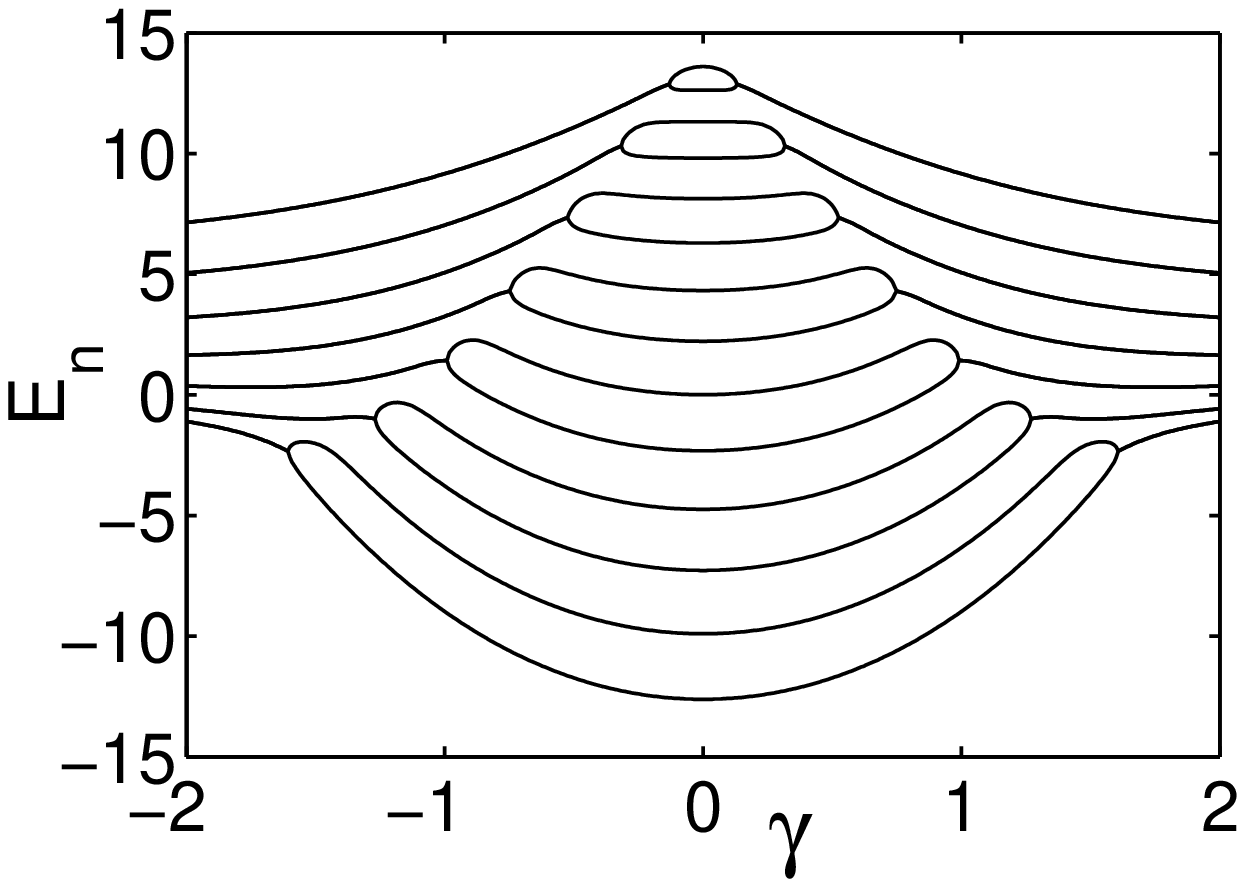}
\includegraphics[width=4cm]{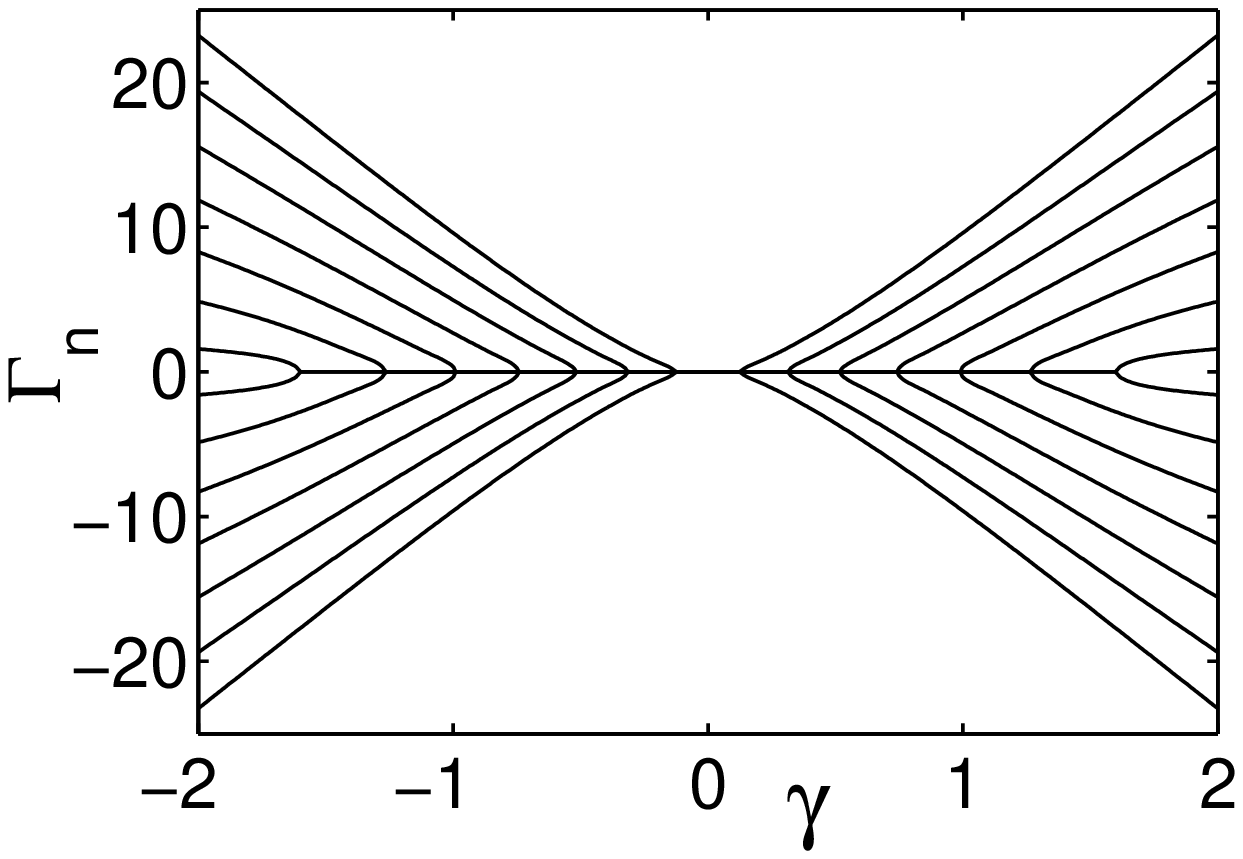}
\caption{\label{fig_BH_PT_EW} Real (left) and imaginary (right) parts of the
eigenvalues $\lambda_n=E_n-\rmi\Gamma_n$ of the Bose-Hubbard
Hamiltonian \rf{Ham1} as a function of the non-Hermiticity
$\gamma$ for $v=1$, $N=13$ particles and $c=0.5/N$ (top) and 
$c=0.9/N$ (bottom).}
\end{figure}

In the limit of vanishing particle interaction, $\U=0$, the eigenvalue 
equation is solvable in closed form and the spectrum 
consists of multiples of the single particle eigenvalues:
\begin{equation}\label{eqn-ev-lin-nherm}
\lambda_n=-\rmi N\gamma+(2n-N)\sqrt{(\epsilon-\rmi\gamma)^2+v^2},\quad n=0,1,...,N.
\end{equation}
Thus, for $\epsilon=0$ at $\gamma=\pm v$ all
eigenvalues degenerate simultaneously. The corresponding 
eigenvectors also coalesce and this configuration thus corresponds to a full Jordan block structure of the Hamiltonian, that is, an EP of higher order \cite{08PT}.
As for the single particle system the unbiased ($\epsilon=0$) 
non-Hermitian Bose-Hubbard dimer can be mapped into a 
$\cP\cT$-symmetric model \rf{Ham1} by an imaginary energy shift $\hat \cH=\hat
\cH_{\cP\cT}-\rmi\gamma \hat N$. For this model the $\cP\cT$-symmetry 
is broken at the EP where all eigenvalues become complex simultaneously.
An arbitrary small interaction strength $c\neq 0$ perturbs 
the system in a manner that leads to a splitting of the EP of higher order into a
series of EPs of second order, that is, degeneracies of pairs of
eigenvalues and the corresponding eigenvectors. The interaction 
thereby always shrinks the region of unbroken $\cP\cT$-symmetry. 
In Fig.~\ref{fig_BH_PT_EW} we show the eigenvalues for $N=13$ particles in dependence on the non-Hermiticity $\gamma$ for two values of the 
interaction strength. It can be seen that the region of purely real 
eigenvalues shrinks with increasing interaction strength. Further 
details concerning the spectral behavior of the $\cP\cT$-symmetric model 
\rf{Ham1} can be found in \cite{08PT}. Some general aspects 
of $\cP\cT$-symmetric models of Lie-algebraic type as the 
present one have been presented in \cite{Assi09}. 

The many-particle dynamics can be conveniently analyzed 
in terms of the angular momentum expectation values. 
The non-Hermitian generalization of the 
Heisenberg equation of motion for an operator $\hat A$ 
(which is not explicitly time dependent)
 is given by \cite{Datt90b,08nhbh_s,09nhclass}
\begin{eqnarray}
\nn \rmi \hbar \frac{\rmd \,}{\rmd\, t}\langle \psi|\hat A|\psi
\rangle &=&\langle \psi|\hat A\hat{\cal H}-\hat{\cal H}^\dagger\hat
A|\psi \rangle\\ &=&\langle \psi|\,[\hat A,\hat H]|\psi \rangle -
i\langle \psi|\,[\hat A,\hat \Gamma]_{\scriptscriptstyle +}|\psi
\rangle, \label{dattoli1}
\end{eqnarray}
where we decomposed the Hamiltonian into Hermitian and
anti-Hermitian parts via $\hat{\cal H}=\hat H -\rmi \hat\Gamma$, with
$\hat H=\hat H^\dagger$ and $\hat\Gamma=\hat \Gamma^\dagger$, and
introduced the notation $[\ ,\ ]_{\scriptscriptstyle +}$ for the
anti-commutator. 
Thus, the equation of motion for the 
expectation value $\langle \hat A\rangle =\langle \psi|\hat A|\psi
\rangle/\langle \psi|\psi \rangle$ reads \cite{08nhbh_s,09nhclass}
\begin{eqnarray}\label{GHE}
\rmi \hbar \frac{\rmd \,}{\rmd\, t}\langle \hat A \rangle
=\langle[\hat A,\hat H]\rangle - 2\rmi \,\Delta^2_{A\Gamma} \,,
\end{eqnarray}
with the covariance
$\Delta^2_{A\Gamma}=\langle {\textstyle \frac12}[ \hat A, \hat
\Gamma]_{\scriptscriptstyle +} \rangle
- \langle  \hat A \rangle \langle \hat \Gamma \rangle\,.$
In the case of the Bose-Hubbard dimer \rf{eqn-BH-Hamiltonian}, 
we find for the dynamics of the angular momentum expectation values:
\begin{eqnarray}\label{KomZerfall1}
{\textstyle \frac{{\rmd}}{\rmd\, t}} \langle \hat L_x \rangle \!\!\!&=&\!\!\!
- 2 \epsilon \langle \hat L_y \rangle - 2\U  \langle [ \hat L_y, \hat L_z]_{\scriptscriptstyle +} \rangle
- 2\gamma  \,\lbrace 2 \Delta^2_{\hat L_x\hat L_z} \!+\! \Delta^2_{\hat L_x\hat N}\rbrace  \nonumber\\
{\textstyle \frac{\rmd}{\rmd\, t}} \langle \hat L_y \rangle \!\!\!&=&\!\!\!
 2\epsilon \langle \hat L_x \rangle \!+ \!2\U  \langle [ \hat L_x, \hat L_z]_{\scriptscriptstyle +} \rangle
\!-\!2\J  \langle \hat L_z \rangle \!-\! 2\gamma  \,\lbrace 2 \Delta^2_{\hat L_y\hat L_z}
\!+\! \Delta^2_{\hat L_y\hat N}\rbrace\nn\\
{\textstyle \frac{\rmd}{\rmd\, t}} \langle \hat L_z \rangle \!\!\!&=&\!\!\!
 2 \J  \langle \hat L_y \rangle - 2 \gamma  \,\lbrace 2\Delta^2_{\hat L_z\hat L_z} \!+\! \Delta^2_{\hat L_z\hat N}\rbrace,
\end{eqnarray}
and the normalization of the many-particle wave function $|\Psi\rangle$ decays
according to
\begin{equation}
\frac{\rmd}{\rmd\, t} \langle \Psi| \Psi \, \rangle = -2\gamma
\,\big\lbrace 2\langle \hat L_z \rangle+ \langle \hat N \rangle
\big\rbrace \langle \Psi | \Psi \, \rangle.
\end{equation}
The many-particle angular momentum dynamics 
becomes identical to the effective Bloch-equations 
for vanishing interaction, $c=0$, if the initial state is coherent, 
as will become clear later. However, to account 
for the particle number the normalization 
of the many-particle wave function has to be
associated with the $N$-th power of the single particle wave
function.
\begin{figure}[tb]
\begin{center}
\includegraphics[width=4cm]{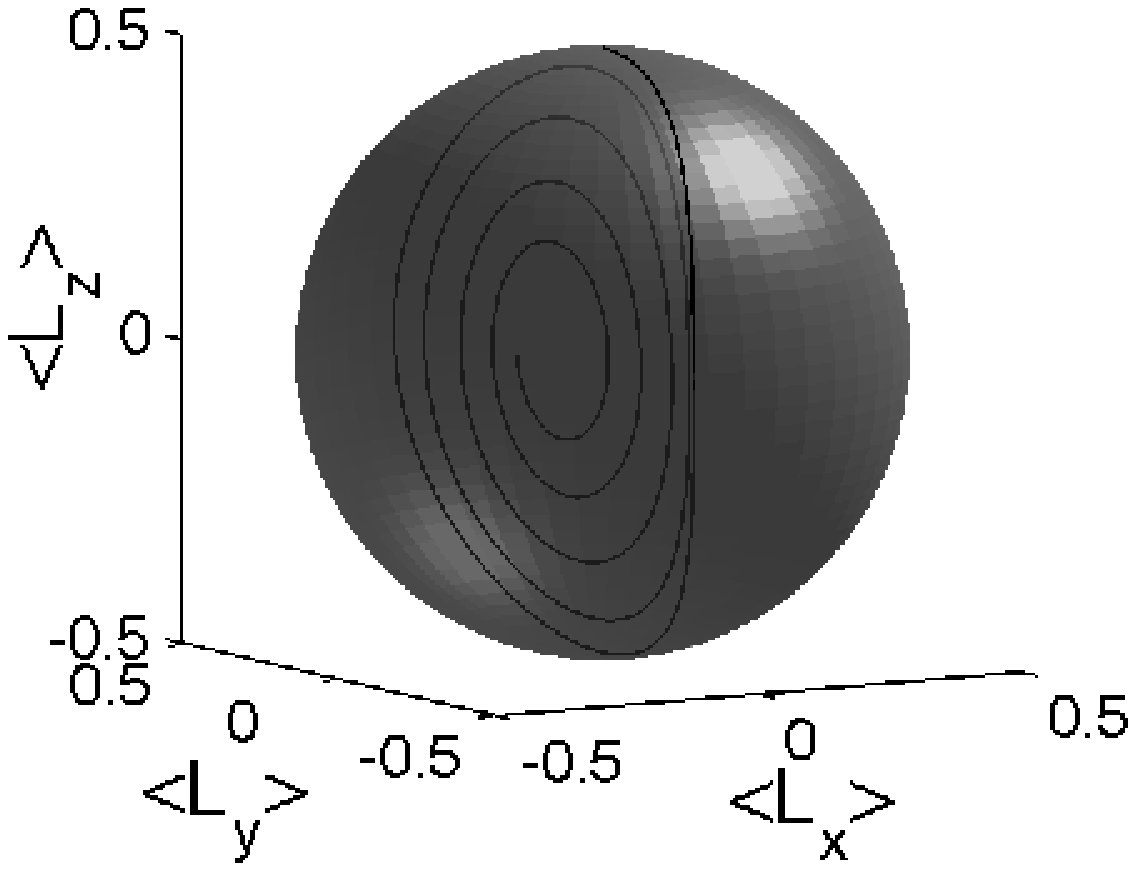}
\includegraphics[width=4.3cm]{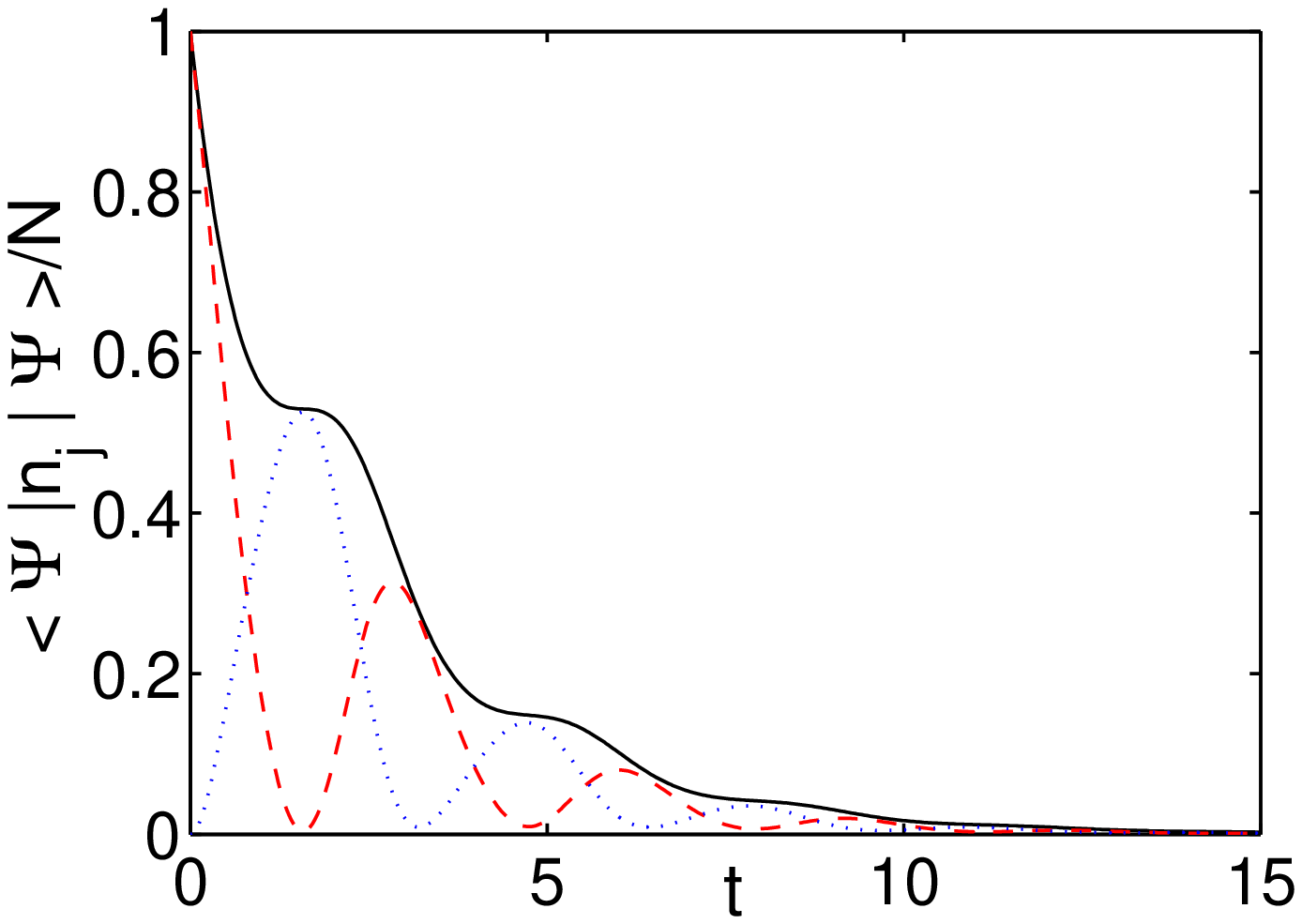}
\caption{\label{pics-MPdyn1} (Color online) 
The left plot shows the dynamics of the expectation 
values of the angular momentum operator for an initial coherent state
located at the north pole of the Bloch sphere (the decaying level) 
for $N=20$ particles, $v=1$, $\gamma=0.01$ and $g=0.5$.
The right plot shows the corresponding decay of the survival
probability (full black curve) and the populations of site $1$
(dashed red curve) and site $2$ (dotted blue curve)}
\end{center}
\end{figure}
\begin{figure}[tb]
\begin{center}
 \includegraphics[width=4cm]{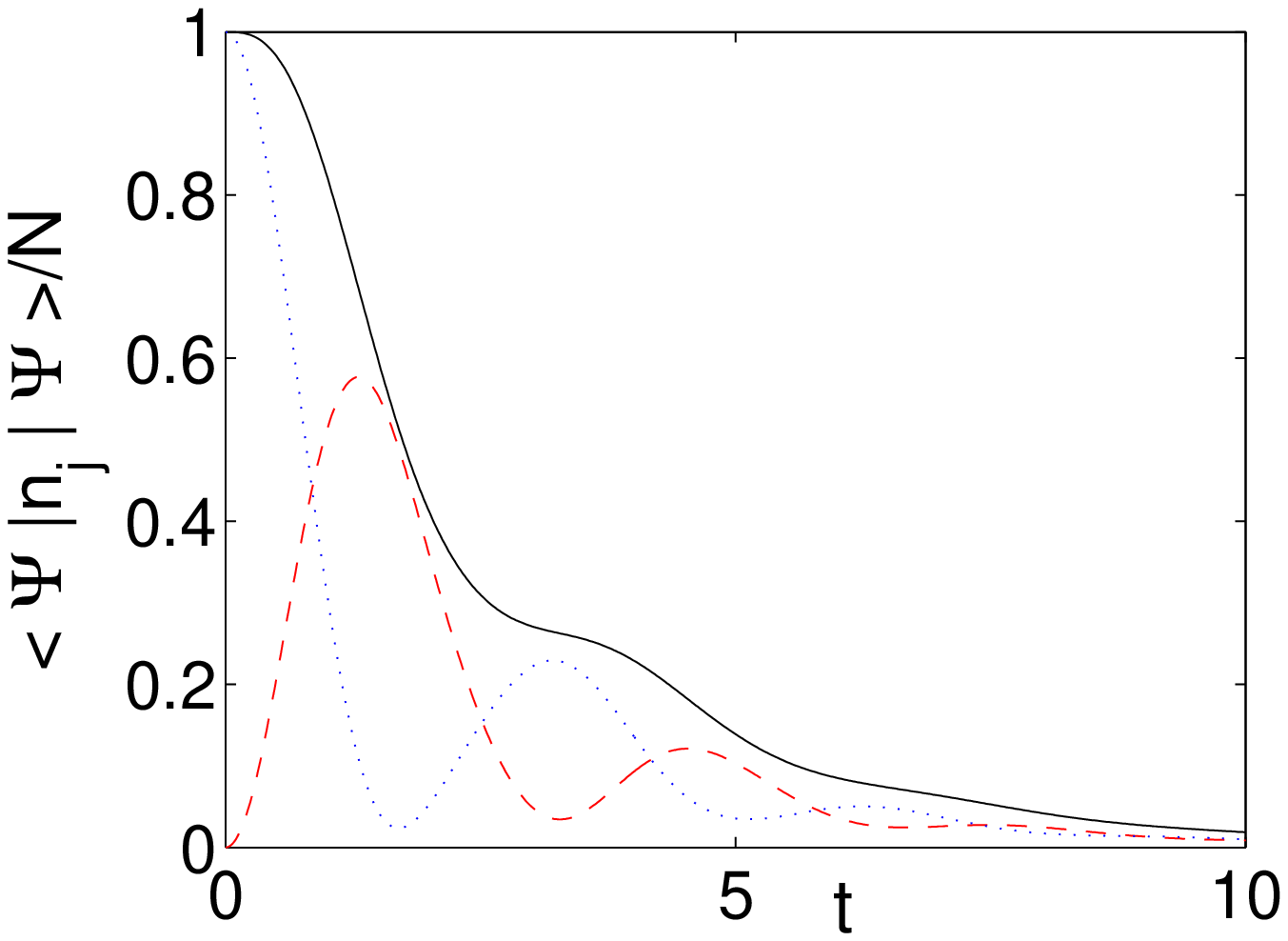}
 \includegraphics[width=4cm]{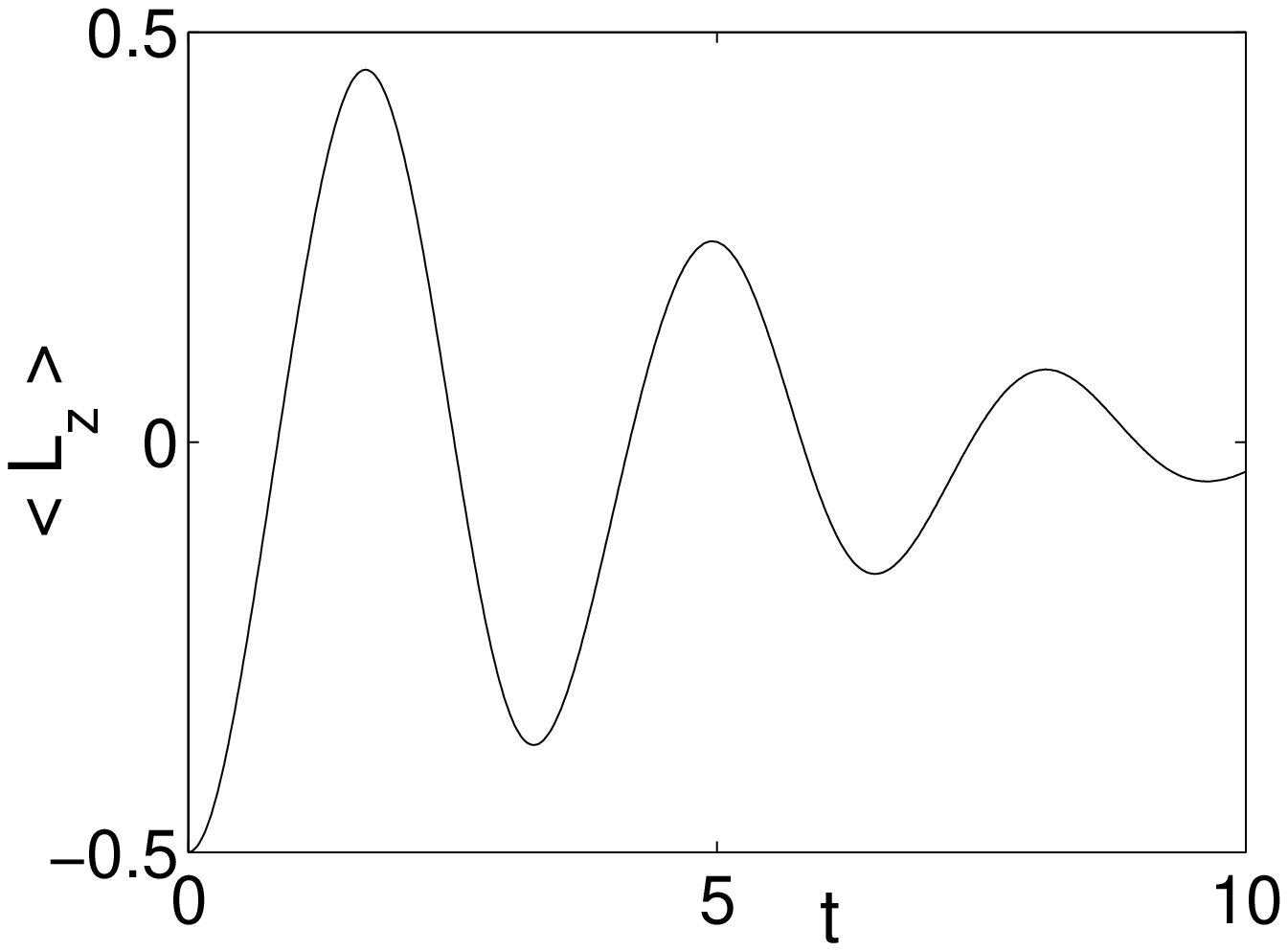}
 \caption{\label{pics-MPdyn2} (Color online) The left plot 
 shows the decay of the survival probability (full black curve) and the populations of
 site $1$ (dashed red curve) and site $2$ (dotted blue curve) 
 for an initial coherent state located at the south pole of the Bloch
 sphere for $N=20$ particles, $v=1$, $\gamma=0.01$ and $g=1$. The 
 right plot shows the corresponding expectation value of $\hat L_z$.}
\end{center}
\end{figure}

To get an impression of the behavior for nonvanishing interaction
strengths we show an example of the many-particle dynamics for a
small value of $\gamma$ and an intermediate value of the
interaction strength $c$ in Fig.~\ref{pics-MPdyn1} for an initial state where 
all particles are in the decaying mode, that is, a state 
located at the north pole of the Bloch sphere. The left plot in
the figure shows the time evolution of the angular momentum
expectation value and the corresponding Bloch sphere. 
Similarly to the Hermitian case \cite{Milb97,Holt01a} the Bloch vector penetrates the
Bloch sphere throughout the time evolution. The right side of the figure
shows the decay behavior captured by the normalization of the
many-particle wave function and the population 
probability $\langle \Psi| \hat n_j|\Psi\rangle/N$ 
of the two levels. The momentary decay rate is proportional to
the expectation value of the $z$-component of the angular
momentum, that is, the population imbalance of the two-modes.
Thus, in comparison with the noninteracting case (that is equivalent
to the behavior of the single particle system investigated in the
previous section) the staircase behavior of the decay is slightly
changed: The steps are not completely flat, having a negative
slope for all times, because the $\hat L_z$ component does not
reach the stable south pole in the depicted time interval. This
behavior becomes more pronounced for stronger interaction
strengths, as depicted for an example in Fig.~\ref{pics-MPdyn2}
where on the right side the $\hat L_z$ expectation value is shown
for comparison. The breakdown behavior in the dynamics of the 
full many-particle observables can be understood as a many particle  
effect on top of the mean-field dynamics which stays confined to the 
Bloch sphere and which we shall introduce in the following. 

\section{The generalized mean-field approximation and a canonical structure}
\label{sec-genMF}
The mean-field approximation in the Hermitian case is often 
formulated in close analogy with the classical approximation 
of single particle quantum mechanics. That is, operators 
are replaced by c-numbers and commutators by 
Poisson brackets, and thus, the Heisenberg equations are replaced 
by Hamiltonian equations. However, this analogy was hitherto 
of little use for non-Hermitian many-particle 
systems, as the classical limit of non-Hermitian 
quantum dynamics itself is still far from being understood. 
Thus, one had to resort to alternative formulations of the 
mean-field approximation. For Hermitian quantum systems the classical analog can be derived in an elegant way using coherent states \cite{Yaff82,Zhan90}. This method has also proven useful in the investigation of the quantum-classical correspondence for cold atoms in optical lattices described by Bose-Hubbard type Hamiltonians where the condensed states are equivalent to $SU(M)$ coherent states \cite{Fran00,Buon08,Trim08}. In \cite{08nhbh_s} a mean-field approximation using generalized coherent states was introduced for the non-Hermitian Bose-Hubbard dimer \eqref{eqn-BH-Hamiltonian}. 
Here we provide details of this generalized mean-field approximation and 
connect it to a recently proposed classical 
approximation for non-Hermitian single particle 
quantum dynamics \cite{09nhclass} where a generalized 
canonical structure arises. Although it has 
only been derived for a flat phase space, it has been shown 
that the mean-field approximation for the present model 
can be formulated in terms of the proposed 
generalized canonical equations of motion. 
From a practical perspective, 
making use of the generalized canonical structure 
strongly simplifies the calculation yielding the 
mean-field dynamics. This is promising for the generalization to 
larger systems involving more than two states.

The underlying idea of the generalized mean-field approximation 
\cite{08nhbh_s} is to describe the whole ensemble of many-particles by only one
macroscopic wave function in the limit of infinite particle
number. In other words, we assume that the particles form a condensate 
throughout the time evolution. For a two-mode system the 
fully condensed states can be expressed in the form
\begin{equation}\label{eqn-condensed-states}
|x\rangle= \frac{1}{\sqrt{N!}} \left( x_1 \hat a^\dagger _1 + x_2
\hat a^\dagger _2 \right)^{N} | 0, 0 \rangle,
\end{equation}
with two complex coefficients that are not necessarily normalized 
to unity, $n=|x_1|^2+|x_2|^2$. The condensed many-particle wave function 
\rf{eqn-condensed-states} is then normalized to 
$\langle x|x\rangle=(|x_1|^2+|x_2|^2)^N=n^N$.
These states are in fact equivalent to the generalized $SU(2)$ coherent
states \cite{Pere86,Zhan90}, often denoted also as atomic coherent states. They can be constructed by an arbitrary $SU(2)$ rotation 
$\hat R(\theta,\phi)= {\rm e}^{{\rm i}\theta(\hat L_x \sin{\phi} - \hat L_y\cos{\phi})}$ 
of an extremal Fock state ,e.g., $|N\rangle$, where all particles are in the first mode:
\begin{equation}
|\theta,\phi\rangle = \hat R(\theta,\phi)|N\rangle.
\end{equation}
This is equivalent to \rf{eqn-condensed-states} if we set
\begin{equation}
x_1 = \sqrt{n}{\rm e}^{-{\rm i}\phi}\cos{\tfrac{\theta}{2}}, \quad x_2 =
\sqrt{n}\sin{\tfrac{\theta}{2}}.
\end{equation}
Thus, the mean-field approximation is equivalent
to the assumption that the many-particle state, initially chosen
as a coherent state, remains coherent for all times of interest.
This assumption is in fact exact if the Hamiltonian is a linear 
superposition of the generators of the dynamical symmetry group \cite{Zhan90}, 
in our case for vanishing interaction $\U=0$. This can be seen by 
calculating the action of the time evolution operator on 
an initially coherent state. For nonvanishing interaction it 
is in general an approximation yielding the mean-field 
dynamics. The mean-field equations of motion 
can thus be obtained from the quantum dynamics 
by replacing all expectation values with their 
values in coherent states and identifying these 
with the mean-field quantities. The resulting mean-field dynamics 
can be interpreted as a special case of constrained 
quantum motion \cite{Brod08b} where the constraint is that the 
many-particle state is coherent. 

Let us now derive the mean-field Bloch dynamics from the 
equations of motion for the many-particle angular momentum 
expectation values \rf{KomZerfall1} using the $SU(2)$ coherent 
state approximation. The expectation values of 
the $\hat L_i$, $i=x, y, z$ in terms of the coherent state coordinates 
$x_1$ and $x_2$ read:
\begin{eqnarray}
\label{Erwdreh}
\frac{\langle x|\hat L_x|x \rangle}{\langle x|x\rangle }&=&\frac{N}{2}\frac{x_1^* x_2+ x_1 x_2^*}{x_1^*x_1+x_2^*x_2},\nn\\
\frac{\langle x|\hat L_y|x \rangle}{\langle x|x\rangle }&=&\frac{N}{2\rmi}\frac{x_1^* x_2- x_1 x_2^*}{x_1^*x_1+x_2^*x_2},\\
\frac{\langle x|\hat L_z|x \rangle}{\langle x|x\rangle }&=&\frac{N}{2}\frac{x_1^* x_1 -  x_2^* x_2}{x_1^*x_1+x_2^*x_2}\nn.
\end{eqnarray}
We can identify these quantities with the components of the
corresponding renormalized mean-field Bloch vector:
\begin{equation}
\s_j=\langle \hat L_j \rangle/N.
\end{equation}
Comparison with the definition of the mean-field Bloch
vector in the single particle case \rf{eqn-bloch-vector} 
reveals that the coordinates of the
coherent state can naturally be associated with the components of
the effective single particle wave function $\psi$. To perform the mean-field
approximation we further need the expectation values of the
anti-commutators appearing in \rf{KomZerfall1} for $SU(2)$ coherent states 
which factorize as
\begin{eqnarray} \label{ErwAntiDrehexakt}
\langle [\hat L_i, \hat L_j]_{\scriptscriptstyle +} \rangle &=& 2
(1-\tfrac{1}{N} ) \langle \hat L_i \rangle \langle \hat L_j
\rangle + \delta_{ij} \frac{N}{2}\,,\nn\\
\langle
[\hat L_i, \hat N]_{\scriptscriptstyle +} \rangle &=& 2N\langle \hat
L_i \rangle,
\end{eqnarray}
with $N=\langle \hat N \rangle$.
Inserting these expressions into (\ref{KomZerfall1}) and
taking the macroscopic limit $N\to\infty$ with $N\U =\g $ fixed we
obtain the desired non-Hermitian mean-field evolution equations:
\begin{eqnarray}\label{eqn-bloch-nherm-nlin}
\begin{array}{rrrrll}
\dot{\s}_x=&-2 \epsilon \s_y&- 4 \g \s_y \s_z & &+ 4 \gamma \,\s_x
\s_z, \\ 
\dot{\s}_y=&+2 \epsilon \s_x&+4 \g \s_x \s_z &-2 \J  \s_z
&+ 4 \gamma \,\s_y \s_z,\\ \dot{\s}_z=& & &+2 \J  \s_y&-\gamma\,
(1-4\s_z^2)\,.
\end{array}
\end{eqnarray}
These nonlinear non-Hermitian Bloch equations
are real valued and conserve $\s^2=\s^2_x+\s^2_y+\s^2_z=1/4$, 
i.e.~the dynamics are regular and confined to the Bloch sphere. 
The total probability $n$ decays as
\begin{equation}\label{Norm}
\dot n =- 2 \gamma \left(2\s_z+1\right) n\,.
\end{equation}

In the limit $g=0$ in which the assumption that the many-particle
state stays coherent in time is exactly fulfilled, these equations
reduce to the equations for the linear single particle two level
system \rf{eqn_bloch_nherm}. Thus, as mentioned before, this
captures the exact many-particle dynamics in this limit. 
Generalized Bloch equations related to \rf{eqn-bloch-nherm-nlin} 
also appear in a different context, where the influence of 
decoherence is investigated \cite{Trim08b,Witt08,Morr08a,Morr08b,Morr08c}. 
It should further be noted that they can be considered a special case of the celebrated Landau-Lifshitz equations with Gilbert damping 
appearing frequently in magnetization dynamics.

Let us now express the mean-field dynamics in the form of a generalized
nonlinear Schr\"odinger equation. In terms of the components $\psi_j$ of
the unnormalized wave function (associated with the coordinates $x_j$
of the many-particle coherent state) this can be
formulated as:
\begin{equation}\label{nlnhGP}
\rmi\frac{\rmd}{\rmd\, t}\begin{pmatrix} {\psi}_1 \\{\psi}_2 \end{pmatrix}
=\begin{pmatrix} \epsilon + g \kappa - 2\rmi \gamma & v \\
  v & - \epsilon - g\kappa \end{pmatrix}\begin{pmatrix}\psi_1\\ \psi_2\end{pmatrix},
  \end{equation}
  with
  \begin{equation}
\quad
  \kappa=\frac{|\psi_1|^2-|\psi_2|^2}{|\psi_1|^2+|\psi_2|^2}.
\end{equation}
The equation of motion for the normalization in this 
formulation is given by $\dot n=-2\gamma(1-\kappa)n$. 
While in the limit $\gamma\to0$ the wave 
function stays normalized and the equations are 
thus equivalent to the usual discrete nonlinear Schr\"odinger
equation of Gross-Pitaevskii type, the nonlinear term gets
modified due to the non-Hermiticity. 
Alternatively we can express the dynamics in terms of the renormalized 
wave function $\varphi_j=\psi_j/\sqrt{n}$:
\begin{equation}\label{nlnhGPn}
\rmi\frac{\rmd}{\rmd\, t}\!\!\begin{pmatrix}\!{\varphi}_1\! \\ \!{\varphi}_2\! \end{pmatrix}
\!=\!\begin{pmatrix} \epsilon \!+\! g \kappa \!- \!\rmi \gamma(1-\kappa) & v\!\! \\
   v & \!- \!\epsilon \!-\! g\kappa\!+\!\rmi\gamma(1\!+\!\kappa)\!\!  \end{pmatrix}
  \! \begin{pmatrix}\! \varphi_1\!\\ \!\varphi_2\!\end{pmatrix},
\end{equation}
with $\kappa=|\varphi_1|^2-|\varphi_2|^2$. This dynamics by
definition conserves the normalization $||\varphi||^2=1$. 

Note that the dynamics induced by the nonlinear non-Hermitian 
Schr\"odinger equation \rf{nlnhGP} differs fundamentally 
from the dynamics of a discrete Gross-Pitaevskii equation 
with an additional imaginary on-site 
energy, where the nonlinearity is determined by
$\kappa=|\psi_1|^2-|\psi_2|^2$. This latter type of non-Hermitian
nonlinear Schr\"odinger equations has attracted considerable 
attention in the context of the description of
scattering phenomena and the influence of leaking boundaries for
Bose-Einstein condensates recently \cite{Scha97b,Hill06,06nlnh,Livi06,07nlres,Fran07}. 
Furthermore, these \textit{ad hoc} nonlinear non-Hermitian equations 
also appear for absorbing nonlinear waveguides \cite{Muss08,Makr08,Elga07,Klai08}. 

In \cite{09nhclass} it has been shown that the mean-field 
approximation of the non-Hermitian Bose-Hubbard dimer 
can also be expressed in terms of a generalized canonical structure, as 
we will review in what follows. 
The generalized canonical equations of motion proposed 
in \cite{09nhclass} are of the form 
\begin{equation}\label{eqn-grad-flow}
\left(\begin{array}{c}
\dot q\\
\dot p\end{array}
\right)= \Omega^{-1}\vec{\nabla}H-G^{-1}\vec{\nabla}\Gamma,
\end{equation}
where $p$ and $q$ are canonical phase space variables and 
$\vec{\nabla}$ denotes the phase space gradient, $\Omega$ 
is the symplectic matrix
\begin{equation}\label{eqn-sympl}
\Omega=\left(\begin{array}{cc}
0 & -1 \\
1 & 0
\end{array}\right)
\end{equation}
and $G$ is the corresponding K\"ahler metric \cite{Arno78,Brod01b} on the 
relevant phase space. The classical Hamiltonian function $\cH=H-\rmi\Gamma$ 
is given by the expectation value of the quantum Hamiltonian 
in the relevant coherent states. 
The dynamics of the normalization of the original wave function
$n=|\psi_1|^2+|\psi_2|^2$ is governed by the equation of motion 
\begin{equation}
\label{eqn-can-norm}
\dot n=-2\Gamma n.
\end{equation}
The dynamical equation (\ref{eqn-grad-flow}) is a combination of a
canonical symplectic flow generated by the real part $H$ of the
Hamiltonian function and a canonical gradient flow generated by the
imaginary part $\Gamma$. The symplectic part evidently gives rise to
the familiar Hamiltonian dynamics of classical mechanics. The gradient
vector with a negative sign points in the direction of the steepest
descent of the function $\Gamma$. Thus, this part of the
dynamics aims to drive the system toward the minimum of 
$\Gamma$ and can naturally be associated with a damping. 

The generalized canonical structure can be used to directly 
calculate the mean-field dynamics without evaluating 
the generalized Heisenberg equations of motion 
and performing the coherent state approximation as follows: 
Our classical phase space is given by the Bloch sphere and can be 
parametrized by the canonical variables $p$ and $q$ 
that are related to the classical Bloch vector via 
\begin{eqnarray}
\s_x&=&\half\sqrt{1-p^2}\cos(2q)\nn\\
\s_y&=&\half\sqrt{1-p^2}\sin(2q)\label{eqn_bloch_pq}\\
\s_z&=&\half p.\nn
\end{eqnarray}
We can express the expectation value of the many-particle 
Hamiltonian \rf{eqn-BH-Hamiltonian} in $SU(2)$ coherent states 
in the variables $p,q$ to find the classical Hamiltonian function:
\begin{eqnarray}\label{eqn-nherm-nlin-ham-fct}
H=\epsilon p+v\sqrt{1-p^2}\cos(2q)+\frac{g}{2}p^2\quad{\rm and}\quad
\Gamma=\gamma p.
\label{classical-E-pq}
\end{eqnarray}
The K\"ahler metric on the Bloch sphere in the variables $q,p$ is given by \cite{09nhclass} 
\begin{equation}\label{eqn-metric-Bloch-pq}
G=\left(\begin{array}{cc}
2(1-p^2) & 0\\
0 & \frac{1}{2(1-p^2)}
      \end{array}\right).
\end{equation} 
Evaluating the generalized canonical equations 
\rf{eqn-grad-flow} of motion yields
\begin{eqnarray}
\dot{q}&=&\epsilon+gp-v\frac{p}{\sqrt{1-p^2}}\cos(2q)\\
\dot{p}&=&-2\gamma(1-p^2)+2v\sqrt{1-p^2}\sin(2q),
\end{eqnarray}
which is equivalent to the nonlinear Bloch equations \rf{eqn-bloch-nherm-nlin}. Similar equations also appear in a related model where a different mean-field approximation is applied \cite{Shch10}.

We note here that the expressions arising for the $\cP\cT$-symmetric
version of the Hamiltonian \rf{eqn-2times2-PT} differ from the
present ones by a complex energy shift. Thus, since the generalized
canonical equations of motion are invariant under a constant energy
shift (as are the usual canonical equations of motion of Hamilton
type), the effective dynamics resulting from the Hamiltonian
functions related to \rf{eqn-2times2-nherm} and \rf{eqn-2times2-PT}, 
respectively, are identical, in agreement with
the previous observations. 

Note also that the nonlinear Schr\"odinger equation \rf{nlnhGPn} can be directly 
formulated as generalized complex canonical equations of motion 
for the coordinates $\varphi_1, \varphi_1^*, \varphi_2, \varphi_2^*$. 
However, here one has to take care of the constraints confining the 
dynamics to the Bloch sphere explicitly and the expression for the metric 
gets more elaborate (see Appendix \ref{chap-metric}). Thus, 
for practical purposes the formulation in real canonical 
variables $p,q$ is more convenient. 

\section{Mean-field dynamics and fixed point structure}
\label{sec-MF-dyn}
In this section we analyze the mean-field behavior arising 
from the interplay of non-Hermiticity and nonlinearity. 
The mean-field dynamics is organized according to fixed 
points, which correspond to stationary solutions of the 
nonlinear complex Schr\"odinger equation \rf{nlnhGP}. 
In contrast to the widely investigated 
behavior of vector fields in $\RR^2$, the general features of 
vector fields on the sphere have rarely been investigated 
in detail. Only recently some interest in polynomial vector 
fields as the present one on the two-sphere ${\mathbb S}^2$ 
has emerged in the mathematical literature \cite{Llib06a,Llib06b,Llib06c,Guti02}. 
In this context it was shown that the upper bound of the 
number of fixed points for a general polynomial vector field 
of degree $2$ on the sphere is equal to 6. 

In the present case there are at most four fixed 
points that can be obtained analytically as the 
roots of a fourth order polynomial 
similar to the Hermitian case \cite{Wu00}. 
To see this we have to study the fixed point equation defined by 
(\ref{eqn-bloch-nherm-nlin}) with $\dot{\vec{s}}=0$, which provides
\begin{equation}\label{eqn-nherm-nlin-fixed-points-y}
 v s_y=2 \gamma \left(\tfrac{1}{4}-s_z^2\right).
\end{equation}
Using this and the normalization condition 
$s_x^2+s_y^2+s_z^2=\tfrac{1}{4}$ shows that the 
$s_z$ coordinates of the fixed points are given by 
the real roots of the fourth order polynomial
\begin{equation}\label{fixedpoints}
4(\g^2\!+\!\gamma^2)\s_z^4\!+\!4\g\epsilon \s_z^3\!+\! (\epsilon^2\!+\!\J^2
\!-\!\g^2\!-\!\gamma^2)\s_z^2\!-\!\g\epsilon\s_z\!-\!\epsilon^2/4\!=\!0.
\end{equation}
In the following we will restrict the discussion to the unbiased 
case $\epsilon=0$ where the polynomial (\ref{fixedpoints}) 
becomes biquadratic and the fixed points are easily found analytically. 
The analysis can in principle be extended to the case 
$\epsilon\neq0$ in a straightforward manner. For $\epsilon=0$ 
the polynomial \rf{fixedpoints} has the four solutions $s_z=0,\,0,\, \pm\frac{1}{2} \sqrt{1-\frac{v^2}{g^2+\gamma^2 }}$. The corresponding 
values of $s_x$ and $s_y$ are then given by 
\rf{eqn-nherm-nlin-fixed-points-y} and 
the normalization condition. In summary, this yields the 
solutions
\begin{equation}\label{Loesung12}
 \vec s_{c\pm}
 \!\!=\!\! \begin{pmatrix}\pm \frac{1}{2} \sqrt{1-\frac{\gamma^2}{v^2}}\\\frac{\gamma}{2 v}\\0\end{pmatrix}\!,\quad 
 \vec s_{f\pm}\!\!=\!\!\begin{pmatrix}
 \frac{g  v}{2 \left( g^2+\gamma^2 \right)}\\
 \frac{\gamma v}{2 \left( g^2+\gamma^2 \right)}\\
 \pm \frac{1}{2} \sqrt{1-\frac{v^2}{g^2+\gamma^2 }}\end{pmatrix}.
\end{equation}
Since the components of the Bloch vector are by definition 
real valued, only the real solutions correspond to actual fixed points.
Due to the non-Hermiticity these are not necessarily 
elliptic fixed points or saddle points, which are the only possibilities 
in Hamiltonian systems. Rather, as we already observed for the 
linear non-Hermitian case, the additional gradient flow can 
lead to a destruction of periodic motion and introduce sinks 
and sources to the dynamics. In principle it can also lead to 
the emergence of limit cycles \cite{Kors08,Guck83} which, 
however, have not been observed in the study of the present 
system. 

For a flow on a two dimensional surface 
(as in the case of the Bloch sphere), information on the type 
of fixed points can be obtained by the surrounding linearized 
fields (apart from special cases at parameter values for 
which bifurcations occur, see e.g., \cite{Kors08} and references therein). 
We will now briefly introduce the classification scheme; 
further details can be found, e.g., in \cite{Arno06,Dumo06,Seyr03,Arno88}.
\begin{table}[tb]
\caption{\label{table_fix_class} Classification of fixed points
according to the eigenvalues $\lambda_{1}$ and $\lambda_{2}$ of
the Jacobi matrix.}
\begin{ruledtabular}
\begin{tabular}{l|l|l|l}
$\lambda_{1},\lambda_{2} \in \RR $&   $\lambda_{1},\lambda_{2} <0$& Stable node (sink) & Index $+1$\\
& $\lambda_{1},\lambda_{2} >0$& Unstable node (source) & Index
$+1$\\ & $\lambda_{1}\lambda_{2}<0$  & Saddle point & Index
$-1$\\\hline
$\lambda_{1,2} = \alpha \pm {\rm i} \beta$ & $\alpha<0$  & Stable focus (sink) & Index $+1$ \\
& $\alpha>0$  & Unstable focus (source) & Index $+1$\\ &
$\alpha=0$  & Center & Index $+1$
\end{tabular}
\end{ruledtabular}
\end{table}

Suppose we have a system of two first order differential 
equations which can be written in the form
\begin{equation}
 \dot q_1=F_1(q_1,q_2), \quad \dot q_2=F_2(q_1,q_2).
\end{equation}
The linearization of this system around an arbitrary point is 
determined by the Jacobi matrix $D_{ij}=\p F_i / \p q_j$, $(i,j=1,2)$ 
of the vector field $\vec F$ at that point. The eigenvalues 
$\lambda_{1,2}$ of this matrix at a fixed point of the 
dynamics, that is, a singular point of the vector field 
$\vec F$, can yield information about the fixed point type. 
These eigenvalues are either real or form a complex 
conjugate pair, due to the reality of the matrix. One 
can distinguish four basic fixed point types (nodes, saddle 
points, foci and centers) and subclasses 
according to the values of $\lambda_{1,2}$, which are 
summarized in Table \ref{table_fix_class}. 

The so-called (Poincar\'e) index, also listed in the table, 
is a further characteristic quantity of a singular point of a 
vector field with respect to an oriented surface 
(see, e.g., \cite{Arno06,Arno88} for details). It is defined 
as the number of revolutions of the vector field 
in traversing an arbitrary curve encircling the (isolated) singular 
point (and no other singular point). The index of a saddle point 
is $-1$, whereas the indices of nodes, foci and centers are all 
equal to $+1$. The number and type of singular points of 
a vector field and the possible bifurcation scenarios for a 
given manifold are restricted by the \textit{index theorem}. 
It states that the sum of the indices of the singular 
points of a vector field on a manifold is independent 
of the choice of the vector field and equals the 
Euler characteristic $\chi_{\rm E}$, which is $\chi_{\rm E}=2$ 
in the case of a sphere.

\begin{figure}[tb]
\centering
\includegraphics[width=8cm]{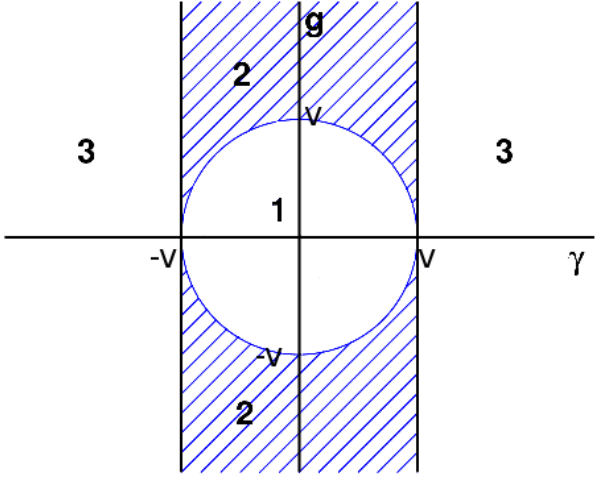}
\caption{\label{fig-fixpointstructure}
Parameter regions belonging to different fixed point configurations of the 
non-Hermitian mean-field dynamics \rf{eqn-bloch-nherm-nlin}.}
\end{figure}

The fixed points of our nonlinear non-Hermitian system 
\rf{eqn-bloch-nherm-nlin} can be categorized completely 
according to the above scheme.  
In summary, we can distinguish three regions in parameter 
space, which are sketched in Fig.~\ref{fig-fixpointstructure}:
\begin{enumerate}
\item In region 1, for $\gamma^2+g^2\leq v^2$, we have
only two fixed points $\vec s_{c\pm}$ which are located at the equator. For 
$g\neq0$ one of them is a sink, the other one is a source. They 
degenerate to centers for $g=0$.
\item In region 2, for $\gamma^2+g^2>v^2$ and 
$|\gamma|<|v|$, there are four coexisting
fixed points, namely, a sink and a source, a center, and a saddle
point. On the line $\gamma=0$ the sink and source become centers,
corresponding to the Hermitian self-trapping states.
\item In region 3, which is defined by $|\gamma|>|v|$, 
only the fixed points $\vec s_{f\pm}$ exist, namely a sink and a source. For positive $\gamma$ we 
have a source on the northern hemisphere and a sink on the southern. 
In general ($\g\neq 0$) they are foci, which 
become nodes in the linear limit.
\end{enumerate}
At the boundaries of these regions, at the critical parameter values, 
bifurcations that necessarily respect the index theorem occur. 
\begin{figure}[tb]
\centering
\includegraphics[width=4cm]{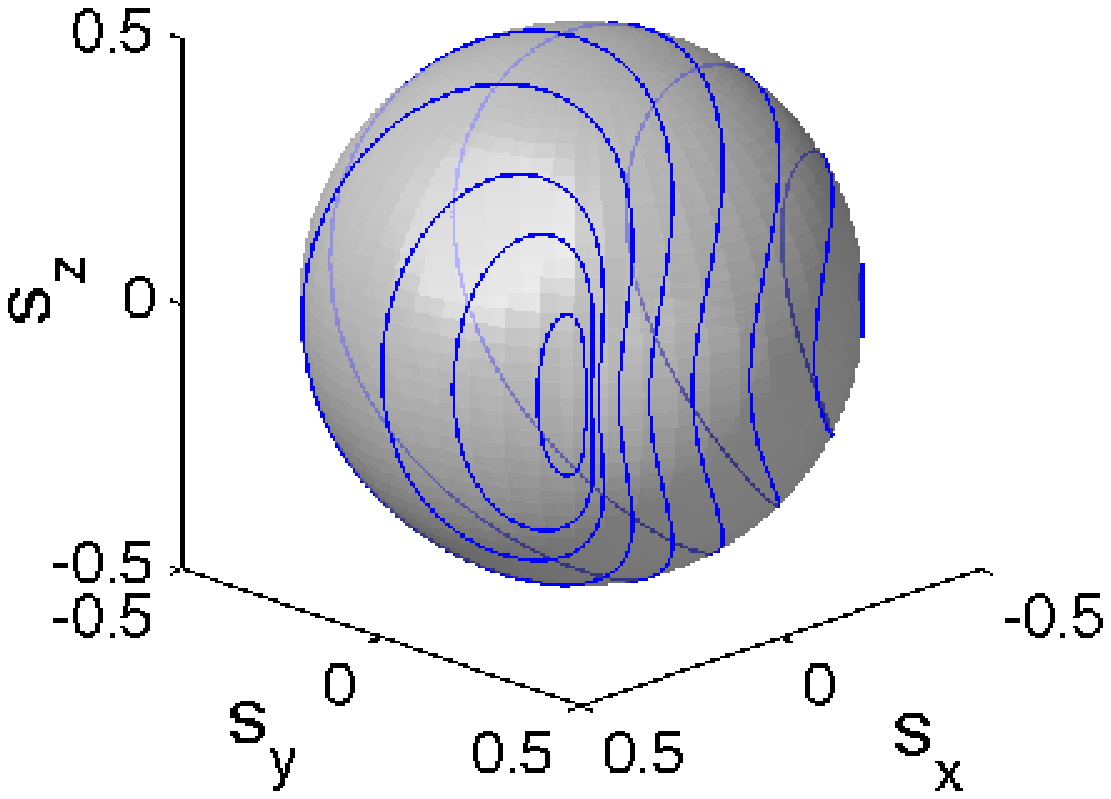}
\includegraphics[width=4cm]{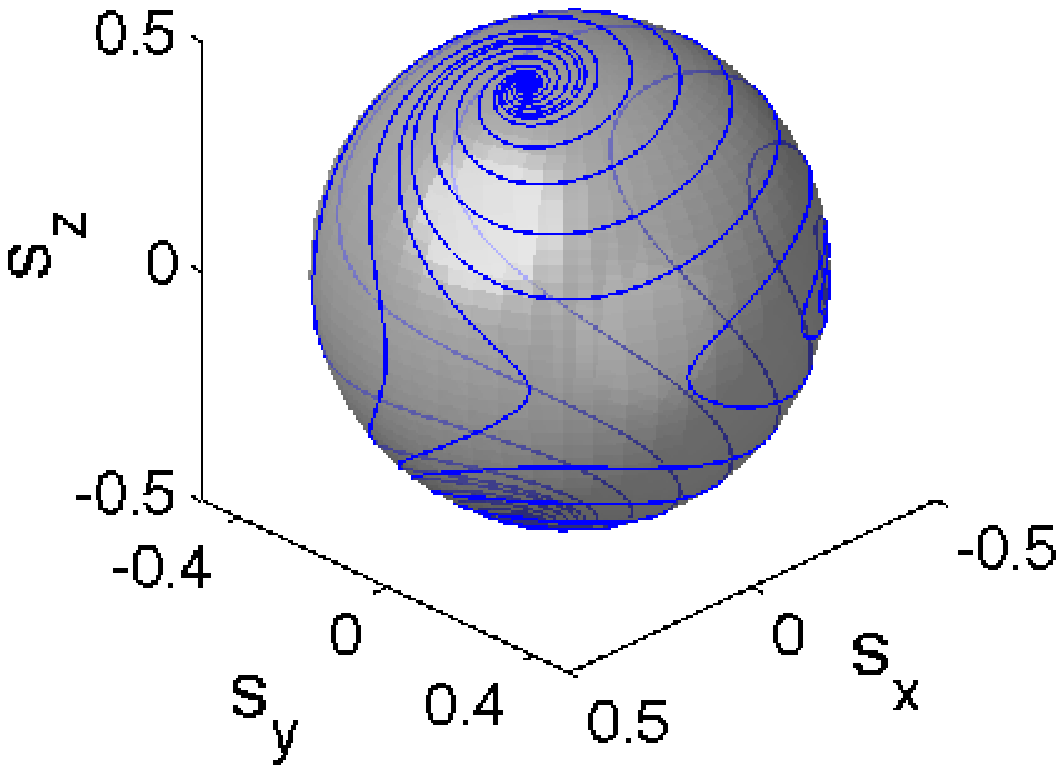}
\includegraphics[width=4cm]{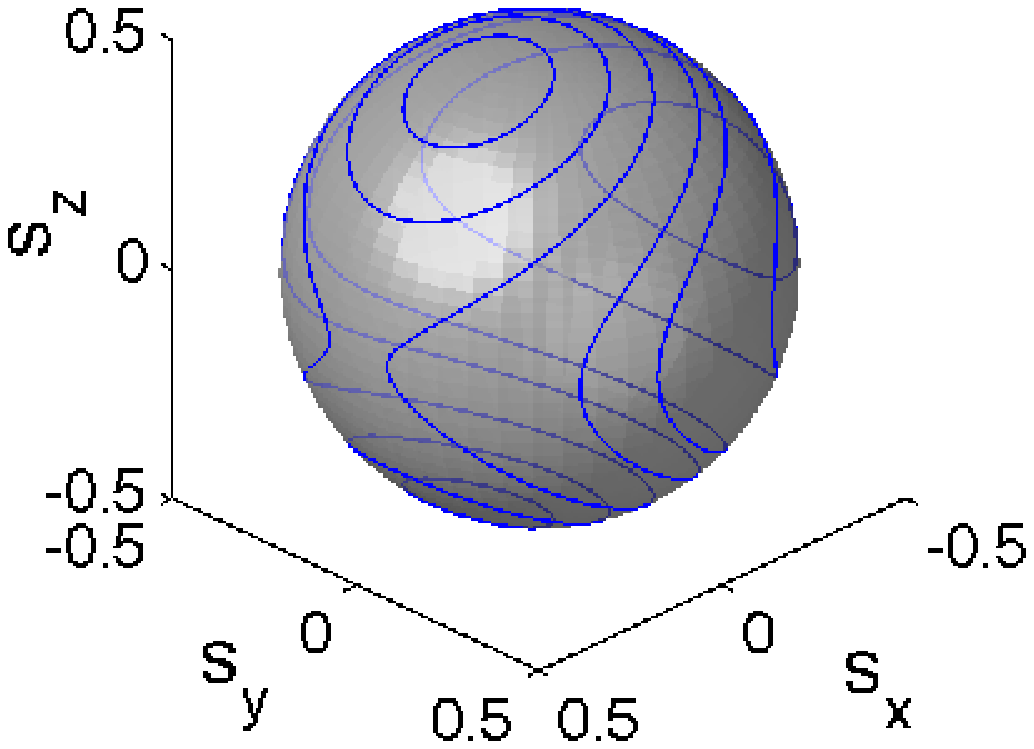}
\includegraphics[width=4cm]{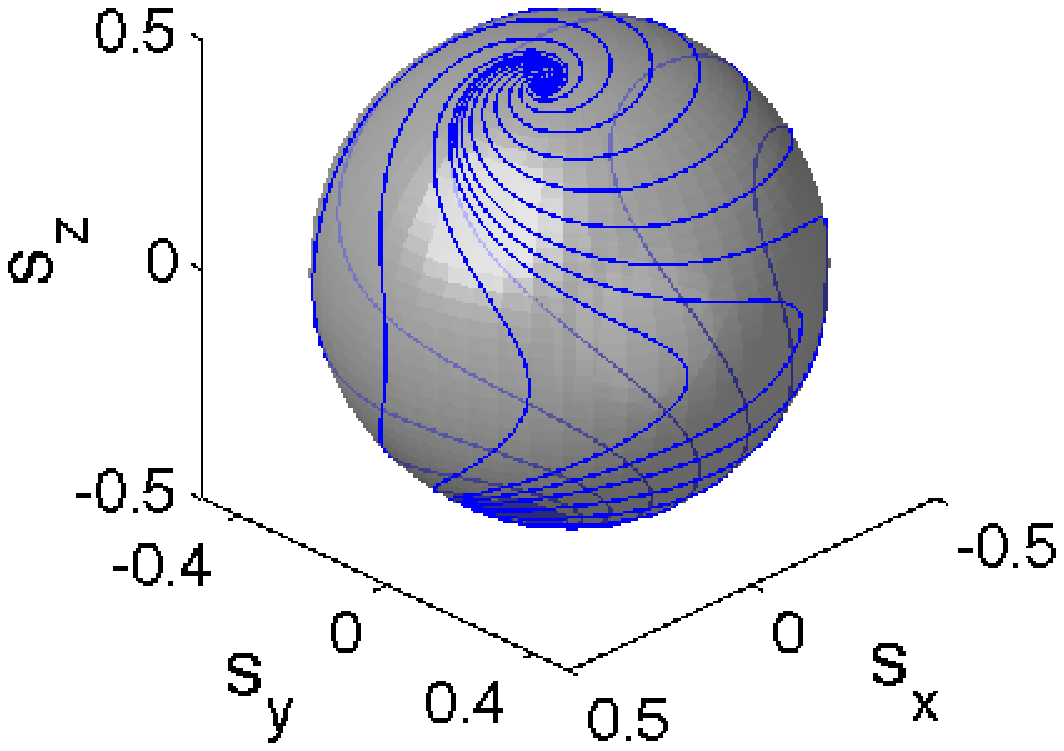}
\caption{\label{fig-bloch-dyn-nherm-nlin} Mean-field dynamics on the Bloch 
sphere for $\epsilon=0$, $\J=1$ and different values of $\gamma$ and $\g$. 
(Left to right and top to bottom: $\gamma=0.7,\, \g=0.7;\ \gamma=0.75,\, g=3;\ \gamma=0, \, g=3;
\ \gamma=1.25,\, g=3$)}
\end{figure}

Figure \ref{fig-bloch-dyn-nherm-nlin} shows examples of 
the Bloch dynamics \rf{eqn-bloch-nherm-nlin} 
in the three different regions and on the Hermitian line. 
In the first plot (left on the top) the dynamics is shown 
for $\gamma=0.7$ and $g=0.7$, that is, in region 1. 
We observe deformed Bloch oscillations surrounding 
the two centers (index $+1$). If the non-Hermiticity 
is increased, the centers approach each other along the equator. 
However, before they meet one of them bifurcates at the
critical circle $\g^2+\gamma^2=\J^2$ (and $\gamma\neq0$) 
into a saddle (index $-1$) and two foci (index $+1$), 
$\vec s_{f\pm}$, one stable (a sink) 
and one unstable (a source). 

The second plot (right on the top) in 
Fig.~\ref{fig-bloch-dyn-nherm-nlin} shows 
the resulting dynamics above this bifurcation, however 
still in region 2. Here we observe four fixed points resulting 
in a mixed dynamics, where besides the periodic motion 
surrounding the remaining center $\vec s_{c\pm}$  there are flows 
from the source to the sink. The appearance of the two fixed 
points $\vec s_{f\pm}$  can be viewed as a non-Hermitian 
self-trapping dynamics, which collapses to the Hermitian case 
in the singular limit $\gamma=0$. This is depicted in the third 
plot (left in the lower panel). Here the foci are replaced by centers. 
The Hermitian self-trapping effect arises as a bifurcation of the center 
(index $+1$) into a saddle (index $-1$) and the additional 
centers (index $+1$) at the critical circle. Thus, 
the critical value $g_{\rm crit}=\sqrt{\J^2-\gamma^2}$ is decreased for 
$\gamma\neq 0$ compared to the Hermitian case. 
In other words, the presence of the non-Hermiticity promotes 
the self-trapping effect, however, the resulting self-trapping oscillations 
are damped due to the non-Hermiticity, as we shall discuss later. 
\begin{figure}[tb]
\begin{center}
\includegraphics[width=4.2cm]{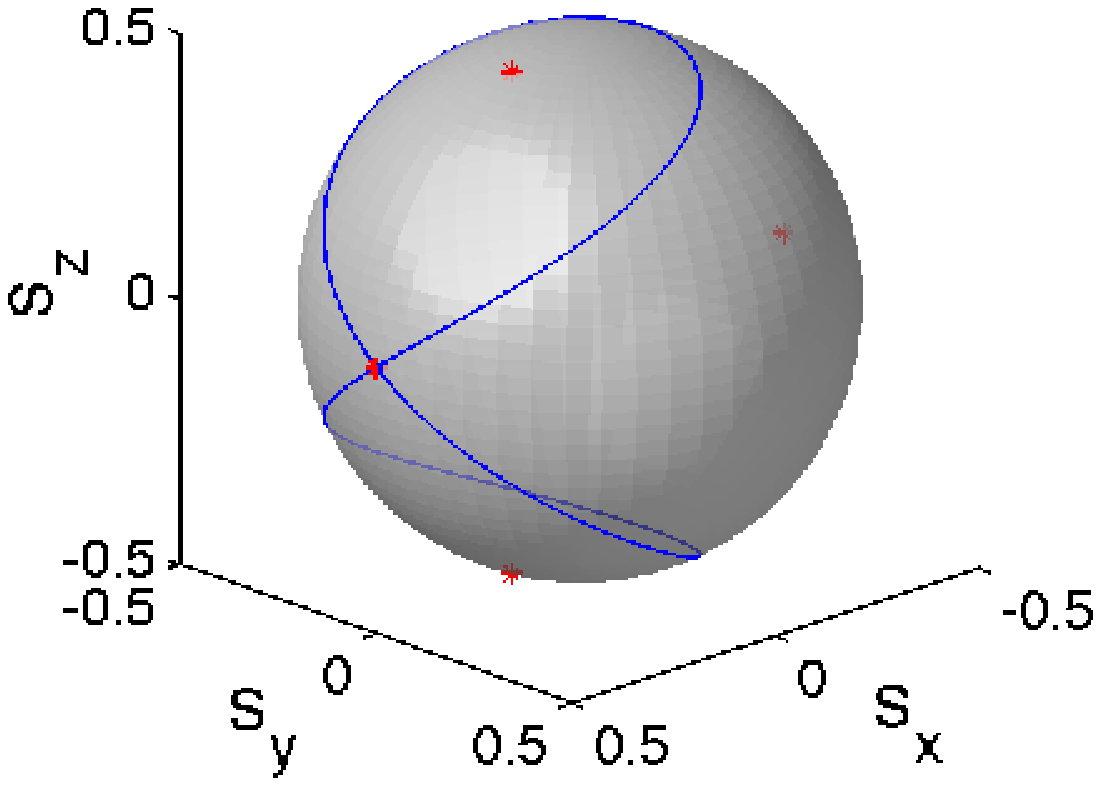}
\includegraphics[width=4.2cm]{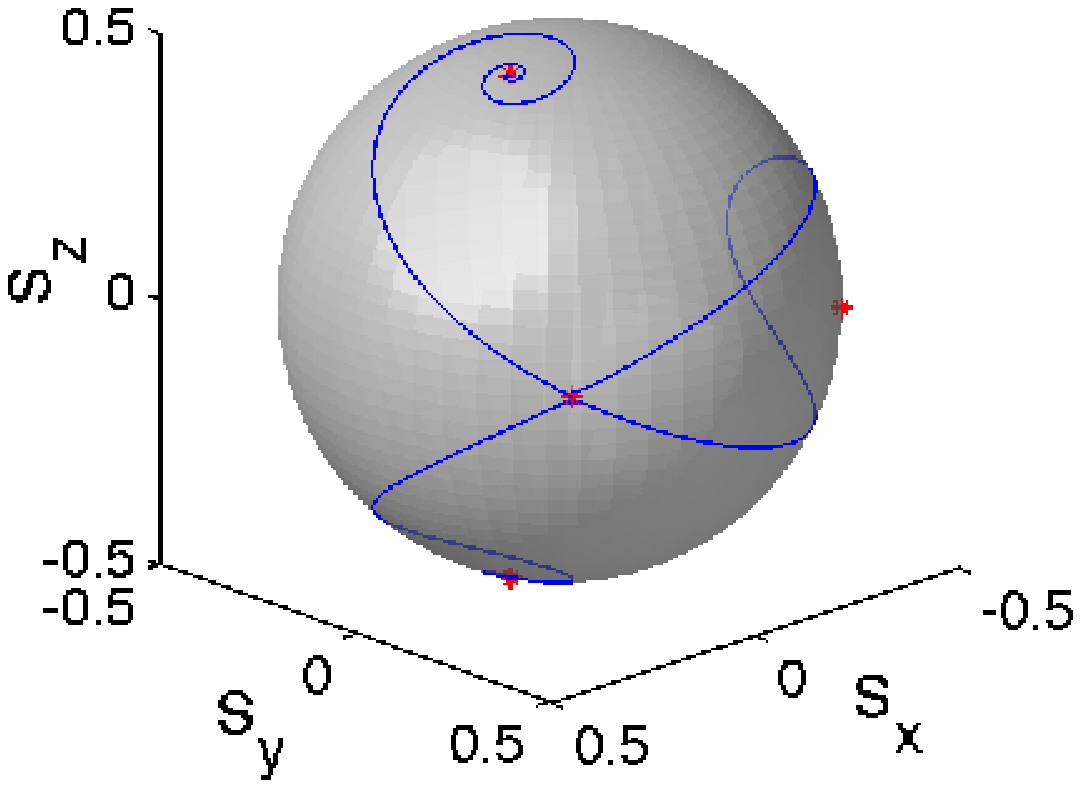}
\caption{\label{fig-manifold} (Color online) Stable and unstable manifolds of the
saddle point $\vec s_{c-}$ for $g=3$ and a Hermitian 
$\gamma =0$ (left) and a non-Hermitian case $\gamma =0.75$ (right). The four
fixed points are marked by red dots.}
\end{center}
\end{figure}

In region 2 the dynamics is mainly organized by the stable 
and unstable manifolds of the saddle point, 
as shown in Fig.~\ref{fig-manifold} for $g=3$ 
and  two values of $\gamma$. In the Hermitian case, $\gamma=0$,
these manifolds form a single figure-eight curve, a separatrix, 
encircling the self-trapping regions around the two centres 
$\vec s_{f\pm}$. In addition, there is a third center localized at the 
equator opposite to the saddle point. 

In the non-Hermitian case the two self-trapping centers change 
into a sink close to the north-pole and a source close to the south-pole. 
The saddle-point and the center at the equator survive and the 
separatrix through the saddle point
transforms into a single curve emanating from the source, passing through
the saddle-point, encircling the center, passing again through the
saddle and, finally, spiralling into the sink, as shown in the left panel of 
Fig.~\ref{fig-manifold} for  $\gamma=0.75$. The surface is 
divided into two regions, an area $A_c$ of oscillatory motion 
encircling the center, and the rest, the basin
of attraction of the sink. With increasing interaction $g$, the
area $A_c$ shrinks into a thin region close to the equator (note
that the positions of the center and the saddle point are independent
of $g$).  Decreasing the interaction $g$,  the sink and the source approach the 
saddle-point and meet at the critical value $g_{\rm crit}$. During this process,
the area $A_c$ grows until it covers the whole sphere at $g_{\rm crit}$.

For increasing $\gamma$, starting from parameter region 2, the saddle 
point (index $-1$) and the center (index $+1$) on the equator 
approach each other along the equator until 
they meet and annihilate for $\gamma=v$ at $\vec s =(0,1/2,0)$. 
For larger values of $\gamma$, that is, in region 3 only the source and 
the sink remain, and the dynamics is fully governed by the 
flow from the former to the latter, as illustrated 
in the last plot (right in the lower panel) in 
Fig.~\ref{fig-bloch-dyn-nherm-nlin}. 
For $g=0$, the transition occurs directly between 
region 1 and 3 in a non-generic bifurcation at 
$\gamma=\pm\J$ (the EP), which is depicted 
in Fig.~\ref{fig_bloch_nhermI} where the two centers meet 
and simultaneously change into a sink and a source.
\begin{figure}[tb]
\begin{center}
\includegraphics[width=4cm]{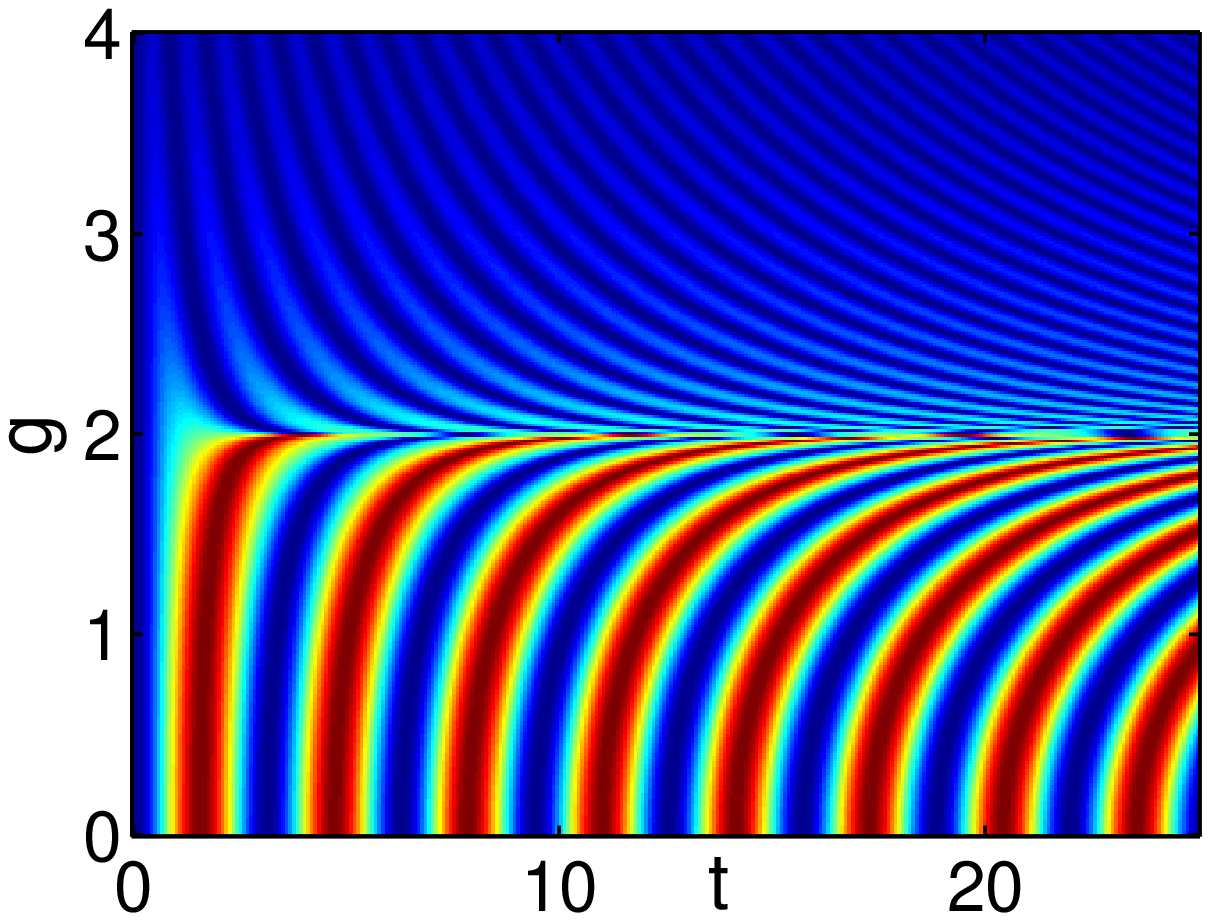}
\includegraphics[width=4cm]{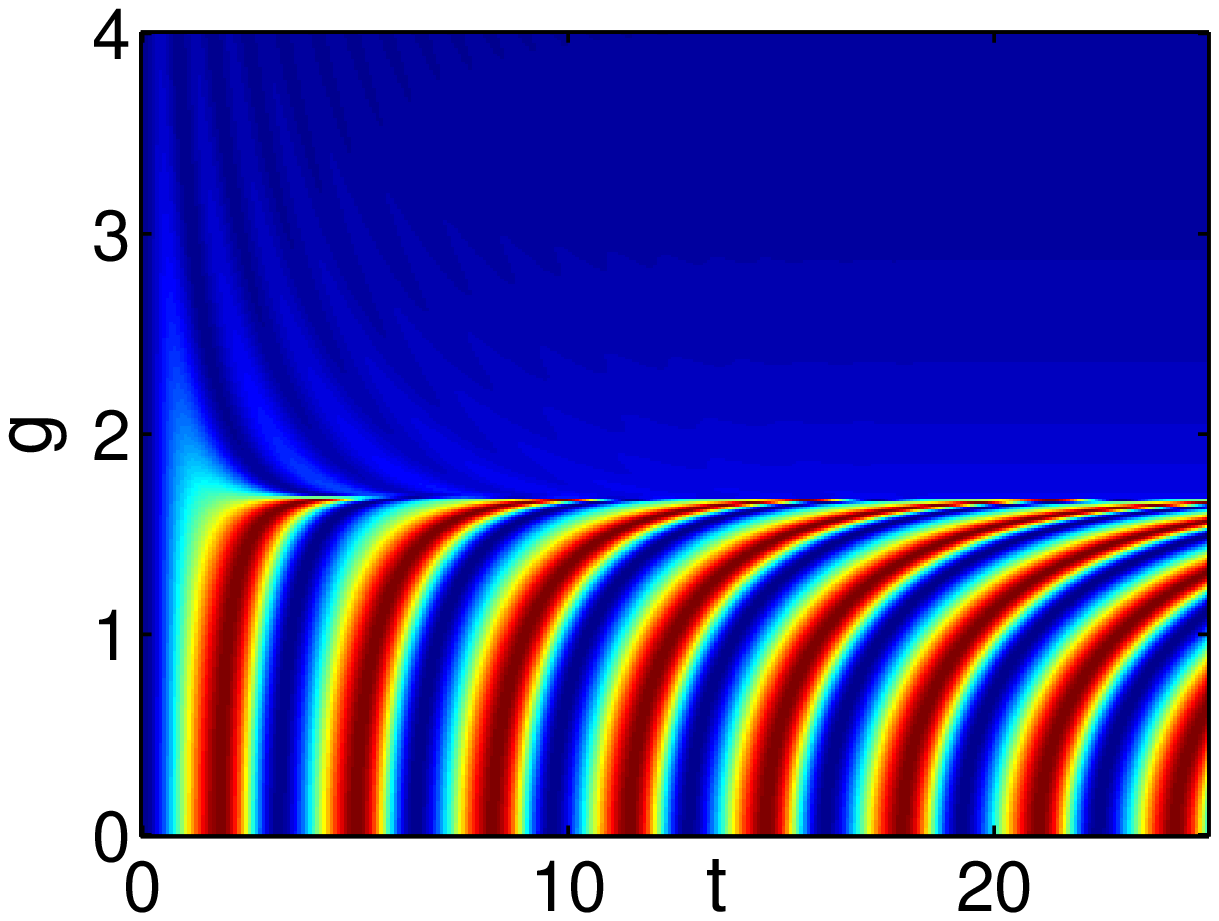}
\includegraphics[width=4cm]{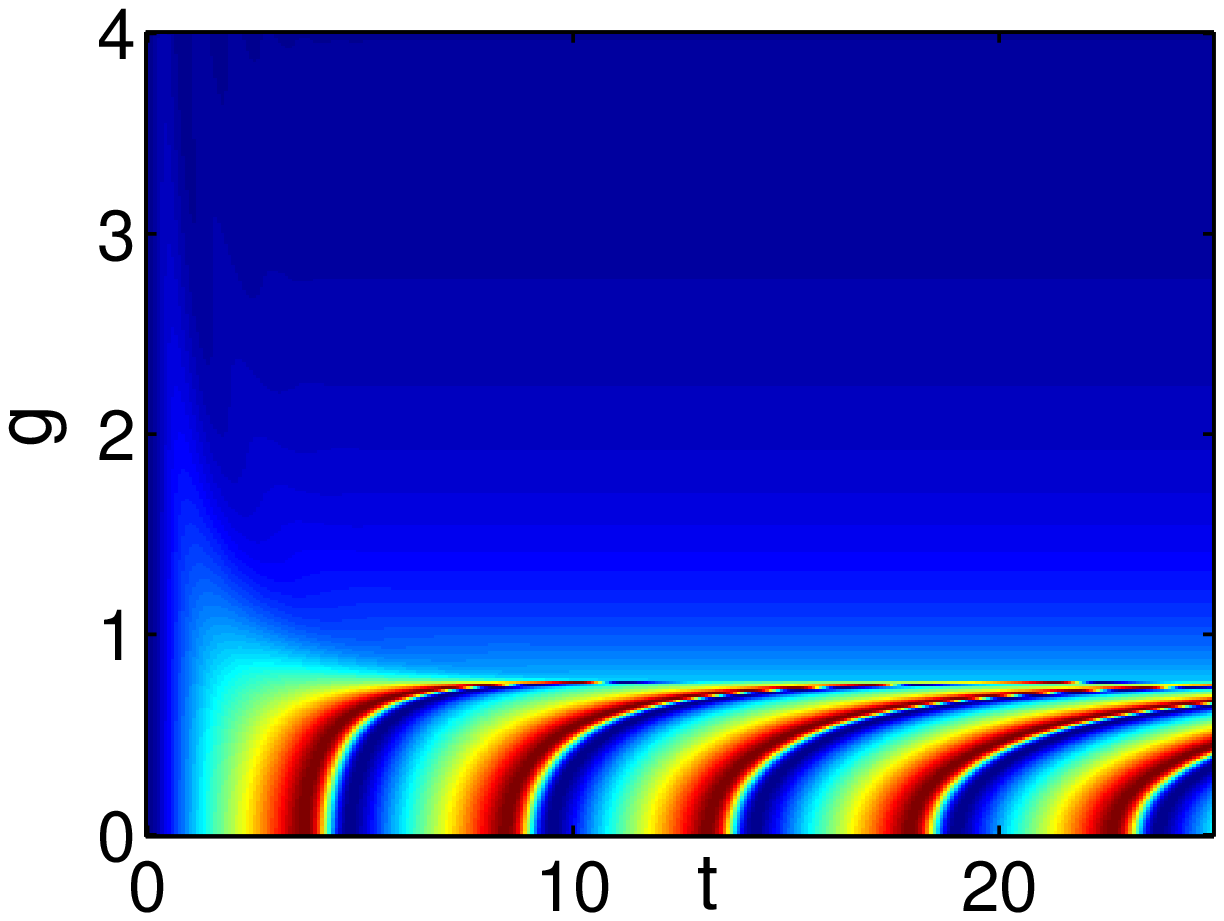}
\includegraphics[width=4cm]{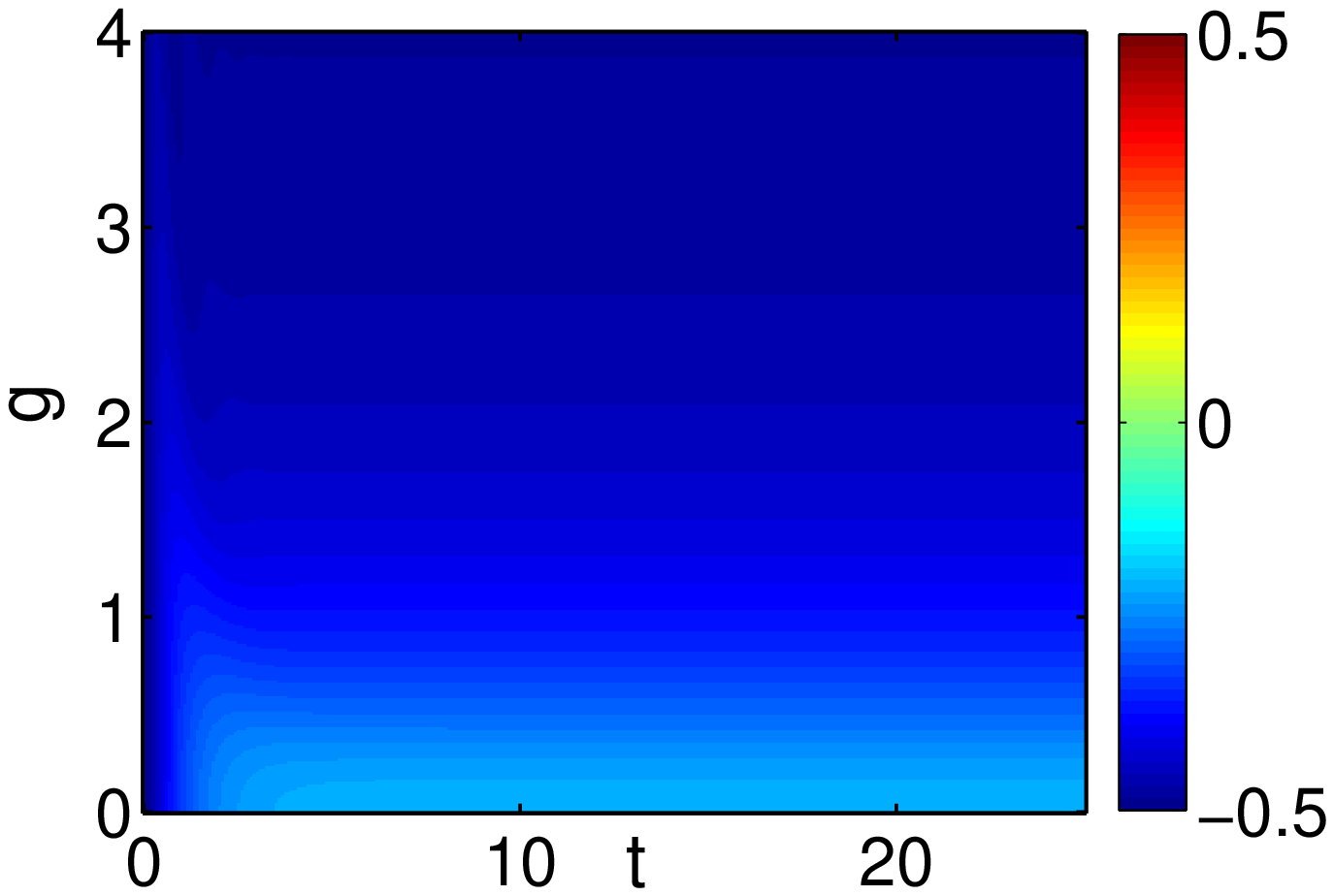}
\caption{\label{Population2} (Color online) Mean-field dynamics of 
$s_z$ plotted in false colors in dependence on the nonlinearity $g$ 
for $v=1$, $\epsilon=0$ where the initial state is the south pole 
of the Bloch sphere, for different values of the non-Hermiticity (top to 
bottom and left to right: $\gamma=0$, $\gamma=0.2<v$, 
$\gamma=0.75<v$, and $\gamma=1.1>v$).}
\end{center}
\end{figure}

In the Hermitian case, for  $g>\g_{\rm crit}=v$, we find
self-trapping oscillations in the vicinity of the fixed points
$\vec s_{f\pm}$. For $\gamma\neq0$ these fixed points 
change into a sink and a source of the dynamics which
results in a damping of the self-trapping oscillations. 
Figure \ref{Population2} illustrates 
this damping effect. Here we plot in false colors the 
time dependence of $s_z$, the population imbalance between 
the two levels, as a function of the nonlinearity $\g$ for an initial state at
the south pole of the Bloch sphere for four different values of 
$\gamma$. The first plot on the left shows the behavior in the case
$\gamma=0$. We observe two distinct regimes: For $g<g_{\rm sep}=2$, the
starting point, and hence the whole trajectory, is inside the area $A_c$
and the motion $s_z(t)$ shows a large amplitude oscillation extending to
the vicinity of the north pole. For $g>g_{\rm sep}=2$. in the self-trapping
region, the motion is confined to the neighborhood of the south pole. 
At $g_{\rm sep}=2$ the separatrix passes through the south pole and the motion
starting there approaches in infinite time the saddle point along the
stable manifold. (Note the increase of the period of oscillation for
$g\rightarrow g_{\rm sep}$ where the period diverges.) This behavior
continues for $\gamma \ne 0$, however with a smaller value of $g_{\rm sep}$.
For small nonlinearities the population is completely
transferred between the two levels, that is, the Bloch vector oscillates between the 
south and the north pole and above $g_{\rm sep}$ 
the oscillation stays closer to the south pole with increasing interaction. 
As observed in Fig.~\ref{fig-bloch-dyn-nherm-nlin}, 
the self-trapping states are then associated with 
a sink and a source of the dynamics. 
Therefore, for a nonvanishing but subcritical non-Hermiticity 
$0<\gamma<v$, the system relaxes to a state with excess population 
in the non-decaying state above a critical value $g_{\rm sep}$ of the
interaction. This appears as a damping in the self-trapping oscillations, 
which is visible in the second and third plot (top right and bottom left) in 
Fig.~\ref{Population2}. A similar observation was reported in \cite{Wang07} where the effect was related to decoherence. 
For even larger values $\gamma>v$ in region 3, 
as shown in the right plot in the lower 
panel in Fig.~\ref{Population2}, the oscillatory motion is already destroyed even in 
the linear case $g=0$, and the dynamics is dominated by the flow from the sink to the source irrespective of the nonlinearity, that is, the system stays confined to the lower half of the Bloch sphere.

\begin{figure}[tb]
\centering
\includegraphics[width=4cm]{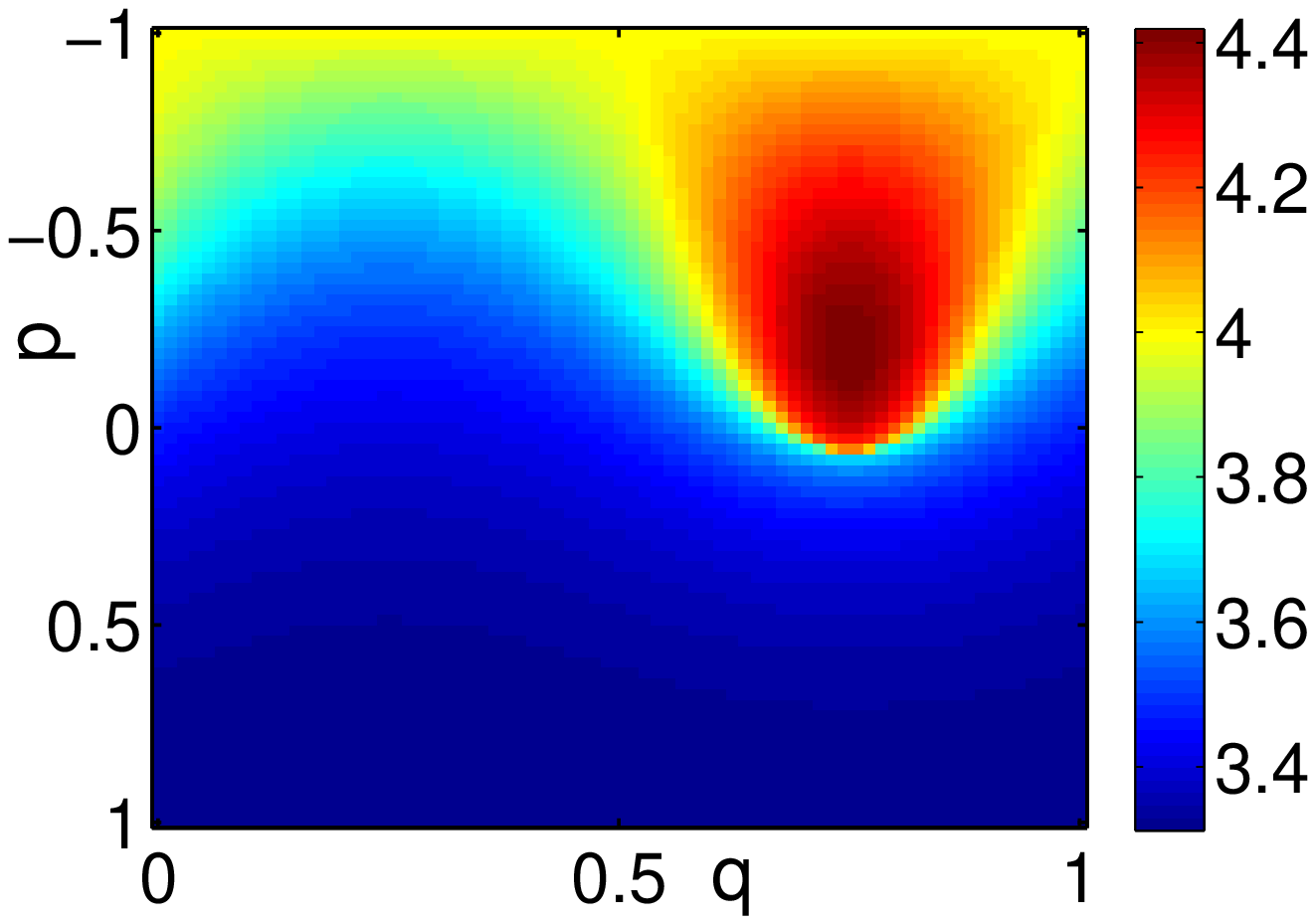}
\includegraphics[width=4cm]{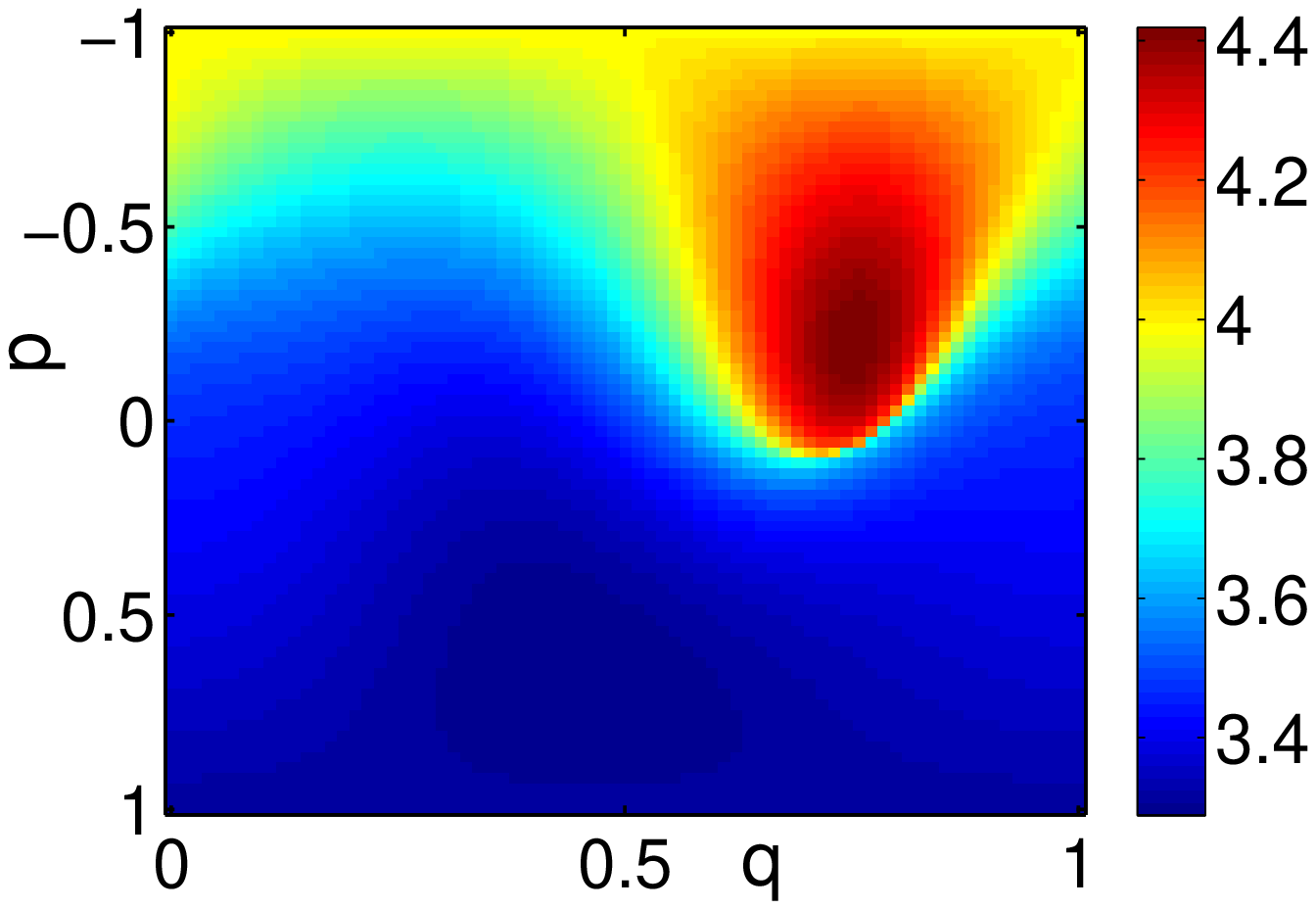}
\includegraphics[width=4cm]{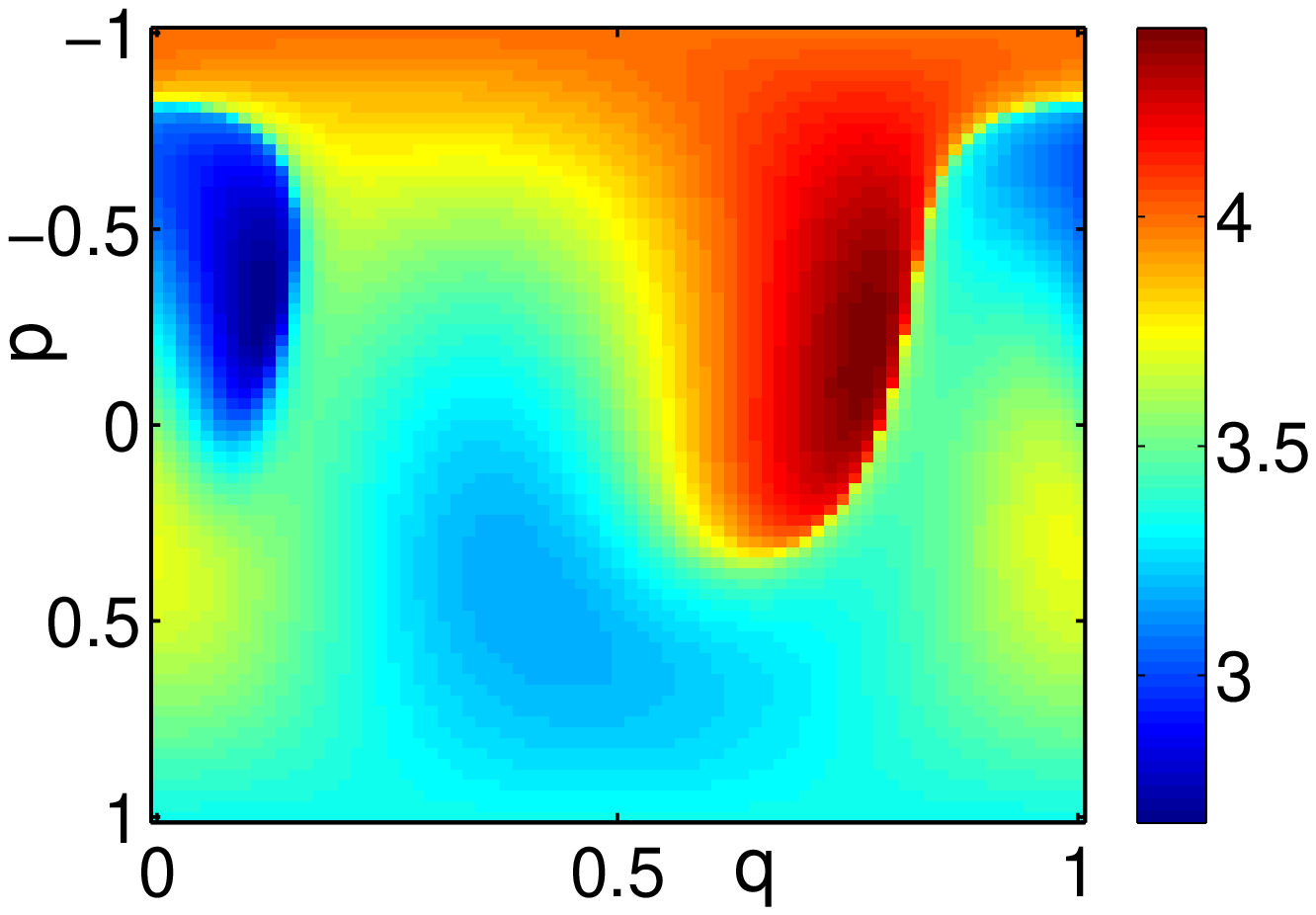}
\includegraphics[width=4cm]{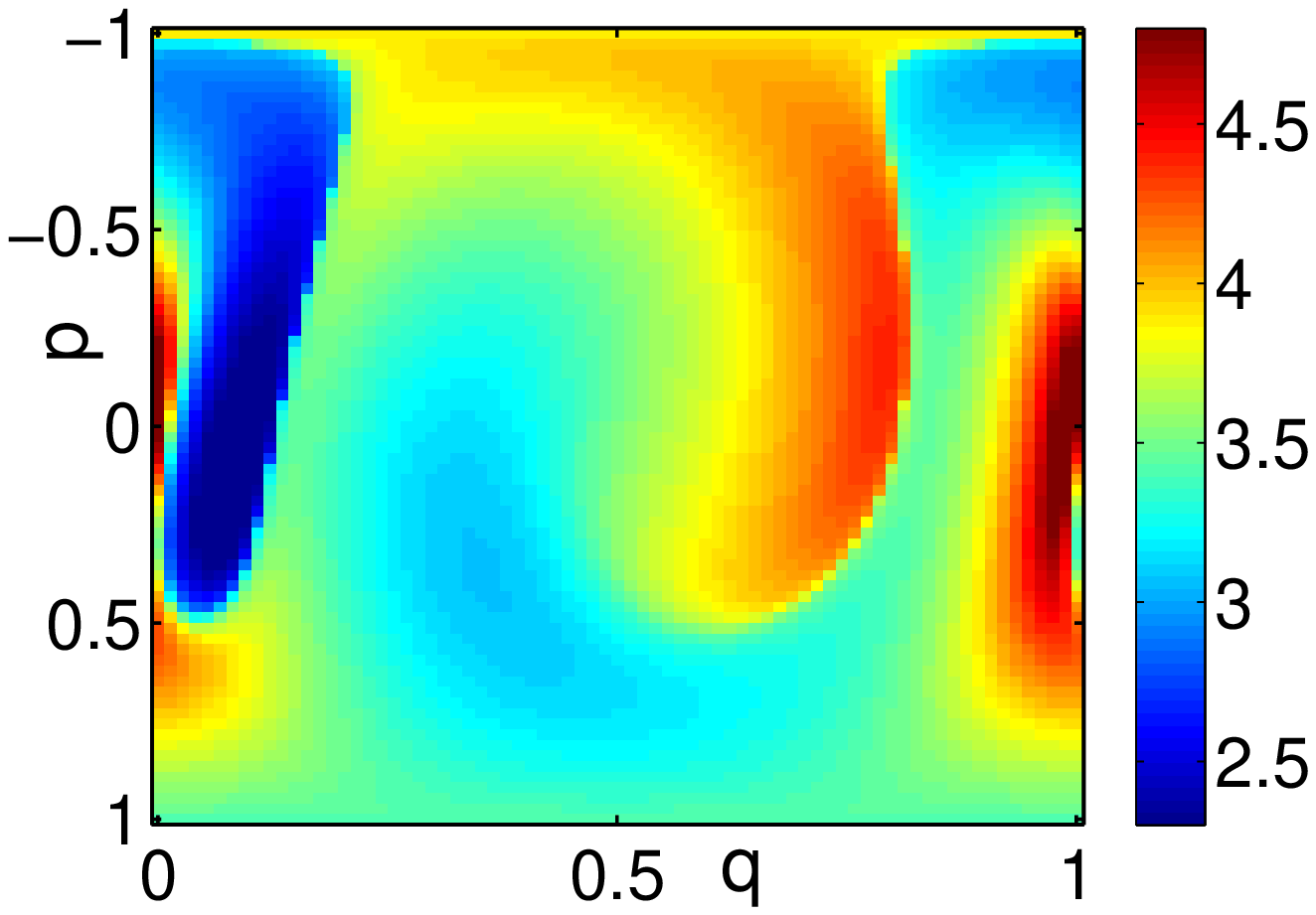}
\includegraphics[width=4cm]{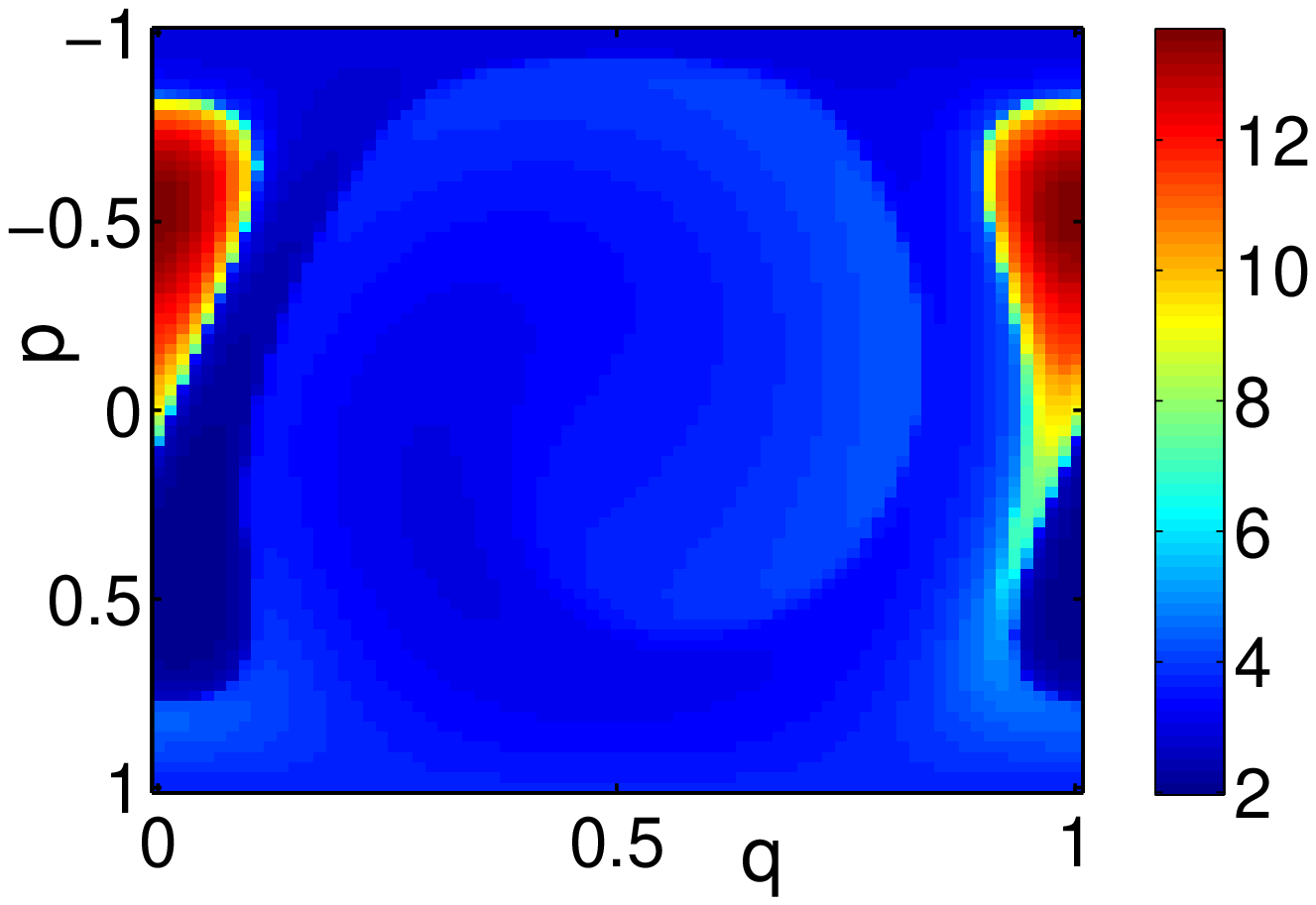}
\includegraphics[width=4cm]{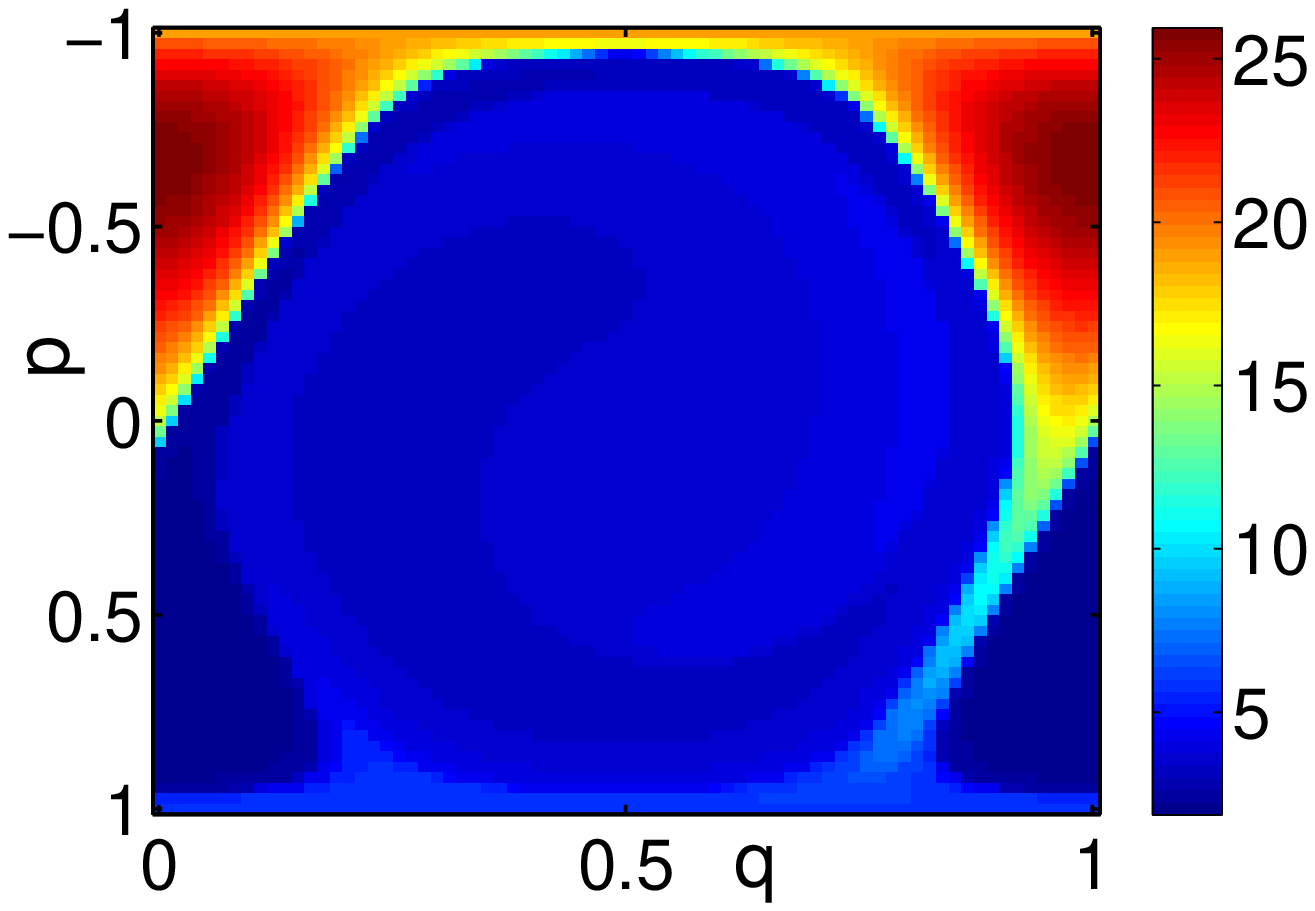}
\caption{\label{fig-MF-decay-bloch} (Color online) Half life time in dependence 
on the initial condition for the non-Hermitian mean-field dynamics for 
$\epsilon=0$ for $v=1$ and $\gamma=0.1$ and different values 
of the nonlinearity, (from left to right and top to bottom, $g=0,\, 0.1,\, 0.5,\, 1,\, 1.5,\, 2$).}
\end{figure}

For the non-Hermitian system the normalization $n$, which can be 
interpreted as the ``survival probability'' of the system, is also time dependent. 
For the non-Hermitian two level system \rf{nlnhGP} the dynamics 
are governed by the equation of motion \rf{Norm}, that is,
$\dot n=-2\gamma(2\s_z+1)n$, which does not explicitly 
depend on the nonlinearity. 
Yet, the instantaneous decay rate is determined by the 
$s_z$ component of the renormalized Bloch vector, 
whose dynamics are sensitively influenced 
by the nonlinear term in the Schr\"odinger equation. 
This is illustrated in Fig.~\ref{fig-MF-decay-bloch} which 
shows the half life time as a falsecolor plot, as a function of 
the initial conditions $(p,q)$ for a weak decay, 
$\gamma=0.1$, and different nonlinearities. It is clearly visible that 
the nonlinearity can stabilize the system significantly for 
certain initial conditions. (Note the different colorscales.)

\begin{figure}[tb]
\centering
\includegraphics[width=4.2cm]{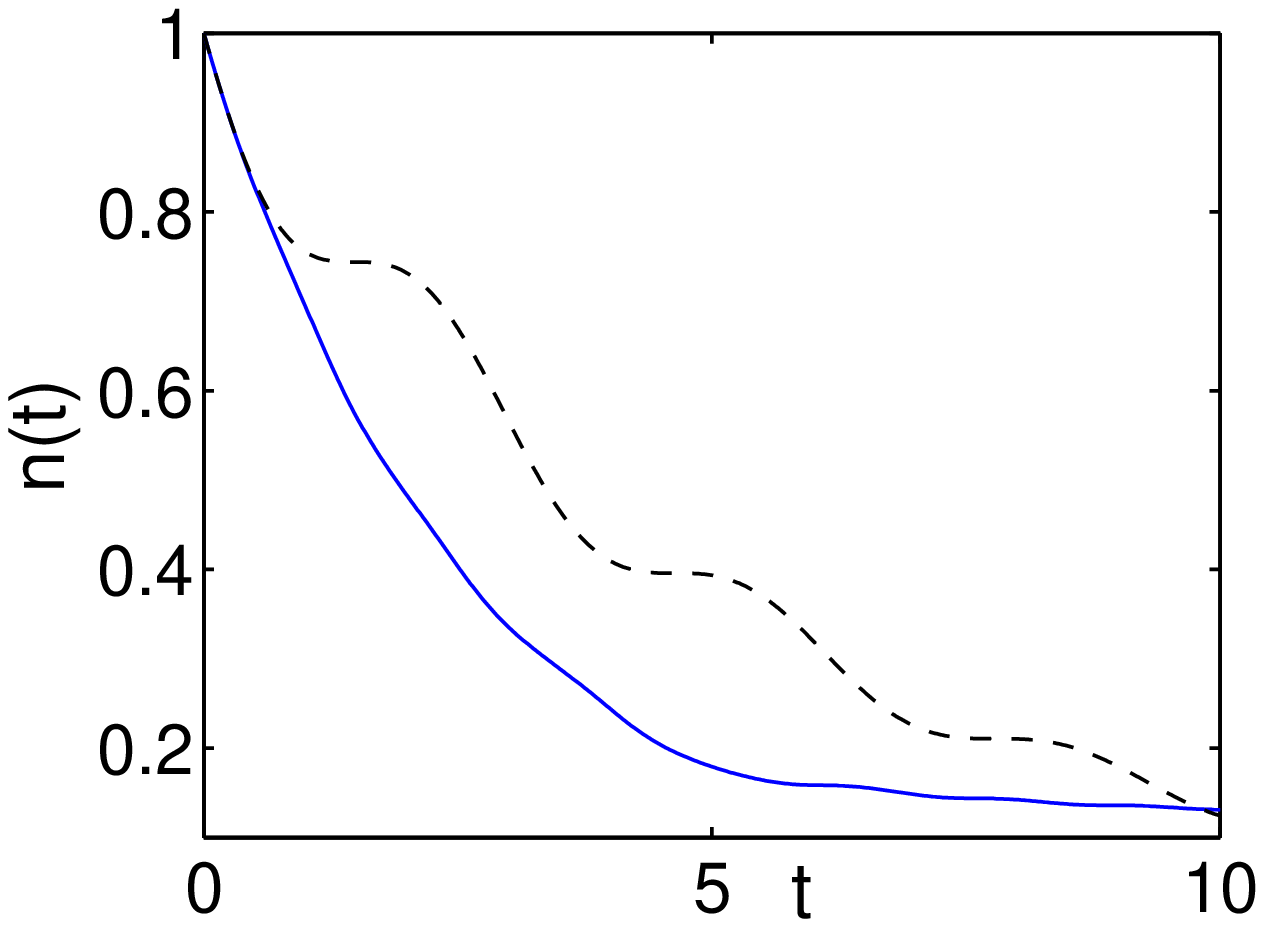}
\includegraphics[width=4.2cm]{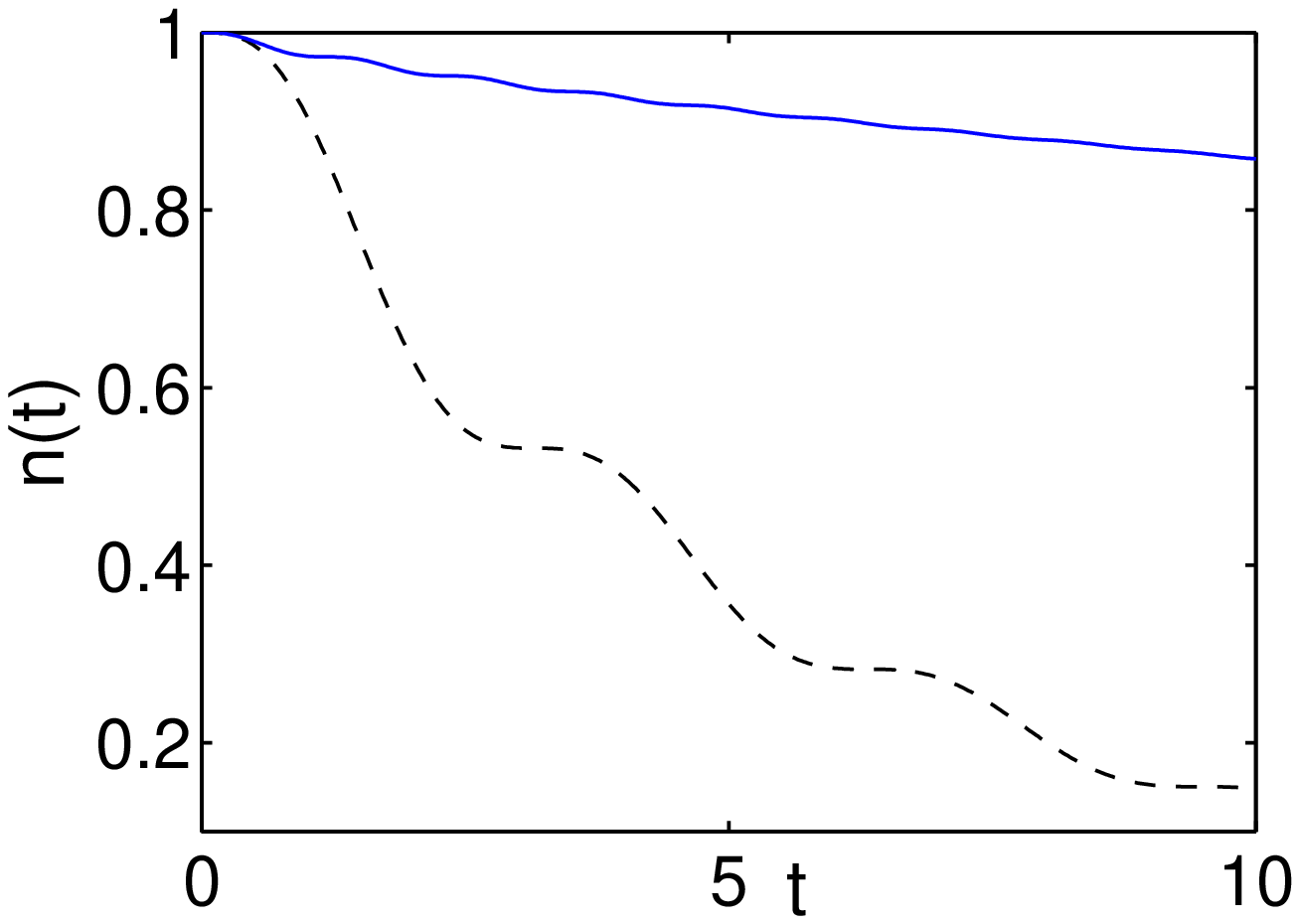}
\caption{\label{fig-comp01} (Color online) Decay of the mean-field normalization $n$ in dependence on the nonlinearity for a small decay $\gamma=0.1$. The evolution for a nonlinearity of $g=3$ (blue solid curves) is compared to the linear case (black dashed curves). The left figure shows the time evolution of $n(t)$ starting in the north pole, the right figure shows the same for an initial state at the south pole.}
\end{figure}

To understand this behavior in more detail, we investigate the 
full time evolution of the normalization for some examples. 
In  Fig.~\ref{fig-comp01} we show the normalization 
$n(t)=|\psi_1|^2+|\psi_2|^2$ of the wave function as a 
function of time for a small non-Hermiticity $\gamma=0.1$ 
and a supercritical nonlinearity $g=3$ (blue lines), in comparison 
to the linear evolution for $g=0$ (black lines). The left plot shows 
the dynamics for an initial state at the north pole of the Bloch 
sphere, and the right plot corresponds to a state initially at the south 
pole. We observe that for an initial condition at the north pole 
(corresponding to the decaying level) the decrease of the 
normalization is slightly faster due to the nonlinearity, although 
from time $t\approx 5$ onward it slows down considerably. 
In the limit $t\to\infty$ the decrease becomes exponential with 
a very small decay coefficient. The modulations present in the 
linear case are much less pronounced here from the very 
beginning. Despite these differences, the overall decay time 
characterized, e.g., by the half life time, is not drastically 
changed here. If we now turn to the right plot and compare 
the nonlinear decay behavior to the linear one for an initial 
condition in the south pole of the Bloch sphere (corresponding 
to the stable level) the induced changes become much more 
pronounced. In fact, the decay is considerably slowed down 
by the nonlinearity. For longer times the modulations nearly vanish and 
the decay becomes approximately exponential with the same decay 
coefficient as for the initial condition in the north pole.

This behavior can be understood in terms of the Bloch dynamics 
discussed before. For large nonlinearities the \textit{source} 
of the Bloch dynamics moves toward the north pole, 
which is connected with the decaying level and thus a 
\textit{sink} for the probability. The \textit{sink} of the Bloch 
dynamics, on the other hand, moves close to the south pole 
which corresponds to the \textit{stable} level. Thus, if we 
start the system at the south pole (right plot in Fig.~\ref{fig-comp01}), 
then due to the nonlinearity it stays on the southern hemisphere 
($-\half\leq s_z\leq 0$) and spirals into the sink of the dynamics 
instead of performing Rabi oscillations extending over all values 
of $s_z$. Hence the instantaneous decay rate is smaller than for 
the linear case and the decay is significantly decelerated. If we start 
the dynamics at the north pole (left plot in Fig.~\ref{fig-comp01}), 
on the other hand, the Bloch dynamics also move toward the 
sink close to the south pole, where they remain. However, 
until the small instantaneous values of the decay coefficient 
associated with the southern hemisphere of the Bloch sphere 
become relevant, the normalization already decayed considerably.

Summarizing, the interplay of nonlinearity and non-Hermiticity 
introduces a qualitatively new behavior to the mean-field dynamics. 
This is manifested in the different types and numbers of fixed points 
generated in the renormalized dynamics, and in the resulting sensitivity 
of the normalization dynamics to the initial conditions.  

\section{Many-particle mean-field correspondence}
\label{sec-MP-MF}
\begin{figure}[tb]
\begin{center}
\includegraphics[width=4cm]{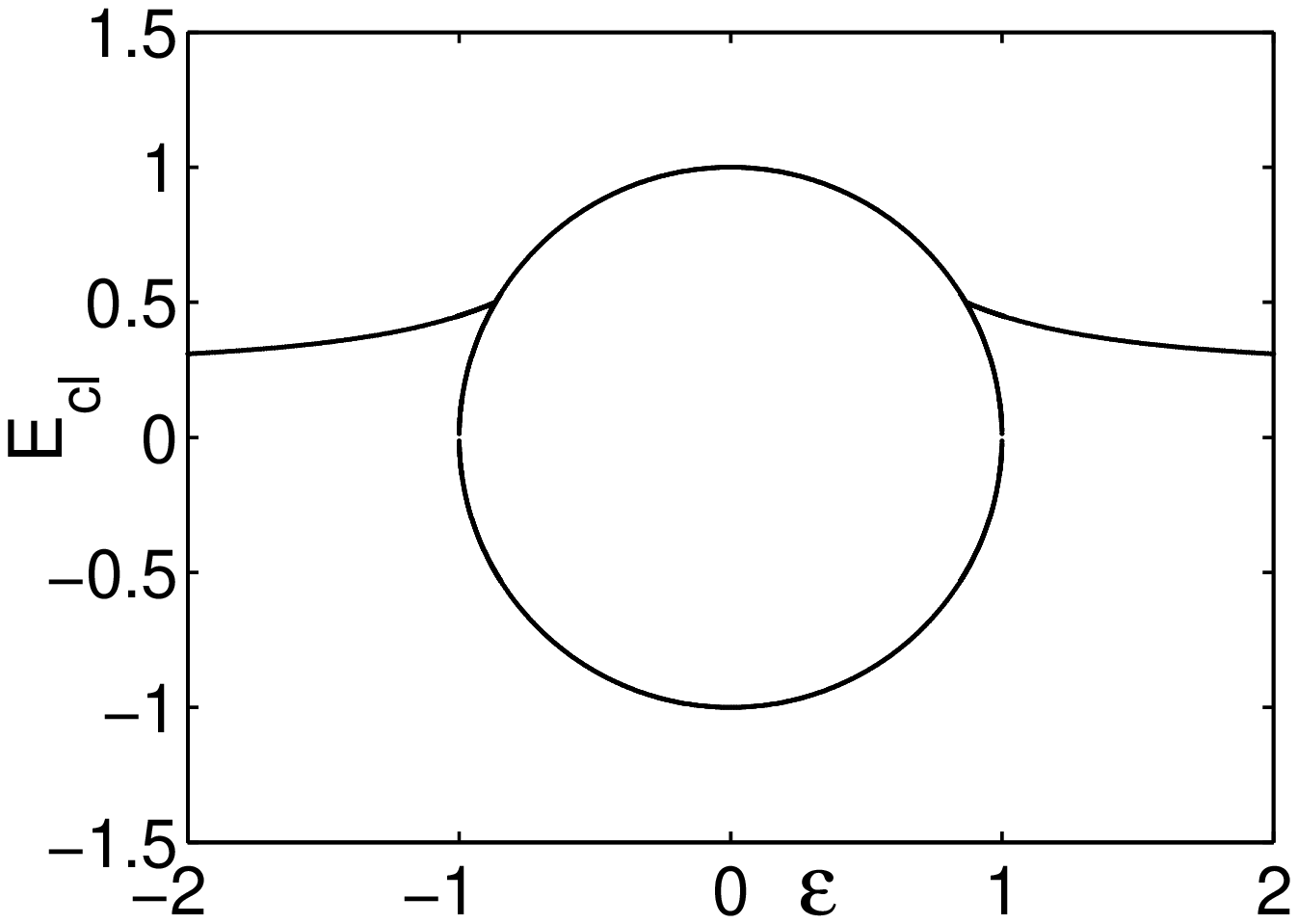}
\includegraphics[width=4cm]{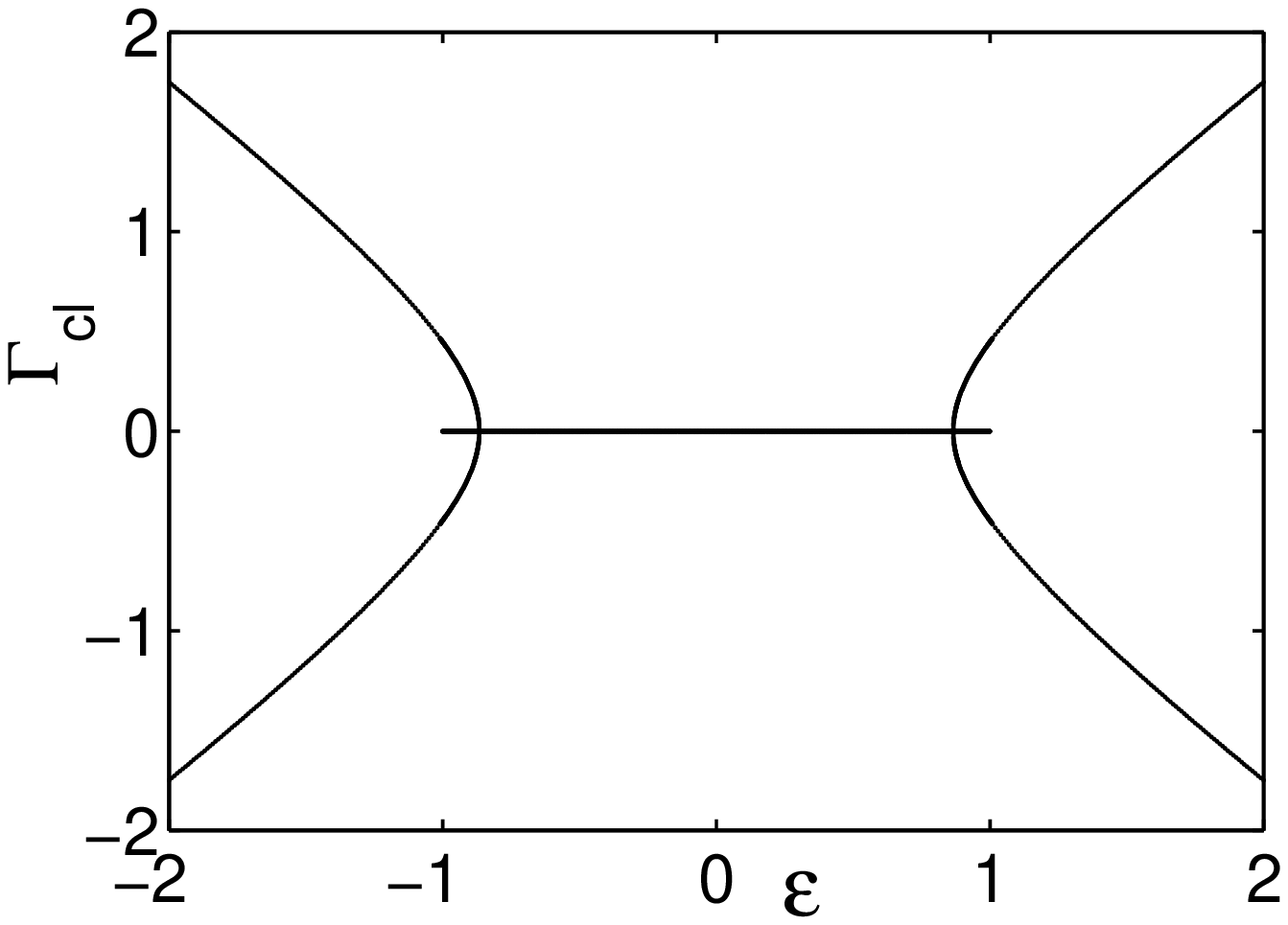}
\includegraphics[width=4cm]{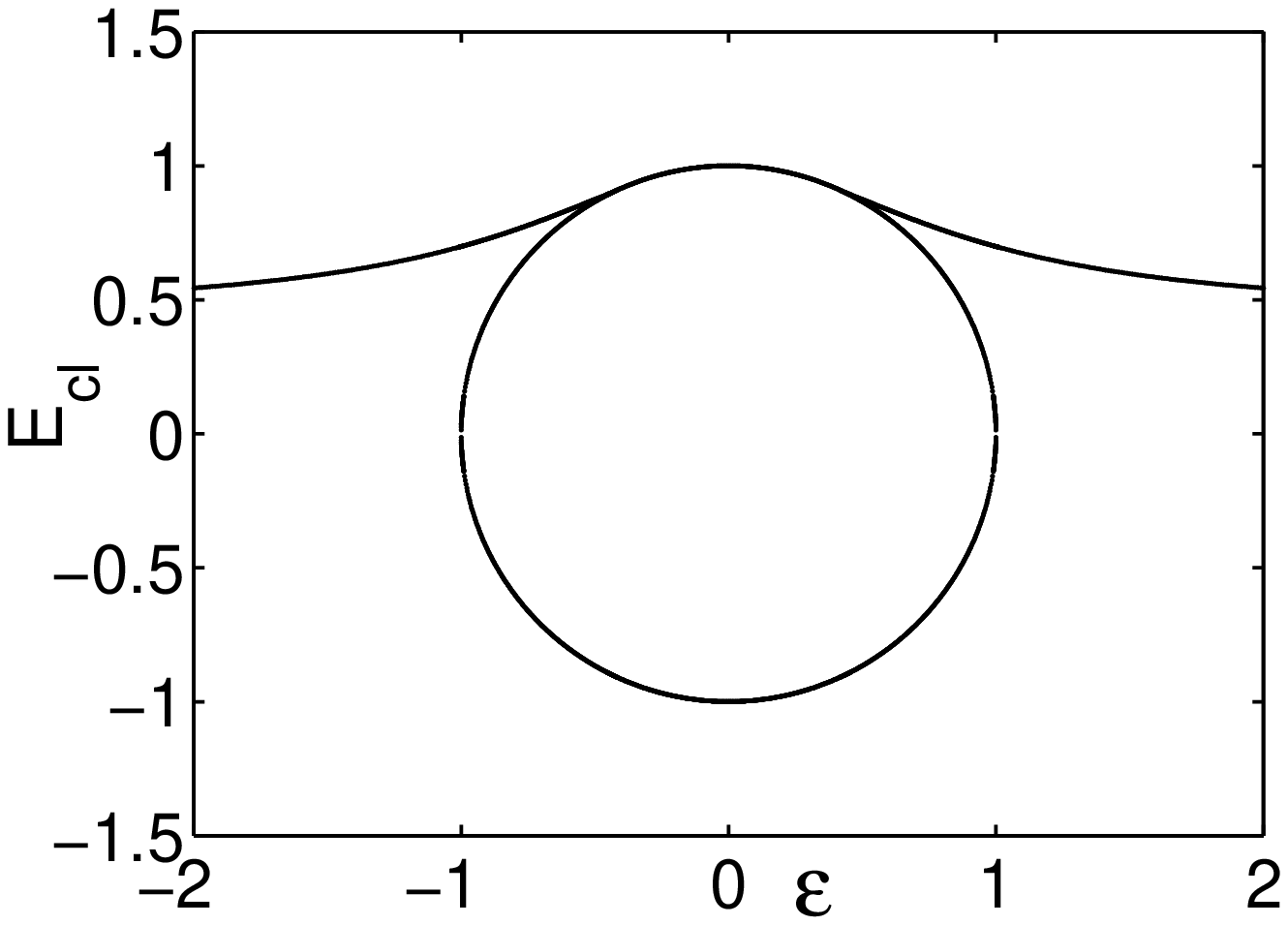}
\includegraphics[width=4cm]{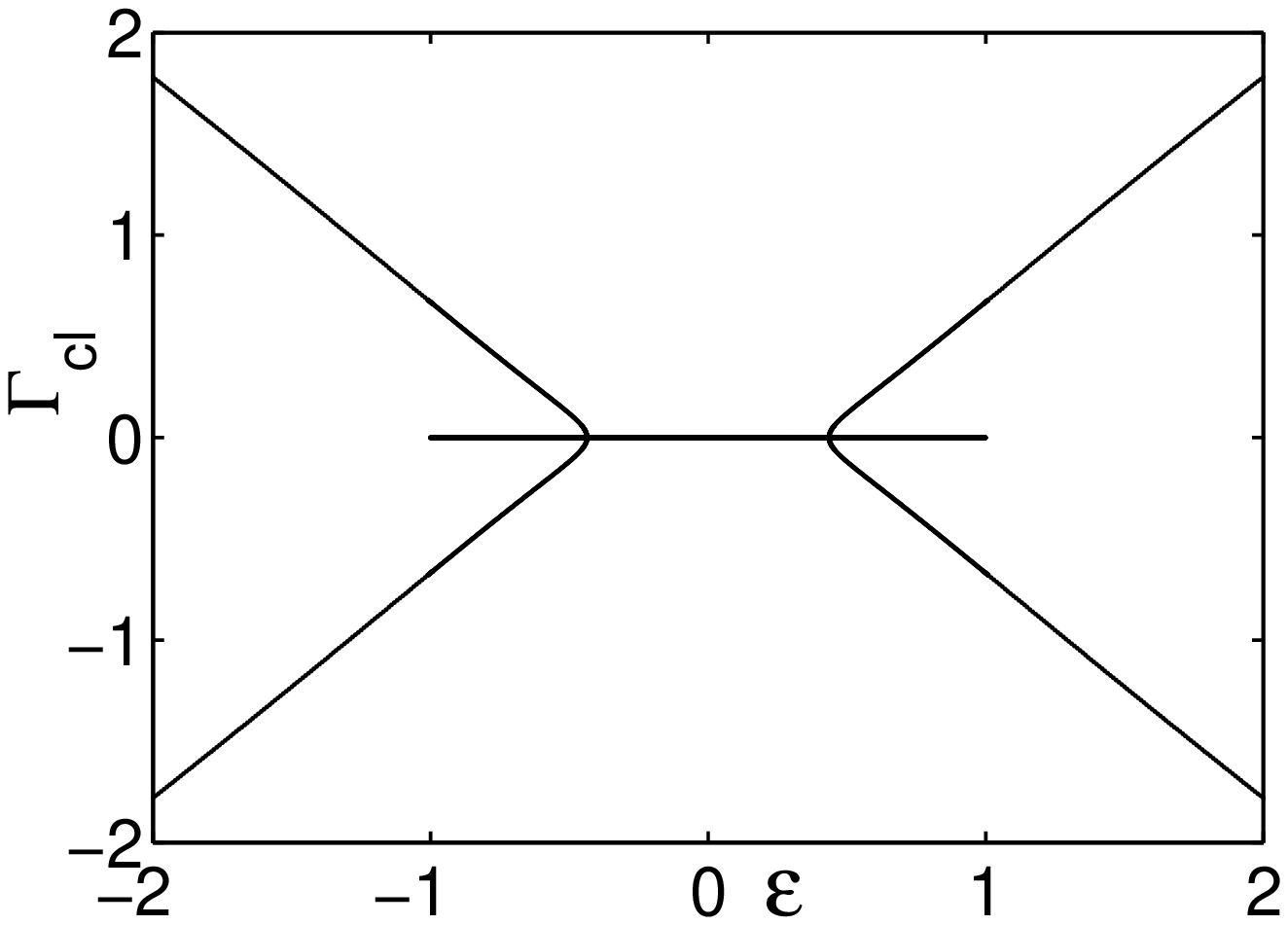}
\end{center}
\caption{\label{fig-2level-nlin-cross-PT}Real (left) and imaginary
(right) part of the mean-field eigenenergies (values of the 
Hamiltonian function at the fixed points) as a function of 
$\gamma$ for $\epsilon=0$, $v=1$ and $\g=0.5$ 
(top) and $\g=0.9$ (bottom).}
\end{figure}

Let us finally compare the mean-field description
with the full many-particle behavior. We begin 
with a comparison of the spectral behavior. 
For this purpose we first have to define the eigenenergies of the 
mean-field system. We will identify them with 
the values of the Hamiltonian function 
at the fixed points of the mean-field dynamics. 
Note that these are different from the generalized 
eigenvalues of the nonlinear non-Hermitian Schr\"odinger operator 
(the chemical potentials) which were investigated 
in some detail for a closely related model in \cite{06nlnh,07nlres}. 
Figure~\ref{fig-2level-nlin-cross-PT} shows the real and 
imaginary parts of the eigenenergies in dependence on 
$\gamma$ for two different values of the nonlinearity. 
For nonvanishing nonlinearity we observe a similar 
behavior as in the linear case, where the two eigenvalues
are purely imaginary until they meet at the critical value 
$|\gamma|=\J$ and turn into a complex conjugate pair. 
Here, however, the two eigenvalues vanish after their ``collision'', 
which is connected to the collision and simultaneous destruction 
of the saddle point with the center in the phase space. 
In particular, the energy values of these two fixed points 
are identical to the linear case. This is evident from the 
fact that they are located at the equator of the 
Bloch sphere, that is, at $\s_z=0$ and thus the nonlinear 
term (proportional to $s_z^2$) in the energy vanishes. However, 
for values of $\gamma$ above the saddle-center collision, we still 
have two eigenvalues associated with the sink and the source that 
result from the bifurcation of one of the original centers at the 
critical value $|\gamma_{\rm crit}|=\sqrt{v^2-g^2}$. Their imaginary 
parts are always nonzero, due to the fact that they are located at 
values $\s_z\neq0$. Thus, the critical value for the emergence 
of the sink and the source defines the border of unbroken 
$\cP\cT$-symmetry for the mean-field system. 
In agreement with the many-particle results, we conclude 
that the nonlinearity $\g$ shrinks the region of unbroken $\cP\cT$-symmetry.

\begin{figure}[htb]
\centering
\includegraphics[width=8cm]{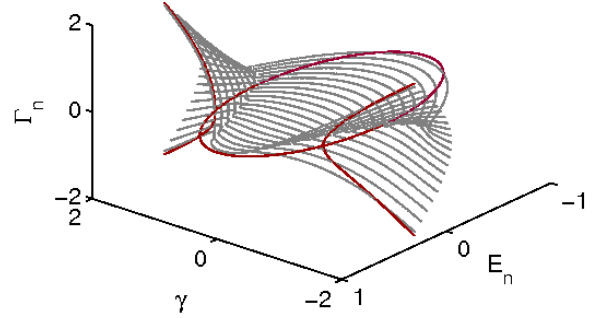}
\caption{\label{fig-comp1} (Color online) 
Many-particle (gray) and mean-field (dark red) energies
for the $\cP\cT$-symmetric system ($\epsilon=0$) for $N=20$ particles 
as functions of $\gamma$, for $v=1$ and $g=0.9$.}
\end{figure}

The observed behavior is evidently the counterpart of 
the pairwise crossing structure and the unfolding of the 
EP of higher order in the many-particle spectrum. For a better 
comparison we show both the many-particle and mean-field 
eigenenergies in Fig.~\ref{fig-comp1} for the $\cP\cT$-symmetric case 
as a function of $\gamma$ for an intermediate interaction 
strength, $g=0.9$. We indeed observe that the qualitative phenomenon 
of the shrinking region of unbroken $\cP\cT$-symmetry is
reproduced by the mean-field energies. However, the critical value
of $\gamma$ that defines this border is different for the two
descriptions. This is not surprising if we account for two facts:
First, we note that the positions of the individual EPs depend on
the particle number $N$, and the large $N$ limit (in which one
assumes the mean-field description to be valid) is not reached 
for $N=20$ particles, as in the present figure. It is in general 
an open question in which manner the mean-field limit is 
approached for non-Hermitian systems. Second, we do not 
expect an individual feature of the spectrum to have an 
impact on the classical limit. This is due to the fact that 
this limit is only defined up to arbitrary orders of $\hbar$ 
(i.e. $1/N$ in the present case), whereas the exact positions 
of individual structures, such as exceptional points, 
is dependent on these additional terms. Therefore, usually
isolated degeneracies do not have counterparts in the associated
classical limit. Only if there is an accumulation of such points 
one expects a direct correspondence.
Nonetheless, in the present case the $\cP\cT$-symmetry itself is
mirrored in the classical system and thus we expect the breaking of
this symmetry to be present as well. This is in agreement with the
observed behavior for which the breaking of the symmetry takes
place both in the mean-field and the many-particle system, and the
influence of the interaction shrinks the region of unbroken $\cP\cT$
symmetry in both cases.

\begin{figure}[tb]
\centering
\includegraphics[width=4cm]{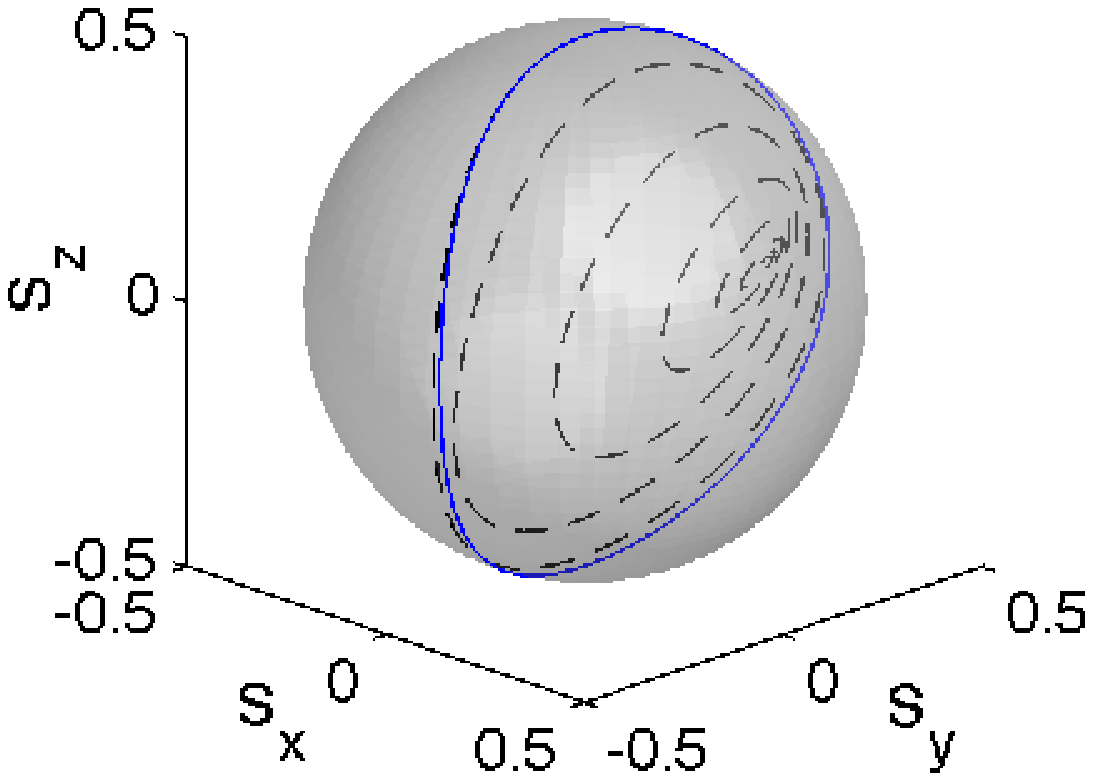}
\includegraphics[width=4cm]{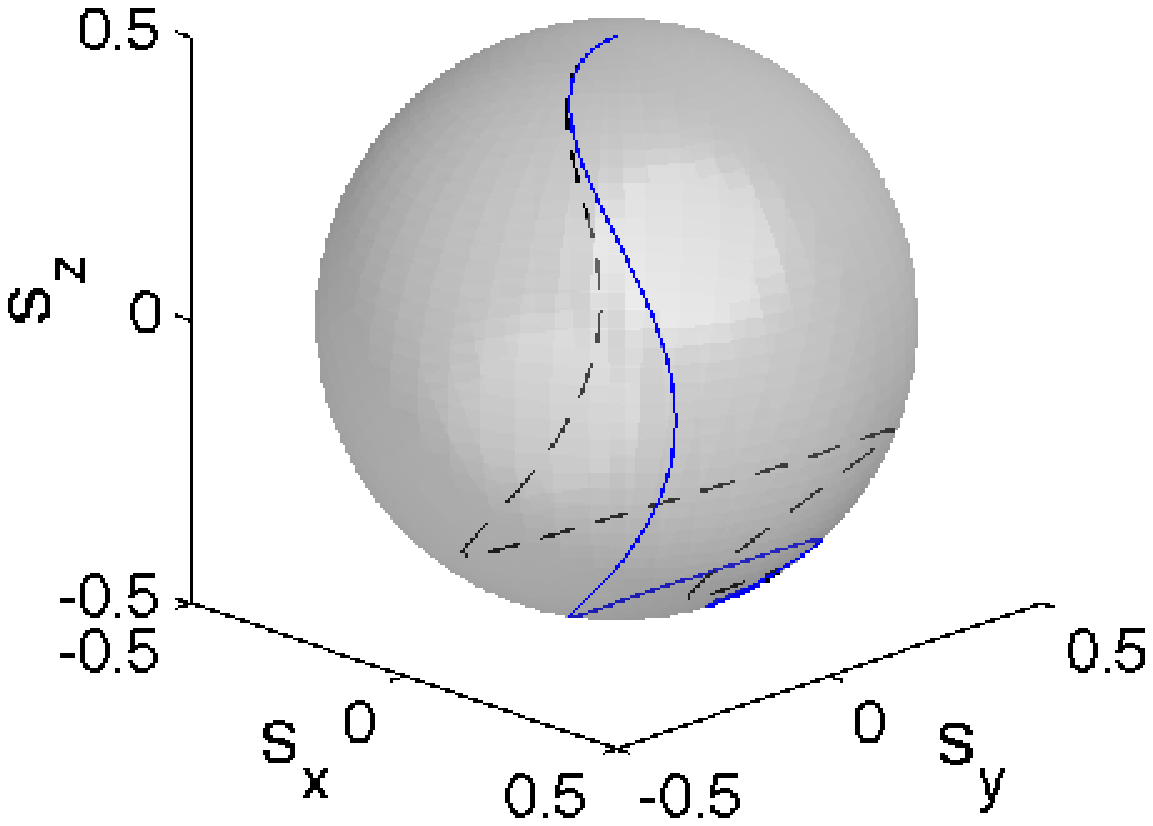}
\includegraphics[width=4cm]{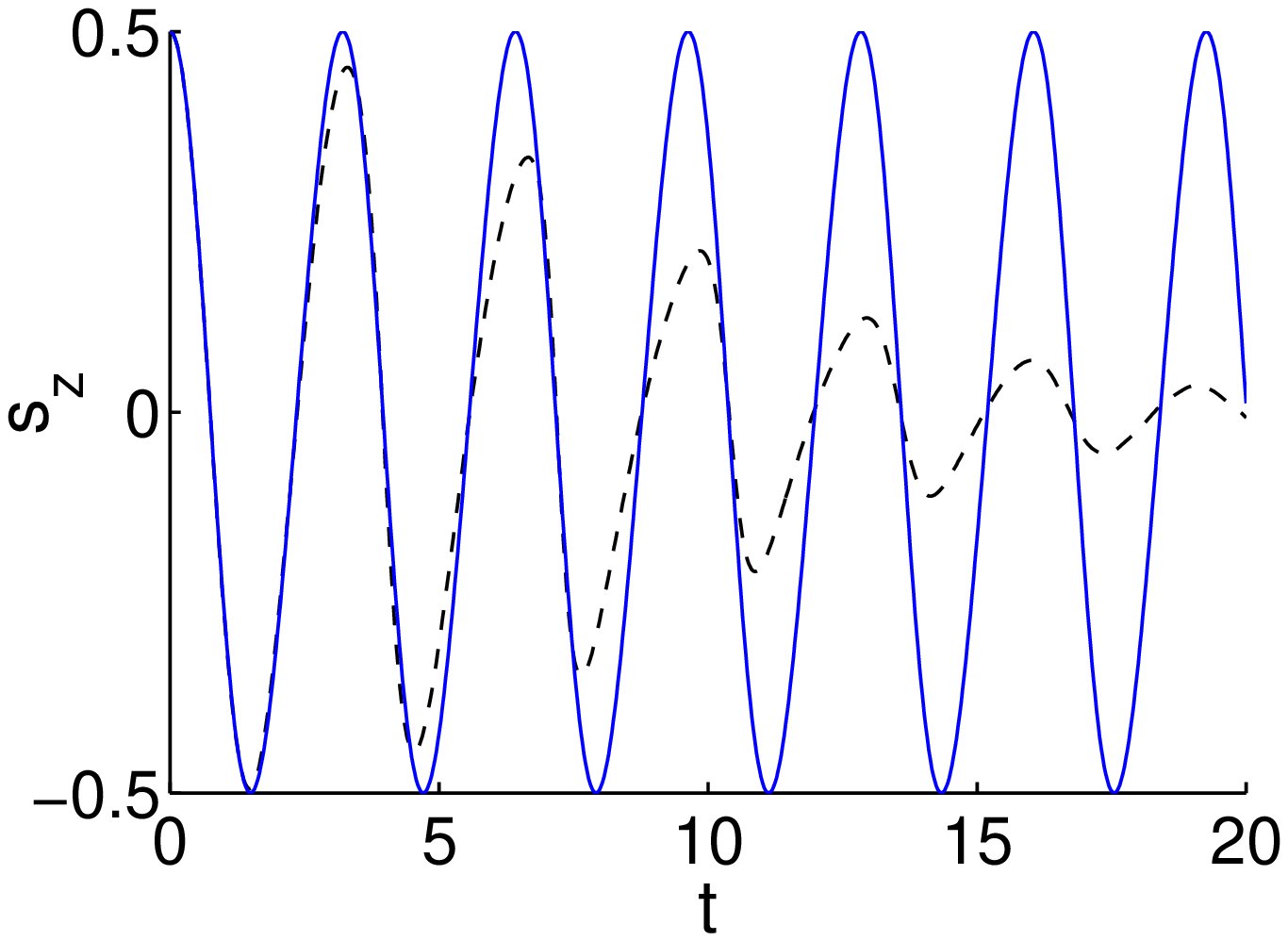}
\includegraphics[width=4cm]{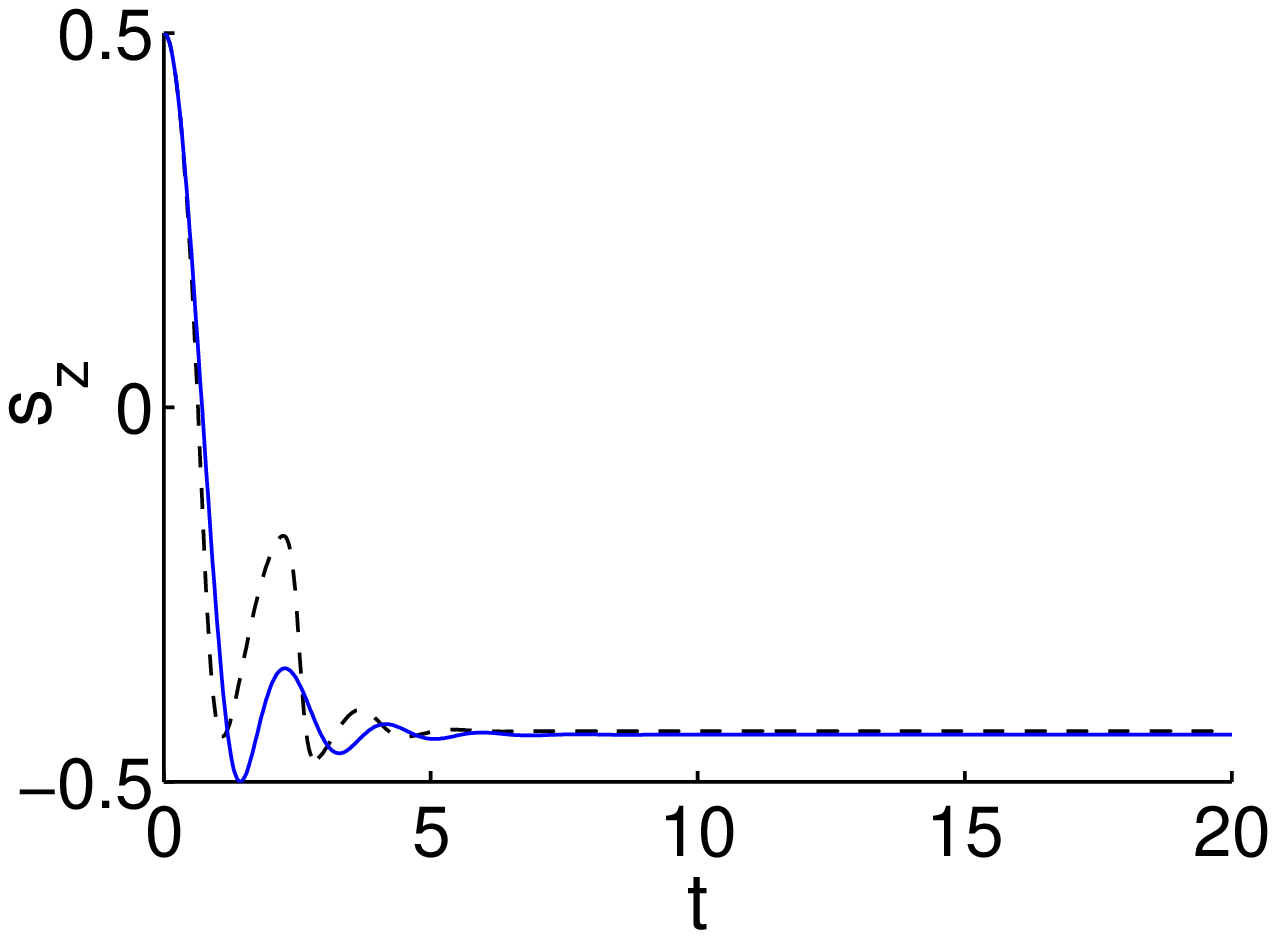}
\includegraphics[width=4cm]{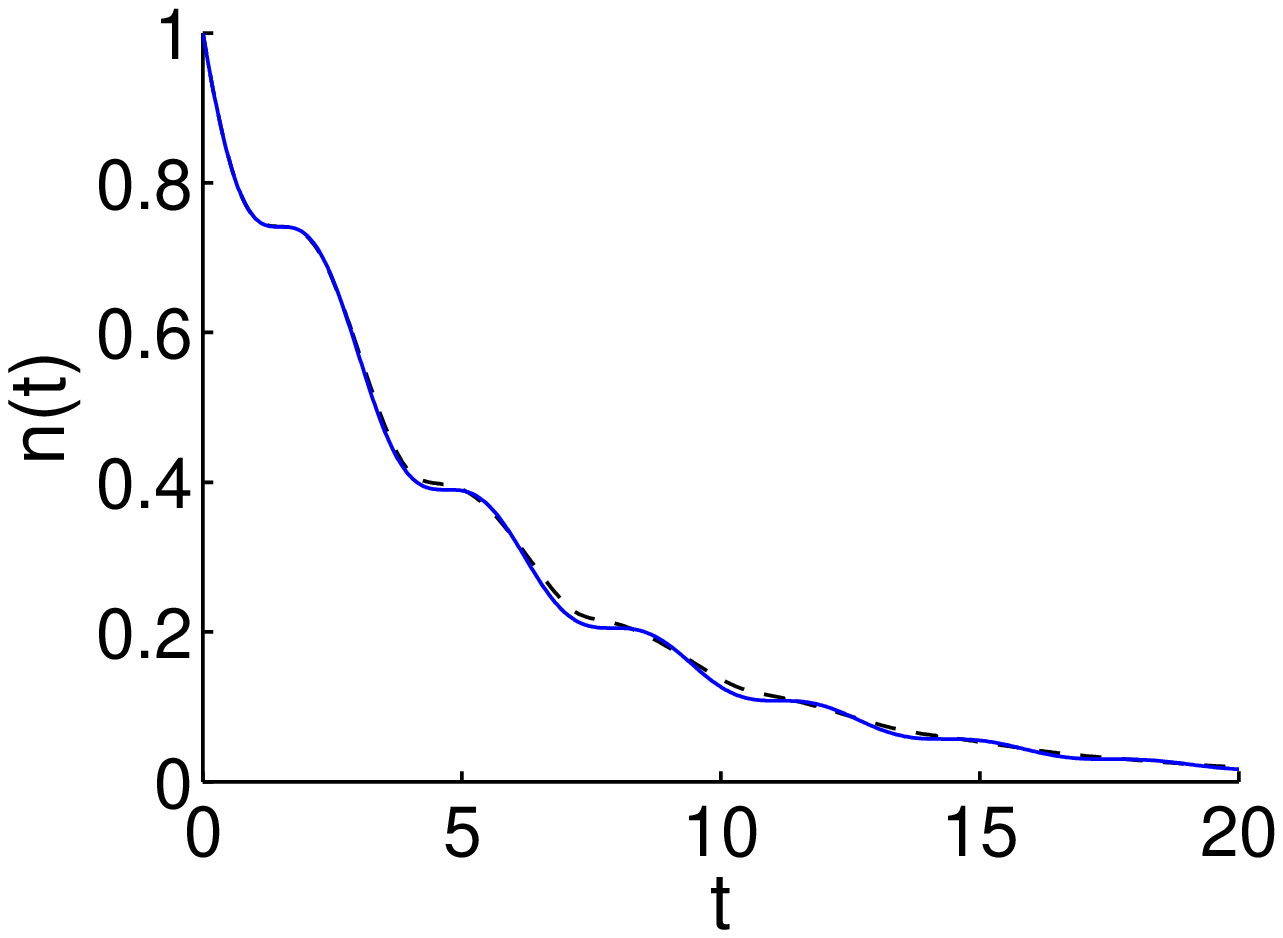}
\includegraphics[width=4cm]{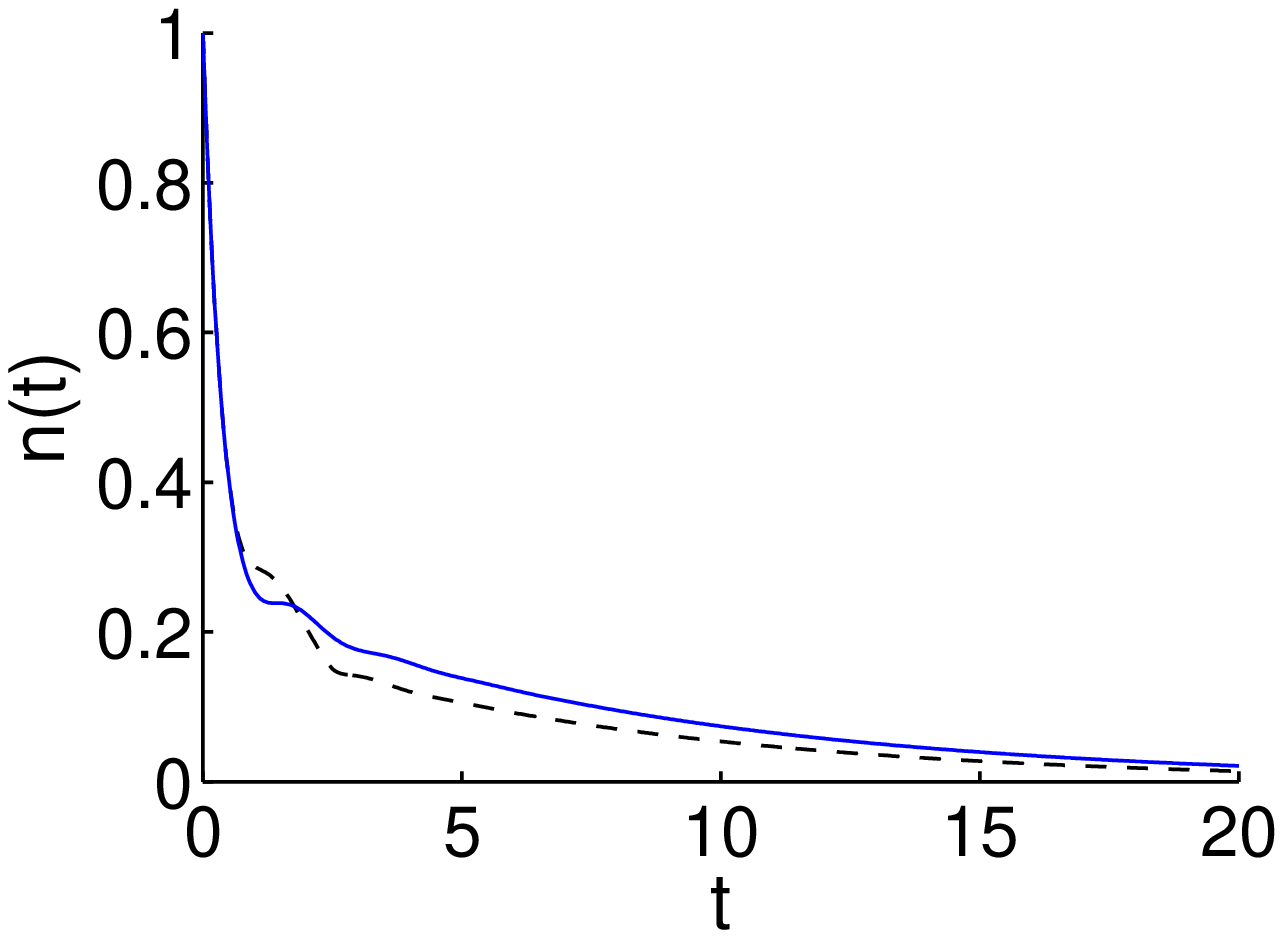}
\caption{\label{fig-MP_MF_dyn_nherm1} (Color online) Many-particle (dashed 
black lines) for  $N=20$ particles and mean-field (solid blue lines) 
dynamics for an initial state in mode $1$, that is, the north pole of the Bloch
sphere with $v=1,\, \epsilon=0$ and $\g=0.5$, $\gamma=0.1$ 
(left plots), $\g=2$, $\gamma=0.5$ (right plots). The upper plots show the dynamics of the angular momentum expectation values and the mean-field Bloch vector, respectively. The middle panels show the corresponding $z$-component 
and the lower plots the evolution of the overall probability $n(t)$.}
\end{figure}

\begin{figure}[tb]
\centering
\includegraphics[width=4cm]{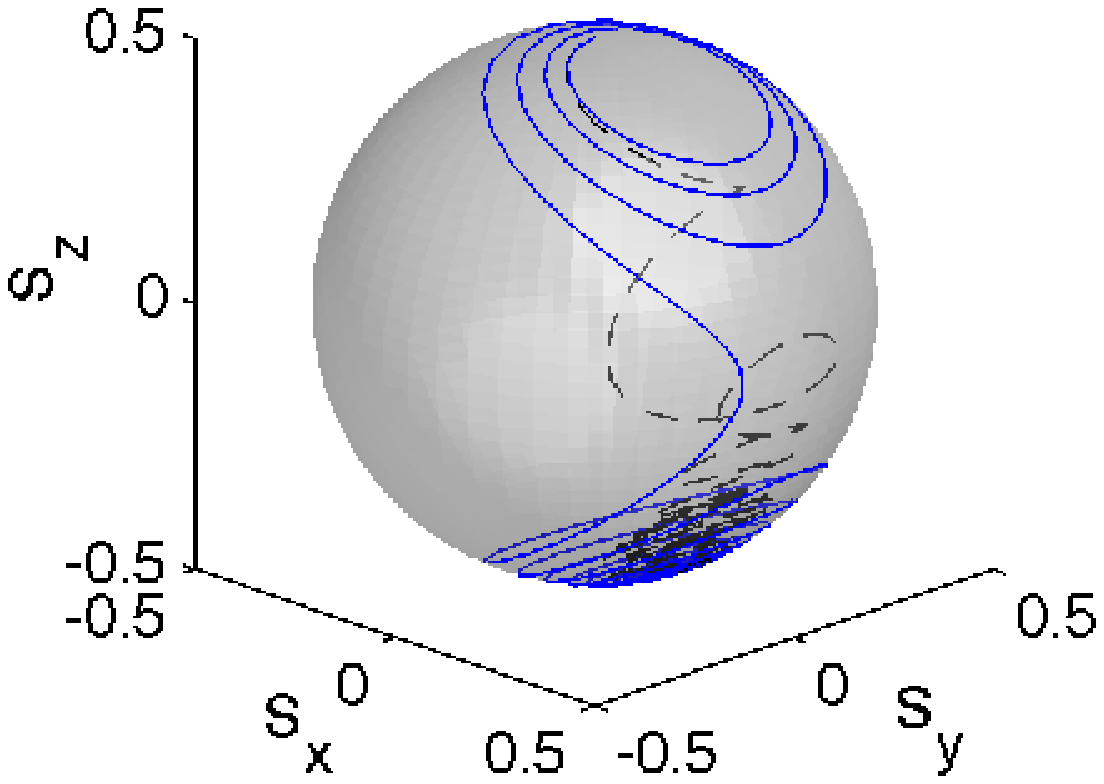}
\includegraphics[width=4cm]{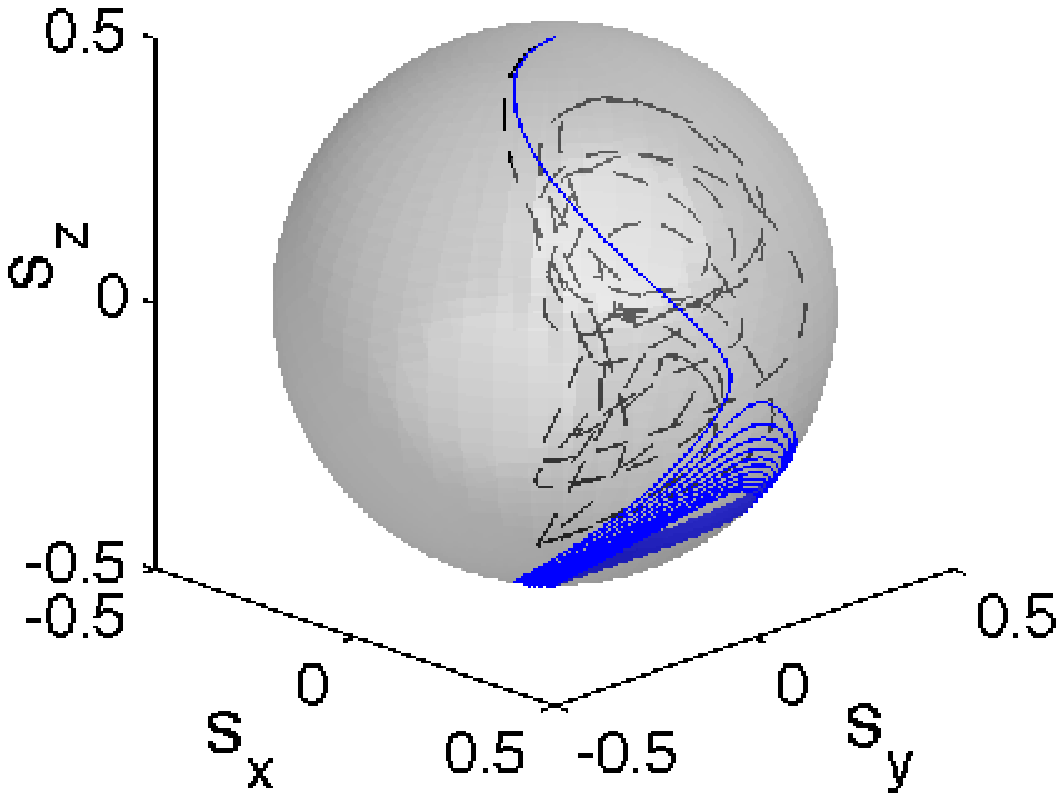}
\includegraphics[width=4cm]{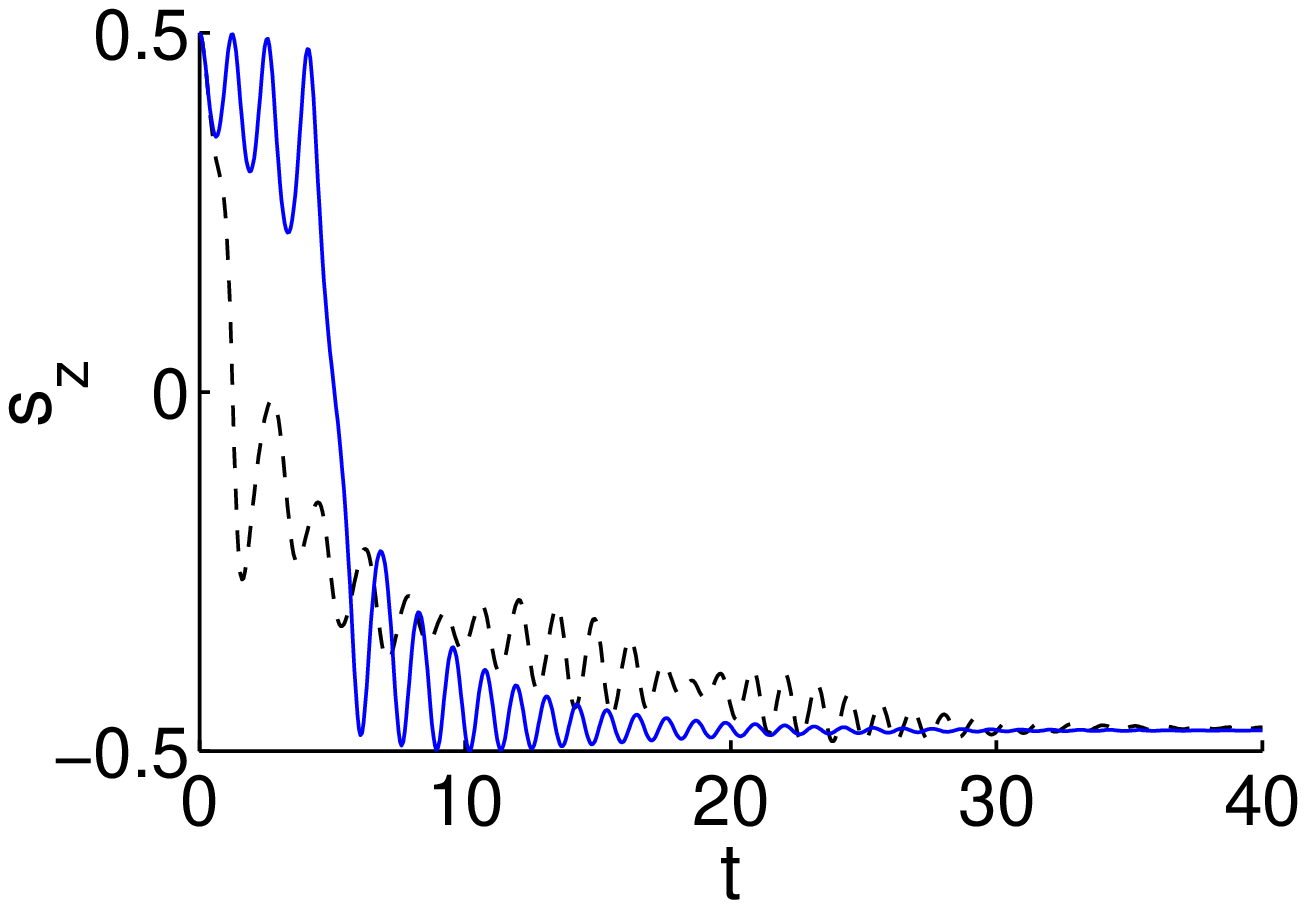}
\includegraphics[width=4cm]{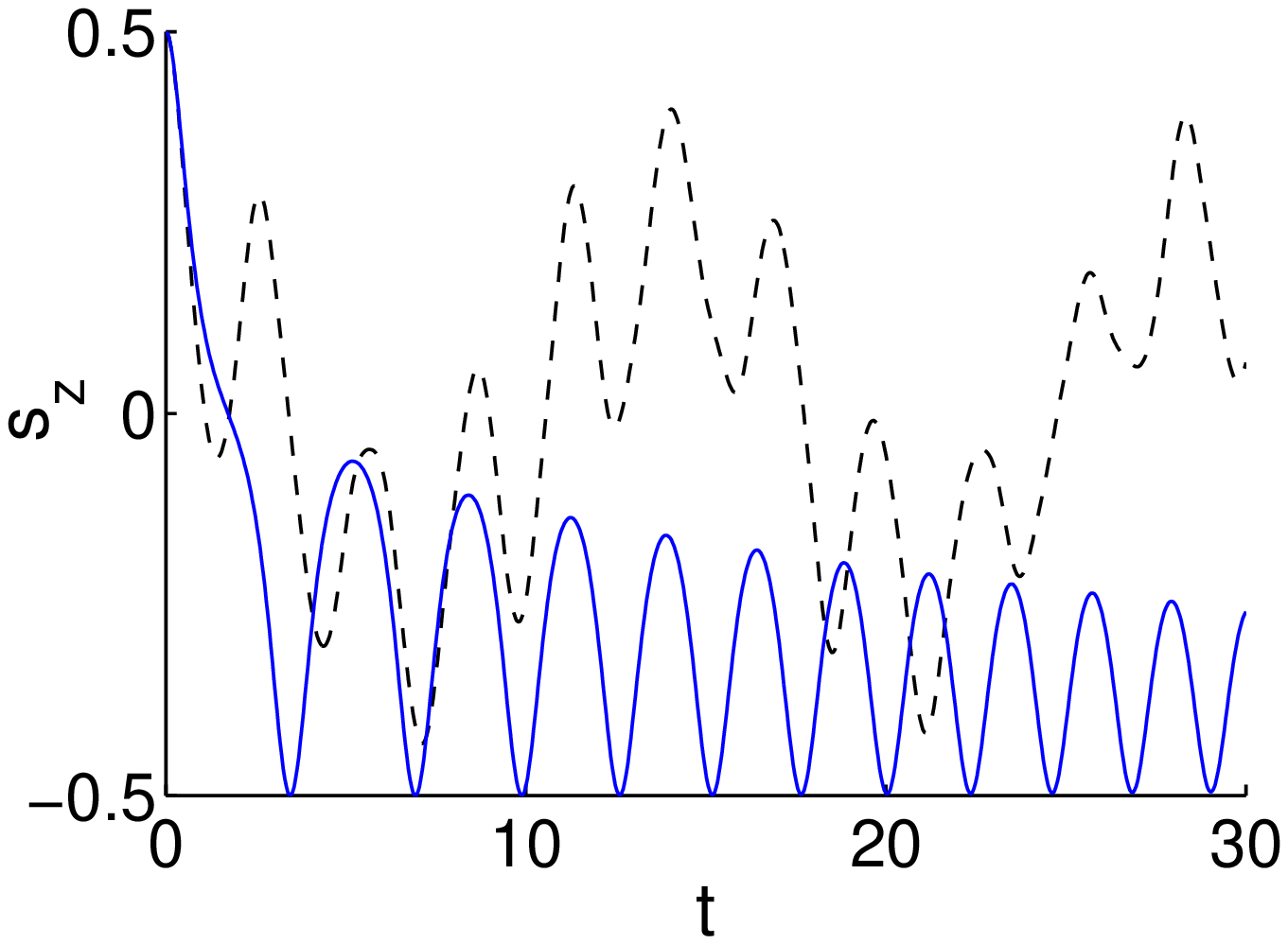}
\includegraphics[width=4cm]{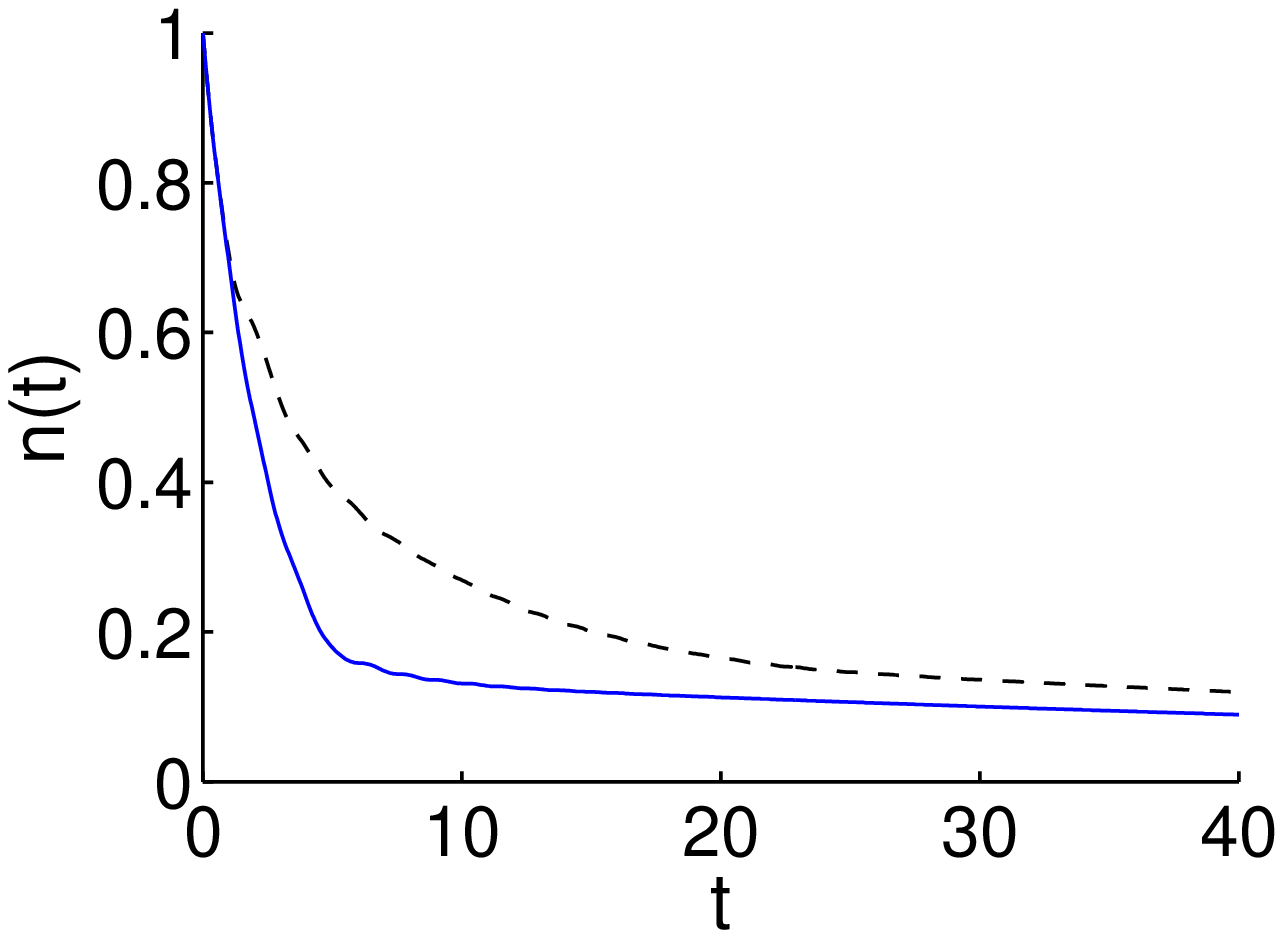}
\includegraphics[width=4cm]{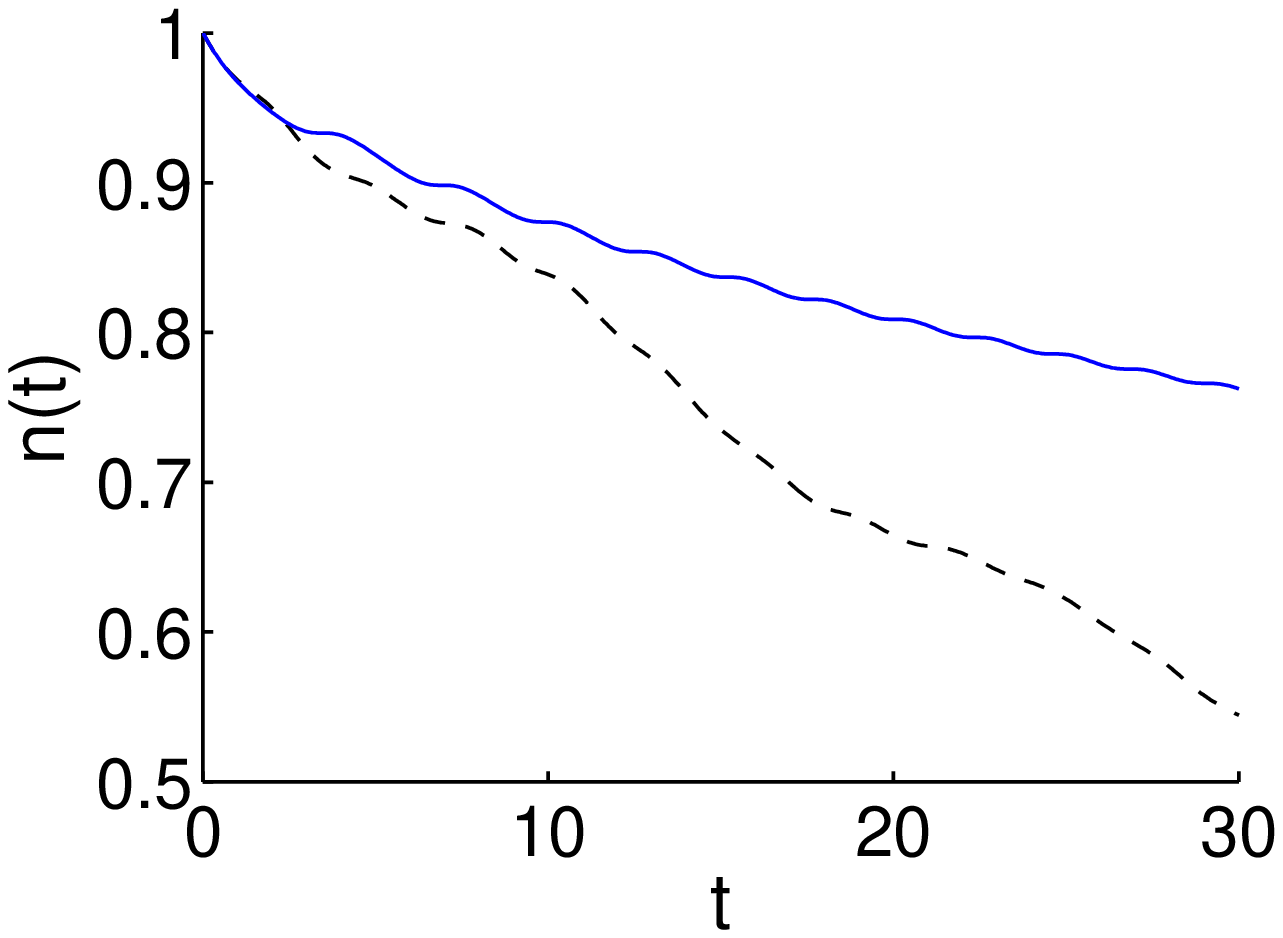}
\caption{\label{fig-MP_MF_dyn_nherm2} (Color online) Many-particle (dashed black lines) and mean-field (solid blue lines) dynamics for different parameters and an
initial state in mode $1$, that is, the north pole of the Bloch
sphere. As in Fig.~\ref{fig-MP_MF_dyn_nherm1} but for the parameters 
$v=1,\, \epsilon=0$ and $\g=3$, $\gamma=0.1$, and $N=20$ particles 
(left plots) and $\g=2$, $\gamma=0.01$, and $N=5$ particles (right plots). }
\end{figure}

To get some insights into the correspondence of the mean-field and many-particle 
dynamics, we show several examples in Figs. \ref{fig-MP_MF_dyn_nherm1} and
\ref{fig-MP_MF_dyn_nherm2}. The figures on the top show the dynamics of
the mean-field and the many-particle Bloch vector for a state
initially located at the north pole of the Bloch sphere. For a
better comparison we depict the dynamics of the corresponding 
$z$-component, that is, the relative population imbalance of the two
modes, in the plots in the middle. The resulting time dependence of the
overall probability is shown in the lower plots. Here we have
to compare the mean-field probability $n(t)$, given by the
normalization of the single particle wave function, to the
normalization of the many-particle wave function in the following way:
\begin{equation}
n(t)=|\psi_1(t)|^2+|\psi_2|^2 \longleftrightarrow \sqrt[N]{\langle \Psi(t)|\Psi(t)\rangle},
\end{equation}
thus accounting for different values of the particle number $N$.

Let us first focus on Fig.~\ref{fig-MP_MF_dyn_nherm1} where we
show the dynamics for $N=20$ particles. The left column shows an
example where both the interaction strength and the non-Hermiticity
are small ($\g=0.5$ and $\gamma=0.1$). The classical mean-field
dynamics shows the typical deformed Rabi oscillations. In the
many-particle system we observe the familiar breakdown behavior.
Numerical results for a longer propagation suggest that the revival
phenomena are strongly suppressed by the non-Hermiticity. The right
column shows the dynamics for a stronger interaction and a stronger 
decay ($\g=2$ and $\gamma=0.5$), i.e.~in the mean-field self-trapping 
region. The mean-field trajectory, commenced from the north pole, 
approaches the fixed point located at $\s_{z}=-0.433$. The full 
many-particle system shows a very similar behavior. For both 
examples the many-particle survival probability, depicted in the 
lower panel, is also reproduced by the mean-field approximation. 
In the regime of strong interaction, we can also observe more 
complicated behavior, including phenomena related to a 
many-particle tunneling from one self-trapping state to the other. 
This is illustrated in Fig.~\ref{fig-MP_MF_dyn_nherm2} where 
we plot the dynamics for large values of the interaction strength 
and comparatively small values of
$\gamma$ for an initial state at the north pole. The left column
shows an example with $N=20$ particles for the parameters $\g=3$ and
$\gamma=0.1$. One clearly observes a tunneling of the many-particle
dynamics between the mean-field stationary states. However, due to
the fact that the stationary state on the south pole of the sphere
is the sink of the mean-field dynamics, the latter approaches the
southern fixed point as well. The right column shows a similar
example with only $N=5$ particles for a slightly smaller interaction
strength $\g=2$ and a very small decay $\gamma=0.01$, to make the
tunneling process apparent. This superimposed many-particle
effect induces a clear mismatch into the correspondence of the
survival-probability evolution, which is illustrated in the lower
panels.

\begin{figure}[tb]
\centering
\includegraphics[width=4cm]{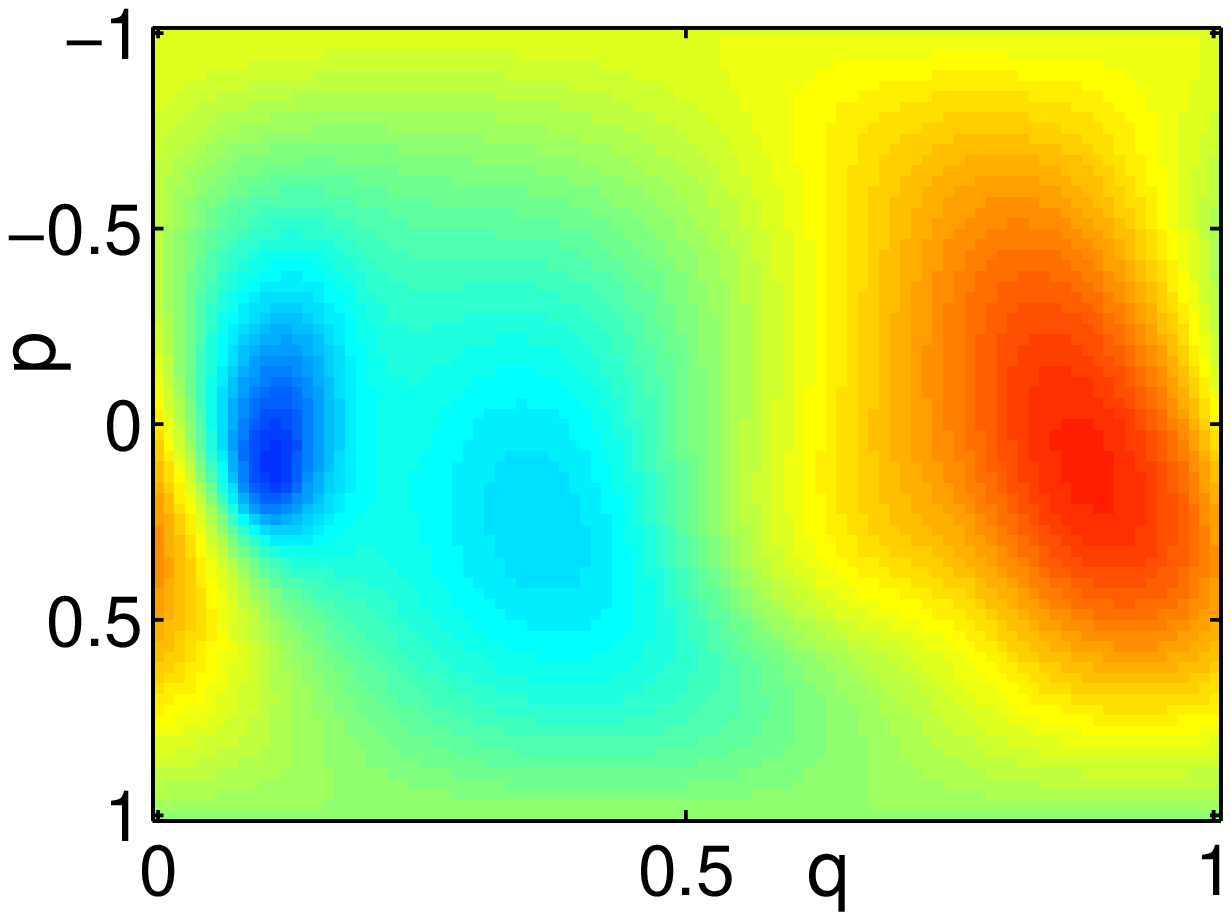}
\includegraphics[width=4cm]{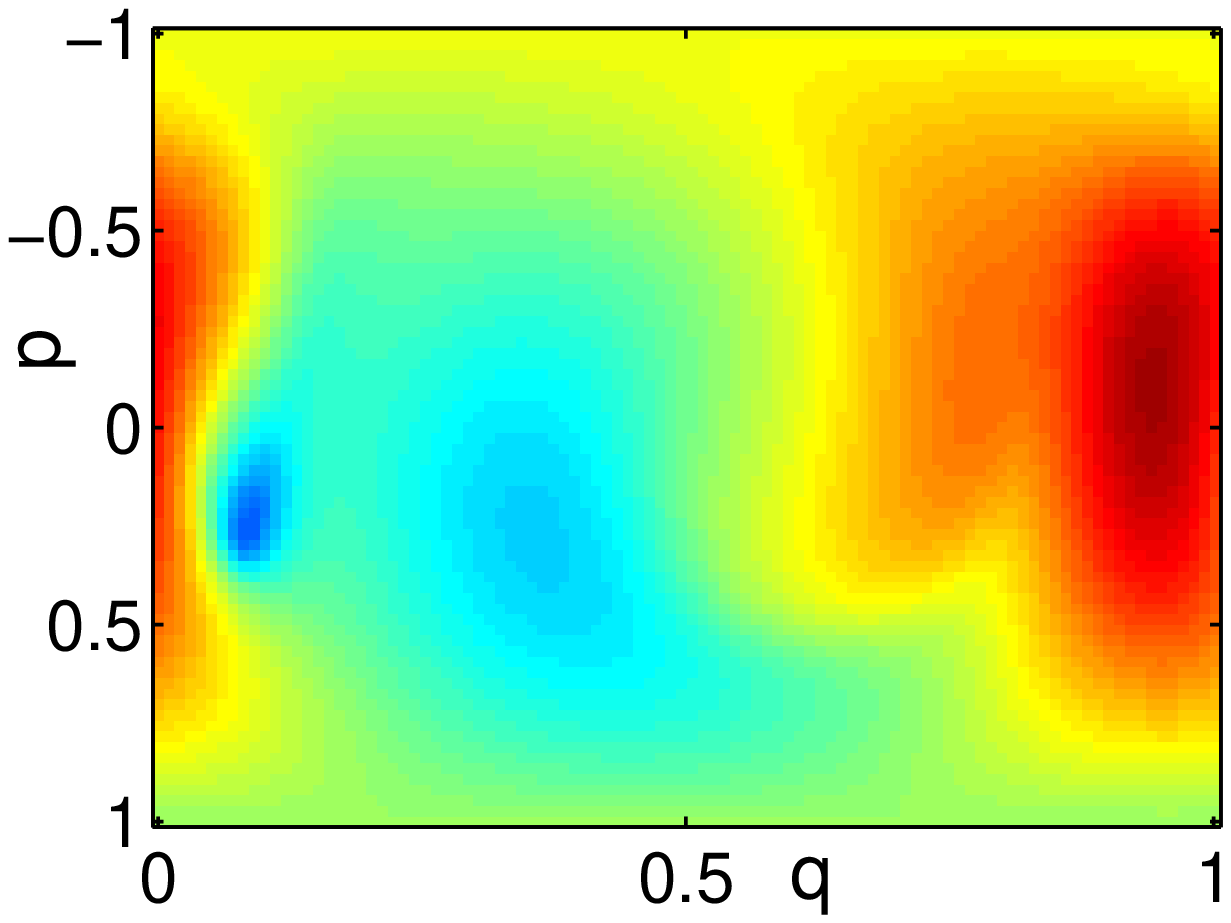}
\includegraphics[width=4cm]{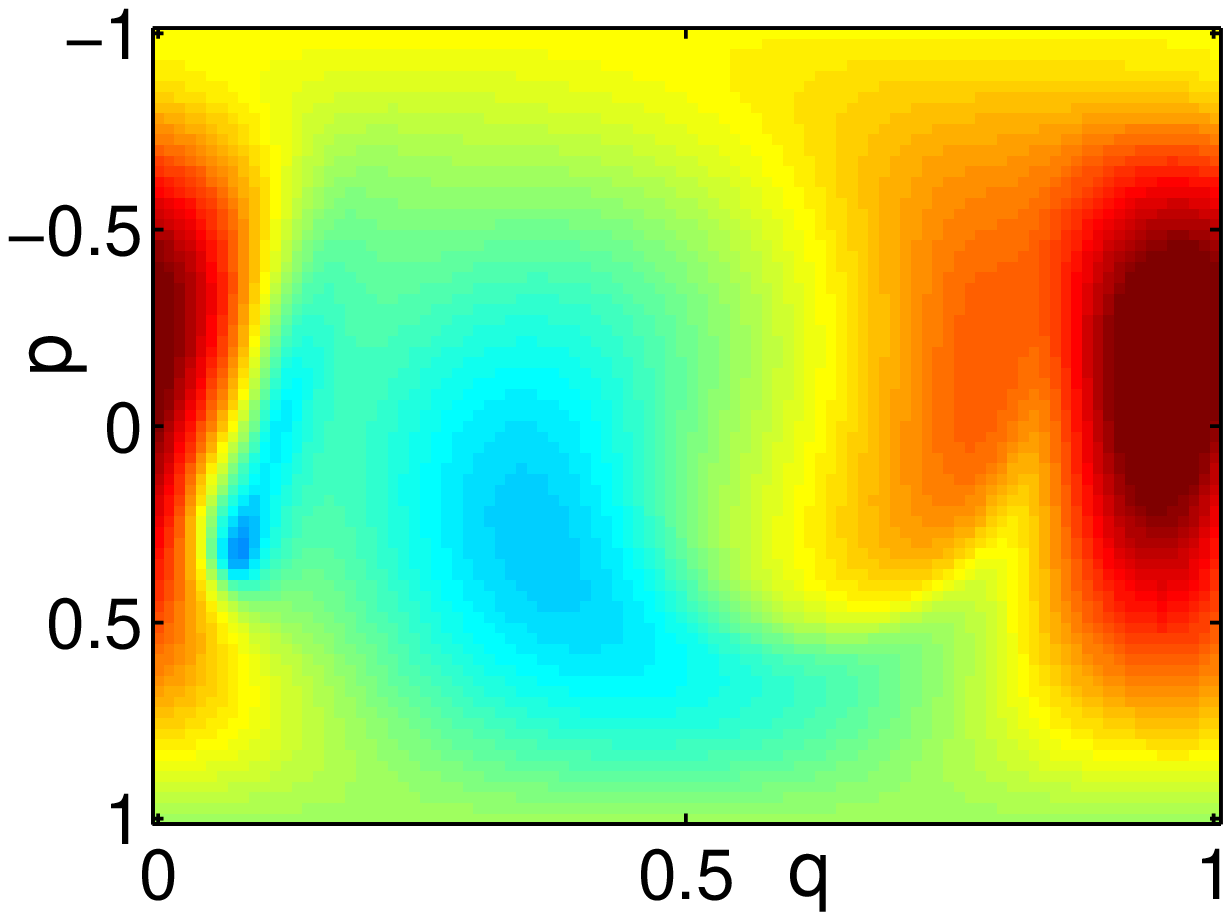}
\includegraphics[width=4cm]{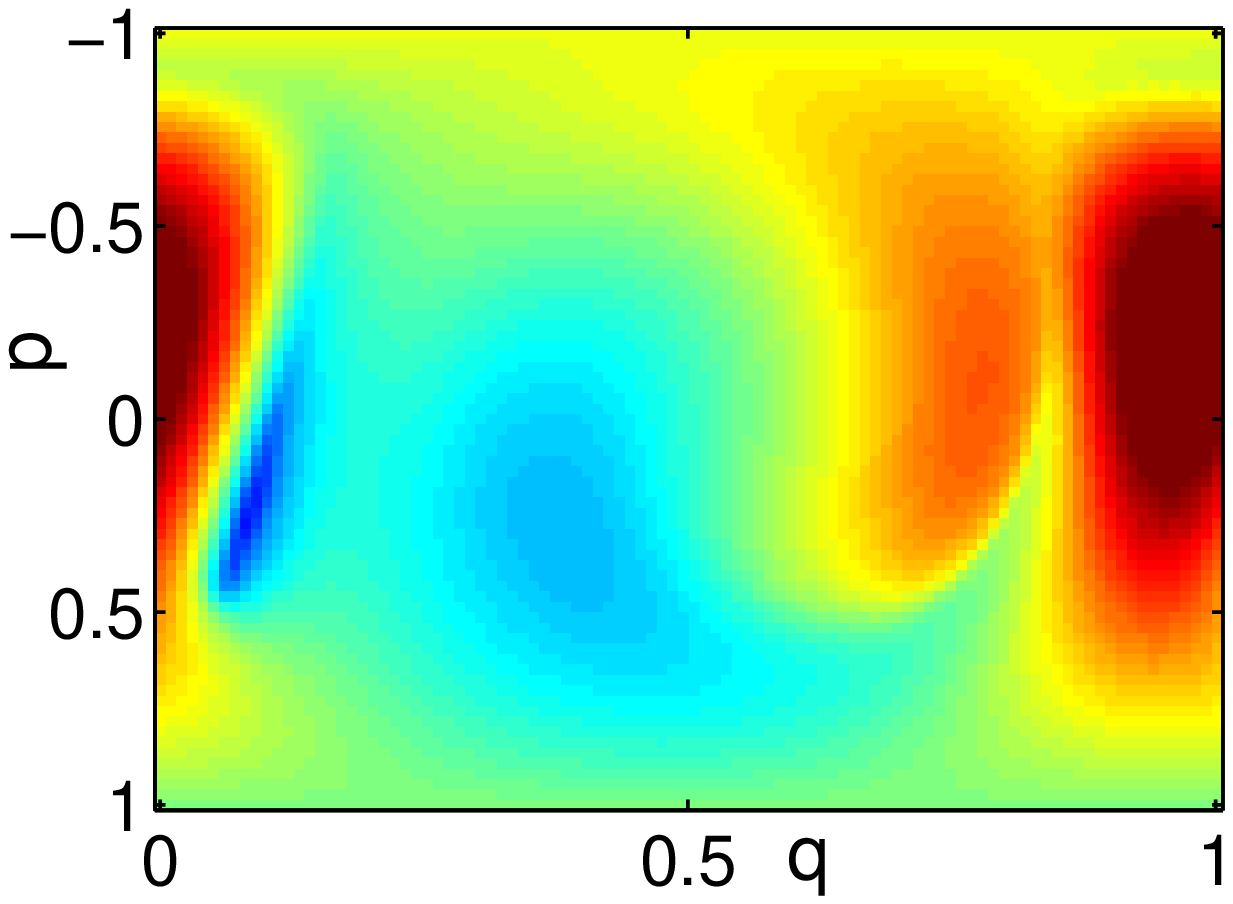}
\caption{\label{fig-MP-decay-bloch} Half life time of the rescaled 
normalization $\sqrt[N]{\langle \Psi(t)|\Psi(t)\rangle}$ of the many-particle wave function for $\epsilon=0$ for $v=1$, 
$\gamma=0.1$, $g=1/N$ and different particle numbers (from left to right and top to bottom: $N=5,\, 10,\, 15,\, 30$).}
\end{figure}

The approach to the mean-field limit with increasing particle number 
can be illustrated by comparison of the half life time of the normalization 
as a function of the initial conditions for different particle numbers. 
In Fig.~\ref{fig-MP-decay-bloch} we show the half life time as a function of 
the initial position on the Bloch sphere for $\gamma=0.1$ and $g=1$ and 
different particle numbers. The corresponding mean-field behavior 
is depicted in the right plot in the middle row of Fig.~\ref{fig-MF-decay-bloch} 
with the same colorscale. It can be nicely seen how the 
mean-field features become more pronounced with increasing particle number. 

The presented results give a first impression on the intricate correspondence 
of the full many-particle description and the mean-field approximation 
for this non-Hermitian system. 
Further investigations of this correspondence and in particular the manner 
in which the mean-field limit is approached are promising topics for future investigations. 
 
\section{Summary and Outlook}
We have studied the dynamics of a non-Hermitian two-mode 
Bose-Hubbard system and a related $\cP\cT$-symmetric model.
We have derived a non-Hermitian mean-field approximation,
which can be expressed in a generalized canonical form,
including a metric gradient flow \cite{09nhclass}, and
demonstrated the close correspondence of the damped 
(pseudo)classical motion in this mean-field description and 
the quantum many-particle evolution. In particular, we have 
analyzed the fixed point structure of the mean-field 
dynamics and its bifurcation arising when the system 
parameters are varied. This results in a rich variety
of phenomena in the many-particle dynamics, as for instance
breakdown and revival, and tunneling, which can be interpreted 
easily in terms of the underlying mean-field structure. 

In conclusion, the combined presence of interaction and
non-Hermiticity introduces a variety of interesting phenomena into
the correspondence between the many-particle dynamics and 
the mean-field description. The understanding of general 
quantum classical correspondence for non-Hermitian systems 
will ultimately require the development of new taylor-made methods, such as 
the Husimi-Schur phase space representation \cite{Kopp10} that was 
recently suggested in the context of open quantum maps. The simple 
model presented here provides an ideal testing ground for 
new methods for non-Hermitian systems. Future investigation 
and categorization of its behavior are thus a promising starting 
point for the formulation of a general framework for quantum classical 
correspondence in the presence of non-Hermiticity.

\section*{Acknowledgments}
Support from the Deutsche Forschungsgemeinschaft 
via the Graduiertenkolleg  ``Nichtlineare Optik und Ultrakurzzeitphysik'' 
is gratefully acknowledged. We thank Dorje Brody for helpful discussions and 
comments. 

\appendix
\section{The generalized canonical equations in terms of the 
coordinates $\varphi_j$}
\label{chap-metric}
The nonlinear complex Schr\"odinger equation \rf{nlnhGPn} can also be 
expressed in terms of a generalized canonical equation of motion, 
in the complex form 
\begin{equation}
\label{eqn_can_phi}
\rmi\left(\begin{array}{c}\!\!
\dot \varphi\!\\
\!\!\dot \varphi^*\!\!\!\!\end{array} \right)= \Omega^{-1}\vec{\nabla}H- \rmi
G^{-1}\vec{\nabla}\Gamma,
\end{equation}
where we have to pick two canonical conjugate variables 
$\varphi$ and $\varphi^*$ from the four variables  
$\varphi_1,\varphi_1^*,\varphi_2,\varphi_2^*$. 
Although the dynamics is apparently governed by all four variables the
normalization is fixed and the dynamics is independent of the 
global phase. Therefore, we have only two independent variables which we can
choose out of the original four. It is convenient to choose
$\varphi_1$ and $\varphi_1^*$ rather than $\varphi_1$ and
$\varphi_2$. They are connected to the coordinates $p,q$ via
\begin{eqnarray}
\varphi_1=\sqrt{\frac{p+1}{2}}\rme^{-2\rmi q},\quad \varphi_1^*=\sqrt{\frac{p+1}{2}}\rme^{2\rmi q}.
\end{eqnarray}
The equation of motion for the other variables are then implicitly
provided. With the choice $\varphi_1$ and $\varphi_1^*$ for the
independent variables we automatically demanded $\varphi_2$ to be
real and fulfill the normalization condition
$\varphi_2=\varphi_2^*=\sqrt{1-\varphi_1^*\varphi_1}$.
The symplectic matrix is the familiar one and for 
the inverse of the K\"ahler metric we find:
\begin{equation}\label{eqn-metric-inv-psi}
\left(G^{(\varphi_1,\,\varphi_1^*)}\right)^{-1}=\left(\begin{array}{cc}
 \frac{\varphi_1^{2}(|\varphi_1|^2-2)}{2(1-|\varphi_1|^2)} &  \frac{2-2|\varphi_1|^2+|\varphi_1|^4} {2(1-|\varphi_1|^2)}  \\[2mm]
 \frac{2-2|\varphi_1|^2+|\varphi_1|^4}{2(1-|\varphi_1|^2)} &  \frac{\varphi_1^{*2}(|\varphi_1|^2-2)}{2(1-|\varphi_1|^2)}
                          \end{array}\right).
\end{equation}
The equations of motion for $\varphi_1$ and $\varphi_1^*$ can then be found from  \rf{eqn_can_phi}, where $H$ and $\Gamma$ are given by the real and imaginary parts of the Hamiltonian function for the non-Hermitian and nonlinear two-level system expressed in terms of $\varphi_1$ and $\varphi_1^*$:
\begin{equation}
\cH=(\epsilon\!-\!\rmi\gamma)(\varphi_1^*\varphi_1\!-\!1)\!+\!v\sqrt{1\!-\!\varphi_1^*\varphi_1}(\varphi_1^*+\varphi_1)\!+\!\frac{g}{2}(\varphi_1^*\varphi_1\!-\!1)^2.
\end{equation}
The equation of motion for $\varphi_2$ can then be deduced from 
the dynamics of $\varphi_1$ via
\begin{eqnarray}
\dot{\varphi}_2=-\frac{\dot{\varphi}_1\varphi_1^*+\varphi_1\dot{\varphi}_1^*}{2\sqrt{1-\varphi_1\varphi_1^*}}.
\end{eqnarray}
The dynamics thus obtained is equivalent to the 
non-Hermitian (nonlinear) Schr\"odinger equation \rf{nlnhGPn}
up to a global phase.

\end{document}